\documentclass[superscriptaddress,aps,prd,twocolumn,showpacs,floatfix,nofootinbib,longbibliography, superscriptaddress,showkeys]{revtex4-2}
\DeclareUnicodeCharacter{02BC}{\textendash}
\usepackage[utf8]{inputenc}
\usepackage[T1]{fontenc}
\usepackage{amsmath,amssymb,amsthm}
\usepackage{easybmat}
\usepackage[colorlinks=true,citecolor=blue,urlcolor=blue, linkcolor= magenta]{hyperref}
\usepackage[pdftex]{graphicx}
\usepackage{times, txfonts}
\usepackage{braket}
\usepackage{color}
\usepackage{ragged2e}
\usepackage{booktabs}
\usepackage{epstopdf}
\usepackage{gensymb}
\usepackage{natbib}
\usepackage{subcaption}
\usepackage{float}
\newcommand{\be}{\begin{equation}}
	\newcommand{\ee}{\end{equation}}
\newcommand{\ba}{\begin{eqnarray}}
	\newcommand{\ea}{\end{eqnarray}}

\usepackage{tikz,xcolor,hyperref}

\definecolor{lime}{HTML}{A6CE39}
\DeclareRobustCommand{\orcidicon}{\hspace{-4pt}
	\begin{tikzpicture}
		\draw[lime, fill=lime] (0,0) 
		circle [radius=0.16] 
		node[white] {\hspace{0.1mm}{\fontfamily{qag}\selectfont \tiny ID}};
		\draw[white, fill=white] (-0.07,0.1) 
		circle [radius=0.01];
	\end{tikzpicture}
	\hspace{-3.2mm}
}

\foreach \x in {A, ..., Z}{\expandafter\xdef\csname 
	orcid\x\endcsname{\noexpand\href{https://orcid.org/\csname orcidauthor\x\endcsname}
		{\noexpand\orcidicon}}
}

\begin{document}

\title{Quantum speed limit time for bipartite entanglement in neutrino oscillations in matter with nonstandard interactions}
 \author{Abhishek Kumar Jha\orcidA{}}\email{abhiecc.jha@gmail.com}
\affiliation{Department of Physics, Indian Institute of Science, Bangalore 560012, India}
\affiliation{School of Technology Management and Engineering, SVKM’s Narsee Monjee Institute of Management Studies (NMIMS) Deemed-to-be-University, Hyderabad 509301, India}
	\author{Lekhashri Konwar\orcidB{}}\email{konwar.3@iitj.ac.in}
    \affiliation{Department of Physics, Indian Institute of Technology Jodhpur, Jodhpur 342030, India}
\author{Rukmani Mohanta\orcidC{}}\email{rmsp@uohyd.ernet.in}
\affiliation{School of Physics, University of Hyderabad, Hyderabad 500046, India}

\begin{abstract}
In the three-flavor neutrino oscillation framework, we investigate the transition probabilities of an initial muon neutrino flavor state in the presence of nonstandard interactions (NSIs) characterized by complex off-diagonal ($|\epsilon_{\alpha\beta}|e^{i\phi_{\alpha\beta}}$) and diagonal parameters ($|\epsilon_{\alpha\alpha}-\epsilon_{\beta\beta}|$), including a $CP$-violating phase and a constant matter potential, under both normal (NO) and inverted mass ordering (IO) scenarios. Within these scenarios and through the lens of mode entanglement, bipartite entanglement measures such as entanglement entropy and capacity of entanglement are quantified in terms of the transition probabilities, which can be measured in neutrino oscillation experiments. Using these two bipartite entanglement measures, we further explore the quantum speed limit (QSL) time, which describes how rapidly bipartite entanglement evolves during neutrino oscillations. We illustrate our results using the baseline lengths and energies corresponding to ongoing long-baseline accelerator neutrino experiments, such as T2K, NO$\nu$A, and the upcoming DUNE experiment. In the presence of a $CP$-violating phase and a constant matter potential, both with and without NSI effects, we compare the QSL time behavior for bipartite entanglement in neutrino oscillations for NO and IO. The most pronounced discrepancies in the QSL time for bipartite entanglement arise from the off-diagonal NSI parameter $\epsilon_{\mu\tau}$ across both the NO and IO scenarios. We emphasize that among all the experiments considered, NO$\nu$A and DUNE exhibit a rapid suppression of bipartite entanglement in neutrino oscillations in the standard oscillation scenario with NO at the end of their baseline lengths for the corresponding best-fit value of $CP$-violating phase. Our results hint at a possible imprint of new physics in neutrino oscillations.
\end{abstract}

\maketitle
\flushbottom

\section{Introduction}
Neutrinos are members of the lepton family that participate only through weak interactions \cite{Giunti:2007ry}. They are electrically neutral, typically ultrarelativistic in nature, and exist in three flavors: electron neutrino ($\nu_e$), muon neutrino ($\nu_\mu$), and tau neutrino ($\nu_\tau$). Neutrino oscillation is a purely quantum mechanical phenomenon in which a neutrino produced in one flavor acquires a nonzero probability of being observed in another as it propagates over time. This oscillation behavior arises from the fact that neutrinos possess distinct, nonzero masses. The idea of neutrino flavor oscillations was initially proposed by  Pontecorvo \cite{Pontecorvo:1957cp,Pontecorvo:1957qd,Pontecorvo:1967fh,Bilenky:1978nj,Bilenky:2004xm} and later formalized by Maki $et \, al.$, which led to the development of the Pontecorvo–Maki–Nakagawa–Sakata (PMNS) matrix \cite{maki1962remarks,Giganti:2017fhf}. This unitary transformation matrix describes how flavor states can be related to the mass eigenstates. A quantum superposition of the three nondegenerate mass eigenstates ($\nu_1$, $\nu_2$, $\nu_3$) forms the initial flavor state. As these mass eigenstates propagate in time, their relative phases evolve, causing a flavor state to propagate into a coherent superposition of different flavor bases, thereby producing the observed oscillation pattern in neutrino experiments \cite{Super-Kamiokande:1998kpq,kajita2016nobel,SNO:2002tuh}. 
Over the past several decades, studies of neutrino oscillations for the initial muon-flavor ($\nu_\mu$) neutrino state have been central to numerous ongoing accelerator-based long-baseline experiments, including T2K \cite{Abe_2023}, NO$\nu$A \cite{Acero_2022}, and others \cite{Abi:2020wmh,DiLodovico:2015nfq,Kudryavtsev_2016}. These experiments provide key insights into various neutrino oscillation parameters, including neutrino masses-squared differences, mixing angles, charge-conjugation and parity ($CP$) symmetry violation \cite{T2K:2019bcf,Minakata:1998bf}, and other issues such as matter effect, normal mass ordering (NO) and inverted mass ordering (IO) scenarios and possible physics beyond the standard model (BSM) \cite{DUNE:2020ypp}. In the current precision era of neutrino physics, such experiments offer a powerful platform to explore BSM, such as nonstandard neutrino interactions (NSIs).

NSIs provide a well-motivated framework to capture possible subleading effects of BSM in neutrino propagation through matter. A model-independent framework for investigating the effects of NSI in neutrino oscillations has been developed and discussed in \cite{Ohlsson:2012kf,Farzan:2017xzy,Miranda:2015dra}. Within the standard three-flavor framework, T2K and NO$\nu$A exhibit a mismatch in the preferred value of the $CP$ phase $\delta_{CP}$ in NO. While this discrepancy could arise from statistical fluctuations or unknown systematic effects, it may also point to BSM. Notably, T2K and NO$\nu$A have different sensitivities to matter effects due to their distinct baselines, suggested that NSI may provide a natural explanation. This discrepancy in the extracted value of the $CP$ phase $\delta_{CP}$ can be alleviated by introducing complex, flavor-changing neutral-current NSIs, in the presence of which both experiments converge to a common value in NO \cite{Chatterjee:2020kkm,Chatterjee:2024kbn}.

In this work, we investigate the quantum superposition characteristic of the time evolution of the initial muon-flavor neutrino state during three-flavor neutrino oscillations in matter, considering the role of quantum entanglement, and further examine how NSIs can modify this quantum behavior.

The weakly interacting nature of neutrinos allows their beams to maintain coherence over long distances, which can have significant implications for quantum information theory. Entanglement and coherence, as intrinsic consequences of quantum superposition \cite{Nielsen_Chuang_2010}, thus naturally motivate their study in neutrino oscillations. Entanglement properties in two- and three-flavor neutrino oscillations using the plane-wave approach have been studied through the lens of "mode entanglement" or "occupation number entanglement" \cite{Blasone_2009,BLASONE2013320,PhysRevD.77.096002,Banerjee:2015mha,Alok:2014gya,Dixit:2017ron,Dixit:2018kev,Dixit:2019swl,Dixit:2020ize,KumarJha:2020pke,Yadav:2022grk,Li:2022mus,wang2023monogamy,Bittencourt:2023asd,Blasone:2023qqf}. The time evolution of an initial muon-flavor neutrino state in vacuum has been investigated using various bipartite and tripartite entanglement measures \cite{KumarJha:2020pke}, revealing that three-flavor neutrino oscillations give rise to genuine tripartite entanglement \cite{Li:2022mus}. A detailed study of the impact of NSI on tripartite entanglement in three-flavor neutrino oscillations has been carried out \cite{Konwar:2024nrd,Konwar:2025ipv}, wherein  various entanglement measures, including entanglement of formation, concurrence, and negativity were studied for both reactor and accelerator-based experimental configurations. Tripartite entanglement measures have also been used to probe neutrino oscillation parameters \cite{Quinta:2022sgq,Banerjee:2024lih}.
Additionally, tools from quantum resource theory have been applied to quantify quantum coherence in experimentally observed neutrino oscillations, demonstrating that neutrinos, like photons, can exhibit quantum properties over macroscopic distances \cite{PhysRevA.98.050302,Ming:2020nyc}. Previous work \cite{Blasone_2009,BLASONE2013320,PhysRevD.77.096002,Banerjee:2015mha,Alok:2014gya,Dixit:2017ron,Dixit:2018kev,Dixit:2019swl,Dixit:2020ize,KumarJha:2020pke,Yadav:2022grk,Li:2022mus,wang2023monogamy,Bittencourt:2023asd,Blasone:2023qqf,Konwar:2024nrd,Konwar:2025ipv,Quinta:2022sgq,Banerjee:2024lih,PhysRevA.98.050302,Ming:2020nyc} has primarily focused on entanglement in neutrino oscillations within qubit systems; however, extensions to higher-dimensional systems, such as qutrits, have also been explored \cite{Jha:2022yik,Alok:2025qqr}. Furthermore, entanglement in neutrino oscillations has been studied using wave-packet \cite{Blasone_2015,Blasone:2021cau,Blasone:2022ete,Ravari:2022yfd,Ettefaghi:2023zsh} and quantum-field–theoretic approaches \cite{Blasone:2014jea,Blasone:2014cub}. These studies have emphasized that multimode entanglement persists in neutrino oscillations. 

Beyond entanglement, several other facets of quantum correlations have been studied in neutrino oscillations, most notably nonlocality \cite{Brunner:2013est} and quantum steering \cite{Uola:2020kps}. Investigations of nonlocality, typically using Bell-type inequalities, test the incompatibility of neutrino oscillation statistics with local hidden variable models. Quantum steering, on the other hand, refers to a scenario in which measurements performed on one neutrino subsystem can nonclassically influence the possible states of another subsystem, reflecting a form of quantum correlation that cannot be explained by classical physics \cite{Konwar:2024pkh}. Leggett–Garg
Inequalities are a measure of temporal quantum correlation analogous to Bell inequalities that concern spatial quantum correlation and measures the correlation in neutrino oscillations at various instances of time  
\cite{PhysRevLett.117.050402,PhysRevD.99.095001,NAIKOO2020114872,Shafaq:2020sqo,Sarkar:2020vob,Blasone:2022iwf,Chattopadhyay:2023xwr,Groth:2025gtf}. In recent work \cite{Konwar:2024nwc} studied Leggett–Garg-type Inequalities violations in three-flavor neutrino oscillations and showed that complex NSIs, particularly $\epsilon_{e\tau}$, can enhance violations, with DUNE offering a distinctive signature, highlighting temporal quantum correlations as probes of new physics. These investigations indicate that neutrino oscillations can be analyzed using the quantum-information–theoretic technique. 

Typically, in quantum information theory, bipartite entanglement represents the simplest form of multipartite entanglement, which characterizes correlations between two subsystems of a composite quantum system. For instance, by mapping neutrino flavor states onto two- and three-qubit mode states, neutrinos in Bell-like superposition states can be realized as bipartite mode (flavor) entangled states \cite{KumarJha:2020pke,Jha:2021itm}. In the present work, we treat a single-neutrino state as an effective bipartite quantum system. In the relativistic limit, the evolution of a flavor neutrino can be described within a two-dimensional Hilbert space. Previous studies have shown that the Hilbert spaces associated with flavor and mass eigenstates are orthogonal and that the flavor vacuum possesses a condensate structure involving neutrino–antineutrino pairs \cite{Blasone:1995zc}. Such vacuum effects become important in the nonrelativistic regime and lead to corrections to the standard oscillation probabilities. In the relativistic domain, where these corrections are negligible, the Hilbert space of flavor neutrinos may be viewed as the tensor product of the Hilbert spaces corresponding to the mass eigenstates (two or three). A complementary viewpoint, discussed in Ref.\,\cite{Blasone:2022quj}, regards particle states as excitations of quantum fields. Since free fields with distinct masses are known to generate unitarily inequivalent Hilbert spaces \cite{Haag:1955ev}, states with different masses cannot be consistently described within a single Hilbert space. Consequently, their description naturally involves a tensor-product structure, and arbitrary superpositions of such states generally exhibit entanglement. By introducing occupation-number states and identifying them with a multiqubit basis, a single flavor neutrino can therefore be interpreted as a multipartite mode-entangled state \cite{PhysRevD.77.096002}. This interpretation is further supported by the extensive body of work devoted to quantum correlations in neutrino oscillations \cite{Blasone_2009,BLASONE2013320,PhysRevD.77.096002,Banerjee:2015mha,Alok:2014gya,Dixit:2017ron,Dixit:2018kev,Dixit:2019swl,Dixit:2020ize,KumarJha:2020pke,Yadav:2022grk,Li:2022mus,wang2023monogamy,Bittencourt:2023asd,Blasone:2023qqf,Konwar:2024nrd,Konwar:2025ipv,Quinta:2022sgq,Banerjee:2024lih,PhysRevA.98.050302,Ming:2020nyc,Jha:2022yik,Alok:2025qqr,Blasone_2015,Blasone:2021cau,Blasone:2022ete,Ravari:2022yfd,Ettefaghi:2023zsh,Blasone:2014jea,Blasone:2014cub,Brunner:2013est,Uola:2020kps,Konwar:2024pkh,PhysRevLett.117.050402,PhysRevD.99.095001,Naikoo:2019eec,NAIKOO2020114872,Shafaq:2020sqo,Sarkar:2020vob,Blasone:2022iwf,Chattopadhyay:2023xwr,Groth:2025gtf,Konwar:2024nwc,Jha:2021itm,Bouri:2024kcl}, which has recently been reviewed in Ref.\,\cite{Blasone:2023qqf}. Collectively, these studies provide a strong theoretical foundation for treating a single-particle neutrino system as an effective bipartite quantum system. A fundamental question then concerns how rapidly bipartite mode (flavor) entanglement is generated or decays in three-flavor neutrino oscillation experiments. In this work, this issue is addressed using the concept of the quantum speed limit (QSL) time for bipartite entanglement measures as the primary analytical tool.

The notion of QSL time arises from the uncertainty relation between conjugate variables in quantum mechanics, which sets a fundamental lower bound on the time required for a quantum system to evolve from an initial to a final state \cite{PhysRev.34.163,PhysRevLett.113.260401,Mandelstam1991,MARGOLUS1998188,PhysRevLett.103.160502,Deffner_2013,Deffner:2017cxz,Thakuria_2024}. QSL time has been shown to hold for both driven closed systems \cite{Thakuria_2024} and open quantum systems \cite{PhysRevLett.111.010402,Baruah:2022zqu}. It has broad applications across many physical systems \cite{PhysRevA.85.052327,Aggarwal:2021xha,PhysRevLett.46.623,Paulson:2021jmi,Paulson:2022lih,Tiwari:2022kxh,Lloyd_2000,PhysRevLett.105.180402,PhysRevLett.103.240501,PhysRevA.82.022318,PhysRevResearch.2.023113,PhysRevA.106.042436,Liegener:2021zhs,Wei:2023jrx,Maleki:2019cqn} and has recently been applied in elementary particle systems, including neutral mesons \cite{Banerjee:2022ckm} and neutrinos \cite{Khan:2021kai}, within the ambit of an open quantum system. However, when neutrinos are treated as a closed quantum system, the QSL time in neutrino oscillations has also been investigated in both flat \cite{Bouri:2024kcl} and curved spacetime \cite{Jha:2024gdq,Jha:2024asl,Jha:2025ekn}. 

Moreover, investigations of QSL times in quantum evolution have also established fundamental limits on how quickly various entanglement measures can change \cite{Mohan:2021vfu,PhysRevA.107.052419,Das:2017nty}. In a closed quantum system, if the driving Hamiltonian is Hermitian and time independent, the dynamical evolution of the system is governed by a unitary transformation. Such a system can generate bipartite entanglement, with the entanglement entropy and the capacity of entanglement serving as key measures. The QSL time for bipartite entanglement in such a system is typically characterized using these two entanglement measures, together with the invariance of the Hamiltonian \cite{PhysRevA.106.042419}. This concept has recently been applied to the framework of two- and three-flavor neutrino oscillations, when the time evolution of an initial muon neutrino flavor state propagates in both vacuum and matter \cite{Bouri:2024kcl}. Their analysis has been performed in various accelerator-based long-baseline neutrino experimental setups for both NO and IO in the presence of a $CP$-violating phase. However, the study of the QSL for bipartite entanglement has not yet been conducted in neutrino oscillations with NSI effects.

In this work, we investigate the QSL time for bipartite entanglement in neutrino oscillations in matter considering the best-fit value of $CP$-violating phase and in the presence and absence of complex off-diagonal and diagonal NSI parameters.
We illustrate our results for the initial muon-flavor neutrino state in both NO and IO scenarios, using the fundamental oscillation parameters from NuFIT data \cite{Esteban:2024eli,NuFIT} as well as the baseline lengths and energies corresponding to ongoing long-baseline accelerator neutrino experiments such as T2K \cite{Abe_2023}, NO$\nu$A \cite{Acero_2022}, and the upcoming DUNE \cite{Abi:2020wmh} experiment. 

The organization of this manuscript is as follows. Section \,\ref{sec2} briefly reviews the concept of the QSL time for bipartite entanglement. In Sec.\,\ref{sec3}, we discuss neutrino oscillations in the presence of a constant matter potential, $CP$-violating phases, and NSI effects, considering the complex off-diagonal and diagonal NSI parameters one at a time. We examine the neutrino transition probabilities for both NO and IO scenarios, with and without NSI effects, with particular emphasis on the initial muon-flavor neutrino state. In Sec.\,\ref{sec4}, the QSL time for bipartite entanglement in neutrino oscillations under these scenarios is explored using bipartite entanglement measures, including the entanglement entropy and the capacity of entanglement. Finally, discussions and conclusions are presented in Sec.\,\ref{sec5}.

\section{QSL time for bipartite entanglement}  \label{sec2}
For a time-independent Hermitian driving Hamiltonian $\mathcal{H}$ ($\mathcal{H}^\dagger=\mathcal{H}$), the time evolution of an initial quantum state $\ket{\Psi}$ is governed (setting $\hbar=1$) by
\begin{eqnarray}
      i\frac{d\ket{\Psi}}{dt}=\mathcal{H}\ket{\Psi}\Rightarrow \ket{\Psi(t)}=e^{-i \mathcal{H} t}\ket{\Psi}=\mathcal{U}_t\ket{\Psi},
      \label{1}
\end{eqnarray}
   where the time evolution operator $\mathcal{U}_t$ is unitary, satisfying $\mathcal{U}_t^\dagger\mathcal{U}_t=I$. 
In a multipartite quantum system, a bipartite quantum state is defined by a composite Hilbert space ${H}_1\otimes{H}_2$ associated with two subsystems \cite{Nielsen_Chuang_2010}. In unitary dynamics, the global state is represented by a density matrix $\rho=\ket{\Psi}\bra{\Psi}$, which describes a pure state that satisfies $\rho^2=\rho$ (idempotent matrix) and $\mathrm{Tr}(\rho^2)=1$. Despite the purity of the global state the correlation between subsystems generally leads to mixed reduced states (by tracing over the other qubit), $\rho_1=\mathrm{Tr}_2(\rho)$ and $\rho_2=\mathrm{Tr}_1(\rho)$, characterized by $\mathrm{Tr}(\rho_1^2)<1$ and $\mathrm{Tr}(\rho_2^2)<1$. A pure bipartite quantum state exhibits entanglement if and only if its density matrix cannot be written in a product (or separable) state, $\rho \neq \rho_1 \otimes \rho_2$. Thus, $\rho$ encapsulates the full information of a nonclassical correlation present in a pure bipartite quantum state. Maximally entangled Bell states constitute paradigmatic examples of such pure bipartite quantum states.

 The entanglement entropy ($S_{EE}$) serves as a measure of entanglement for a pure bipartite quantum state, and is defined as the von Neumann entropy of the reduced state
\begin{equation}
S_{EE}=S(\rho_1)=-\mathrm{Tr}\left(\rho_1\log_2\rho_1\right),
\label{2}
\end{equation}
where $S_{EE}=0$ indicates that the initial state $\ket{\Psi}$ is unentangled (i.e, separable), $S_{EE}=1$ denotes a maximally entangled state, and $0<S_{EE}<1$ signifies a partially (or intermediate) entangled state. The QSL time for bipartite entanglement in the time-evolved state $\ket{\Psi(t)}$ is then given by \cite{PhysRevA.106.042419}
\begin{equation}
 \tau ^E_{\mathrm{QSL}} =
\frac{\left|S_{EE}(\tau)-S_{EE}(0)\right|}
{2\Delta \mathcal{H}\frac{1}{\tau}\int_0^\tau \sqrt{C_E(t)}\,dt} \hspace{0.2cm} \text{and}\hspace{0.2cm} \frac{\tau^E_{\text{QSL}}}{\tau}\leq 1
\label{3}
\end{equation}

is the QSL time bound ratio for bipartite entanglement, where $\tau$ is the propagation time of the initial state $\ket{\Psi}$. Here, the capacity of entanglement ($C_E$), defined as the variance of $S_{EE}$, takes the form
\begin{equation}
C_E=\sum_i \mathcal{\eta}_i \log_2^2\mathcal{\eta}_i - S_{EE}^2,
\label{4}
\end{equation}
with $\mathcal{\eta}_i$ denoting the eigenvalues of $\rho_1$ (or $\rho_2$). The variance of the Hamiltonian (or energy fluctuation) is given by 
   \begin{equation}
   \Delta{\mathcal{H}}= \sqrt{\langle   \mathcal{H}^2\rangle - \langle 
\mathcal{H}\rangle ^2}.
\label{5}
\end{equation}
The physical significance of ${T}^E_{\text{QSL}}$ can be understood by the following interpretation:
 
\begin{itemize}
    \item If the QSL time-bound ratio for $S_{EE}$ is  ${\tau^E_{\text{QSL}}}/{\tau}=1$, the bipartite entanglement in the time-evolved state cannot be speed up. In other words, the bipartite entanglement is already at its maximum speed for the given time-evolved state.
    \item If the QSL time-bound ratio for $S_{EE}$ is ${\tau^E_{\text{QSL}}}/{{\tau}}<1$, the bipartite entanglement in the time-evolved state could be speed up. Furthermore, the smaller ${{\tau}^E_{\text{QSL}}}/{\tau}$ is, the larger is the potential for bipartite entanglement to further speed up in the time-evolved state.
\end{itemize}

Equation~(\ref{3}) thus describes a fundamental bound on the rate at which bipartite entanglement can be generated or degraded during the unitary evolution of a pure bipartite quantum state $\ket{\Psi(t)}$.

\section{Neutrino oscillation in matter with NSI} \label{sec3} 
Neutrinos, while propagating through matter, undergo both charged-current (CC) and neutral-current (NC) interactions with the matter particles in the medium. These interactions modify the effective potential experienced by the neutrinos and can significantly influence the flavor oscillation pattern. Taking CC\footnote{Neutral-current (NC) interactions are flavor-independent and contribute only a common phase that does not affect neutrino oscillation probabilities; therefore, the NC term is ignored in $H_{SO}$ without loss of generality \cite{Giunti:2007ry}.} interactions into account, the Hamiltonian describing standard oscillations (SOs) is given by adding the matter-induced potential to the vacuum Hamiltonian \cite{Giunti:2007ry,Mikheyev:1985zog},
\begin{equation}\label{2.1}
    \mathcal{H}_{SO}= \mathcal{H}_{vac} + \mathcal{H}_{mat},
\end{equation} 
which, in the mass basis, takes the form
\begin{eqnarray}\label{2.2}
\mathcal{H}_{SO}&=&
\begin{pmatrix}
E_{1} & 0 & 0 \\ 
0 & E_{2}& 0 \\
0 & 0 & E_{3}
\end{pmatrix}+U^{\dagger } V_{\rm CC}\begin{pmatrix}
1&0 &0 \\ 
0& 0& 0\\
0 & 0 & 0
\end{pmatrix} U,
\end{eqnarray}
 where $E_{j} = \sqrt{p^{2}+m_{j}^{2} }$ ($j=1,2, 3$) are the energies of the neutrino mass eigenstates with masses $m_{j}$, assuming the same momentum $p$ for all eigenstates, $\ket{\nu_{j}}$. $U$ is the PMNS matrix, and the elements of the PMNS matrix are represented as,
 \begin{equation}\label{2.3}
\begin{pmatrix}
c_{12}c_{13} & s_{12}c_{13}&s_{13}e^{-\iota\delta_{CP}}\\ 
-s_{12}c_{23}-c_{12}s_{13}s_{23}e^{\iota\delta_{CP}}& c_{12}c_{23}-s_{12}s_{13}s_{23}e^{\iota\delta_{CP} }&c_{13}s_{23}\\ 
s_{12}s_{23}-c_{12}s_{13}c_{23}e^{\iota\delta_{CP} }& -c_{12}s_{23}-s_{12}s_{13}c_{23}e^{\iota\delta_{CP}}&c_{13}c_{23}
\end{pmatrix},
 \end{equation}
where $U^\dagger U=I\Longrightarrow U^\dagger=U^{-1}$. Here $s_{ij} = \sin \theta _{ij}$, $c_{ij} = \cos \theta _{ij}$ with $i, j= 1, 2, 3$ and $\delta_{CP}$ is the $CP$-violating phase. $V_{\rm CC}$ denotes the matter potential generated by the coherent forward scattering of electron neutrinos with electrons via CC interaction in matter and expressed as $V_{\rm CC}=\pm \sqrt{2}G_{F}N_{e}$. $G_{F}$ is the Fermi constant and $N_{e}$ is the electron number density. For neutrinos, the effective matter potential ($V_{CC}$) is positive, whereas for antineutrinos it is negative. In the present analysis, the matter potential is taken as \cite{Bouri:2024kcl} $V_{CC}$= $1.01  \times  10^{-13}$ eV.

Neutrinos, being promising windows to new physics, have motivated a wide range of searches BSM in oscillation experiments. One such possibility arises from NSI \cite{Ohlsson:2012kf,Farzan:2017xzy,Miranda:2015dra,Proceedings:2019qno,ESSnuSB:2025vsf,Kopp:2007ne}, an effective four-fermion interaction in the low-energy regime. Within this framework, neutrinos propagating through Earth matter can interact with NC-NSIs, which is parameterized by the 6-dimensional effective operators in the Lagrangian of the form

\begin{eqnarray}\label{2.5}
\mathcal{L}_{NSI}^{NC}=-2\sqrt{2}G_F\sum\limits_{\alpha, \beta, C} \epsilon_{\alpha\beta}^{f,C}(\bar{\nu}_\alpha \gamma^{\mu} P_{L} \nu_\beta)(\bar{f} \gamma_\mu P_{C} f).
\end{eqnarray}
The parameters $\epsilon_{\alpha\beta}^{f,C}$ represent the strength of the NC-NSI effects, where $\alpha$ and $\beta$ take on values $e$, $\mu$, and $\tau$, which correspond to the three neutrino flavors. $f$ corresponds to the fermions on the Earth matter, i.e.,  $e$ (electron), $u$ (up quark), and $d$ (down quark), with which neutrinos interact. $P_{C}$ with $C= L, R$ denote the left- or right-handed chirality of the interaction, respectively. Including both the SO contribution and the NC-NSI effects, the effective Hamiltonian in the mass basis governing neutrino propagation is defined as
\begin{equation}\label{2.6}
    \mathcal{H}_{tot}= \mathcal{H}_{SO}+ \mathcal{H}_{NSI},
\end{equation}
and can be written as \cite{Ohlsson:2012kf,Farzan:2017xzy,Miranda:2015dra,Konwar:2024nwc}
\begin{eqnarray}\label{2.7}
\mathcal{H}_{tot}&=&
\begin{pmatrix}
E_{1} & 0 & 0 \\ 
0 & E_{2}& 0 \\
0 & 0 & E_{3}
\end{pmatrix}+U^{\dagger } V_{\rm CC}\begin{pmatrix}
1+\epsilon_{ee} &\epsilon_{e\mu}  &\epsilon_{e\tau} \\ 
\epsilon_{ e\mu}^{*} & \epsilon_{\mu \mu} & \epsilon_{\mu \tau}\\
 \epsilon_{e \tau}^{*} & \epsilon_{\mu \tau }^{*} & \epsilon_{\tau \tau}
\end{pmatrix} U.
\end{eqnarray}
 The NSI parameters $\epsilon _{\alpha \beta }$ are given by
 \begin{equation}\label{2.8}
    \epsilon _{\alpha \beta }=\sum_{f=e,u,d}\frac{N_{f} }{N_{e}} \epsilon _{\alpha \beta }^{f},
\end{equation}
where $\epsilon _{\alpha \beta }^{f}$ are the NSI parameters, which describe the interaction strengths between neutrinos of flavors $\alpha$ and $\beta$ with fermions of type $f$, and $N_{f}$ is the number density of $f$-type fermions. The six independent matter NSI parameters related for propagation include the diagonal terms $\epsilon_{ee}$, $\epsilon_{\mu \mu}$ and $\epsilon_{\tau \tau}$ that are real and give flavor-conserving modifications to the effective matter potential, whereas the off-diagonal parameters $\epsilon_{e\mu}$, $\epsilon_{e\tau}$ and $\epsilon_{\mu \tau}$ in general, are complex and can be parametrized as $\epsilon_{\alpha\beta}=|\epsilon_{\alpha\beta}|e^{i\phi_{\alpha\beta}}$ ($\alpha\neq\beta$).

\begin{table}[htb!] 
\centering
    \begin{tabular}{|c | c| c|}
        \hline
     Input parameters & Mass ordering & Best fit$\pm 1\sigma$\\
    \hline
    \hline
    $\Delta m_{21}^2 / 10^{-5} \, \text{eV}^2$ & NO & $7.49_{-0.19}^{+0.19}$ \\
    & IO & $7.49_{-0.19}^{{+0.19}}$ \\
    \hline
    $\sin^2{\theta_{12}}$ & NO & $0.307_{-0.011}^{+0.012}$ \\
    & IO & $0.308_{-0.011}^{+0.012}$ \\
    \hline
    $\Delta m_{3l}^2 / 10^{-3} \, \text{eV}^2$ & NO & $2.534_{-0.023}^{+0.025}$ \\
    & IO & $-2.510_{-0.025}^{+0.024}$ \\
    \hline
    $\sin^2{\theta_{23}}$ & NO & $0.561_{-0.015}^{+0.012}$ \\
    & IO & $0.562_{-0.015}^{+0.012}$ \\
    \hline
    $\sin^2{\theta_{13}}$ & NO & $0.02195_{-0.00058}^{+0.00054}$ \\
    & IO & $0.02224_{-0.00057}^{+0.00056}$ \\
    \hline
    $\delta_{CP}/^{\circ}$ & NO & $177_{-20}^{+19}$ \\
    & IO & $285_{-28}^{+25}$ \\
    \hline
    \end{tabular}
\caption{\justifying{The standard neutrino oscillation parameters with NO and IO used in our analysis are taken from Refs.\,\cite{Esteban:2024eli,NuFIT}.}}
 \label{Tab1}
\end{table}

In the three–flavor framework, the evolution of the neutrino mass eigenstate is described by
\begin{equation}\label{2.9}
     \ket{\nu_{j}(t)}=e^{-\iota  \mathcal{H}_{tot}t}\ket{\nu_{j }},
\end{equation}
where $\ket{\nu_{j}}$ with $j=1,2,3$ are the neutrino mass eigenstates and $\mathcal{S}^{m}(t)=e^{-\iota  \mathcal{H}_{tot}t}$ is the evolution operator in the mass basis. At time $t=0$, the flavor eigenstates $\ket{\nu_{\alpha}}$, $\alpha= e, \mu, \tau$ are linear superpositions of the mass eigenstates, defined as
$\ket{\nu_{\alpha}}= \sum_{j}U_{\alpha j} \ket{\nu_{j }}$ and vice versa, $\ket{\nu_{j}}= \sum_{\alpha}U^{\ast }_{\alpha j}\ket{\nu _{\alpha}}.$
The time evolution of the flavor state can be written as
\begin{align}\label{2.10}
    \ket{\nu_{\alpha}(t)} 
        &= \sum_{j}^{} U_{\alpha j}\, e^{-\iota  \mathcal{H}_{tot}t}\ket{\nu_{j}} \nonumber\\
    &= \sum_{j,\beta}^{} U_{\alpha j}\, e^{-\iota  \mathcal{H}_{tot}t}\, U^{\ast}_{\beta j}\ket{\nu_{\beta}} \nonumber\\
    &= \mathcal{S}^{f}_{\alpha\beta}(t)\ket{\nu_\beta},
\end{align}
where $\mathcal{S}^f_{\alpha\beta}(t)= U e^{-\iota \mathcal{H}_{tot}t}U^{-1}$ is the time evolution operator in the flavor basis.

In the ultrarelativistic limit when $t=L$, the evolution operator in the mass basis is expressed as \cite{Ohlsson:1999xb,Konwar:2024nwc}:
\begin{align}\label{2.12}
\mathcal{S}^m(L)
&= e^{-\iota\mathcal{H}_{tot}L} \nonumber\\
&= \phi \sum_{a=1}^3 e^{-\iota L\lambda_a}
\frac{1}{3\lambda^2_a+c_1}
\left[(\lambda_a^2+c_1)I+\lambda_a T+T^2\right],
\end{align}
and similarly, the flavor evolution operator can be written as 
\begin{align}\label{2.13}
\mathcal{S}^f_{\alpha\beta}(L)
&= U e^{-\iota\mathcal{H}_{tot}L} U^{-1} \nonumber\\
&= \phi \sum_{a=1}^3 e^{-\iota L\lambda_a}
\frac{1}{3\lambda^2_a+c_1}
\left[(\lambda_a^2+c_1)I+\lambda_a\tilde{T} +\tilde{T}^2\right] \nonumber\\
&=
\begin{pmatrix}
\mathcal{S}^f_{ee}(L) & \mathcal{S}^f_{e\mu}(L) & \mathcal{S}^f_{e\tau}(L) \\
\mathcal{S}^f_{\mu e}(L) & \mathcal{S}^f_{\mu \mu}(L) & \mathcal{S}^f_{\mu\tau}(L) \\
\mathcal{S}^f_{\tau e}(L) & \mathcal{S}^f_{\tau \mu}(L) & \mathcal{S}^f_{\tau\tau}(L)
\end{pmatrix}.
\end{align}
Here $\phi = e^{-\iota L (\text{Tr}[\mathcal{H}_{tot}])/3}$, $\tilde{T}=UTU^{-1}$ and $c_{1}=\text{det}[T] \times \text{Tr}[T^{-1}]$ and  $\lambda _{1}$, $\lambda _{2}$ and $\lambda _{3}$ are the eigenvalues of the matrix $T$, which is given as $T \equiv \mathcal{H}_{tot}- (\text{Tr}[\mathcal{H}_{tot}])I/3. $ The evolution operator $\mathcal{S}^{f}_{\alpha\beta}(L)$ encapsulates the flavor transformation of neutrinos as they propagate over a baseline $L$, taking into account standard matter effects as well as NSIs. 
Thus, the flavor state $\ket{\nu_{\alpha}(L)}$ at length $L$ of Eq.\,\eqref{2.10} can  be expressed in terms of the flavor basis as:
\begin{eqnarray}\label{2.11}
    \ket{\nu _{\alpha }(L)}=\mathcal{S}^f_{\alpha e}(L)\ket{\nu_{e}}+\mathcal{S}^f_{\alpha \mu}(L) \ket{\nu_{\mu}}
    +\mathcal{S}^f_{\alpha \tau}(L) \ket{\nu_{\tau}}.
\end{eqnarray}

\begin{figure*}[!htbp]
  \centering
  \begin{subfigure}[b]{0.33\textwidth}
    \centering
    \includegraphics[width=\textwidth]{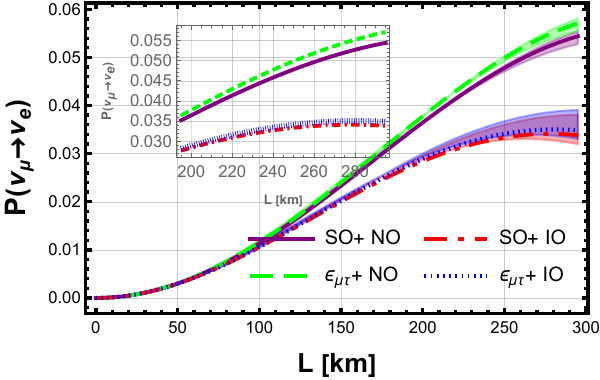}
    \caption{T2K}
    \label{1a_fig:sub1}
  \end{subfigure}
  \hfill
  \begin{subfigure}[b]{0.33\textwidth}
    \centering
    \includegraphics[width=\textwidth]{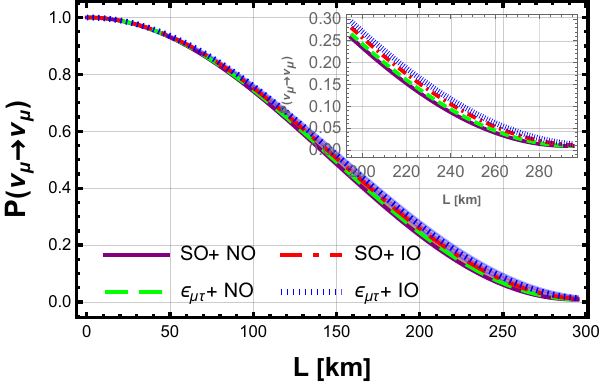}
    \caption{T2K}
    \label{1a_fig:sub2}
  \end{subfigure}
   \begin{subfigure}[b]{0.33\textwidth}
    \centering
    \includegraphics[width=\textwidth]{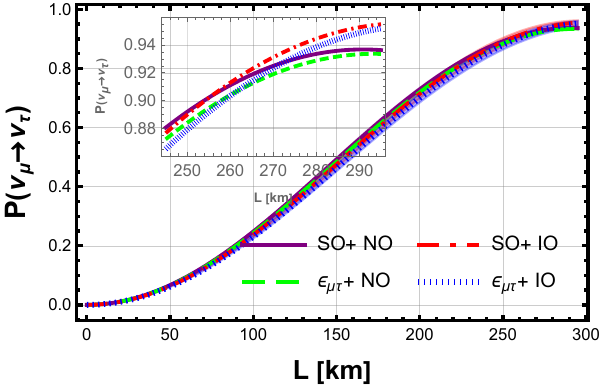}
    \caption{T2K}
    \label{1a_fig:sub3}
  \end{subfigure}
  \hfill
  \\
  \begin{subfigure}[b]{0.33\textwidth}
    \centering
    \includegraphics[width=\textwidth]{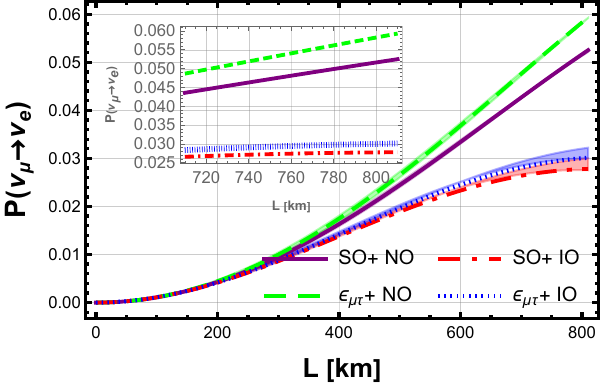}
    \caption{NO$\nu$A}
    \label{1a_fig:sub4}
  \end{subfigure}
  \hfill
  \begin{subfigure}[b]{0.33\textwidth}
    \centering
    \includegraphics[width=1.001\textwidth]{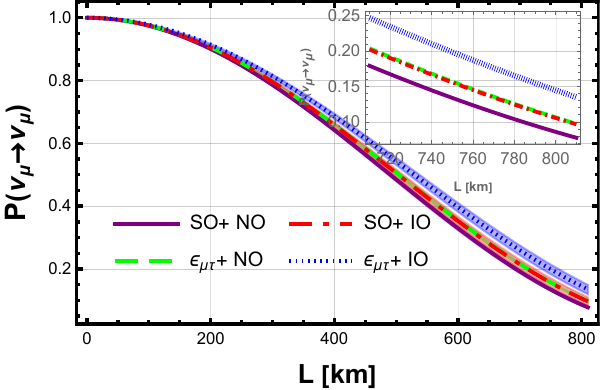}
    \caption{NO$\nu$A}
    \label{1a_fig:sub5}
  \end{subfigure}
  \hfill
  \begin{subfigure}[b]{0.33\textwidth}
    \centering
    \includegraphics[width=\textwidth]{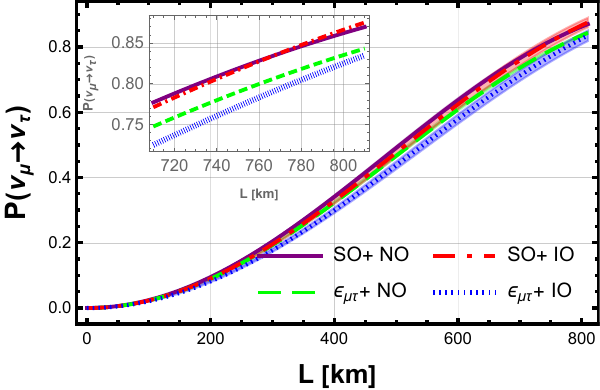}
    \caption{NO$\nu$A}
    \label{1a_fig:sub6}
  \end{subfigure}
\hfill
\\  
\begin{subfigure}[b]{0.33\textwidth}
    \centering
    \includegraphics[width=\textwidth]{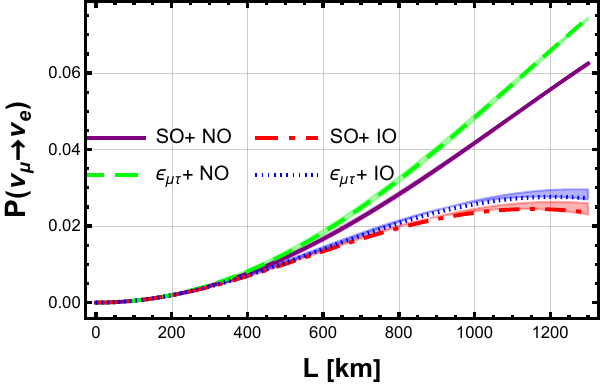}
    \caption{DUNE}
     \label{1a_fig:sub7}
  \end{subfigure}
  \hfill
  \begin{subfigure}[b]{0.33\textwidth}
    \centering
    \includegraphics[width=\textwidth]{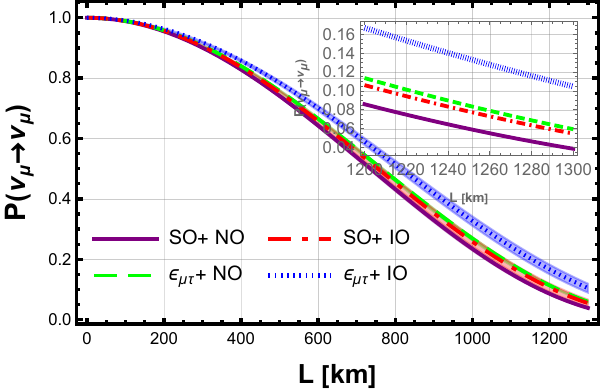}
    \caption{DUNE}
    \label{1a_fig:sub8}
  \end{subfigure}
  \hfill
  \begin{subfigure}[b]{0.33\textwidth}
    \centering
    \includegraphics[width=\textwidth]{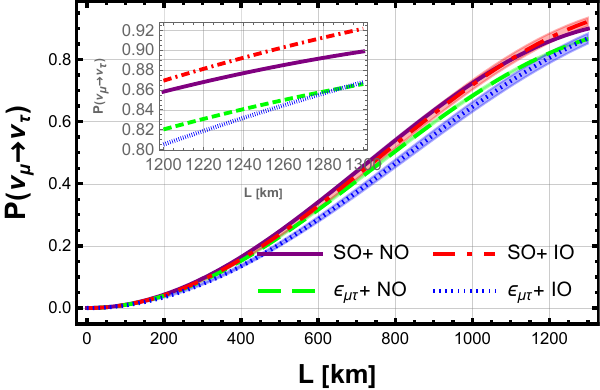}
    \caption{DUNE}
    \label{1a_fig:sub9}
  \end{subfigure}
  \caption{\justifying{From top to bottom, the left, middle, and right panels show the three-flavor-transition probabilities $P({\nu_\mu\rightarrow \nu_e})$, $P({\nu_\mu\rightarrow \nu_\mu})$, and $P({\nu_\mu\rightarrow \nu_\tau})$, respectively, as functions of the baseline length $L\,\text{(km)}$ of the initial muon-flavor neutrino state $\ket{\nu_\mu}$, evolved under four scenarios: SO+NO (purple solid line), SO+IO (red dot-dashed line), NSI+NO (green dashed line), and NSI+IO (blue dotted line), using the best-fit$\pm 1\sigma$ range of $CP$-violating phases $\delta_{CP}$ in NO and IO. The transition probabilities are compared across these scenarios for the off-diagonal NSI parameter $\left|\epsilon_{\mu\tau}\right|$ with complex phase $\phi_{\mu\tau}$, evaluated at the baselines and energies of the T2K (a)-(c), NO$\nu$A (d)-(f), and DUNE (g)-(i) experiments. The best-fit$\pm 1\sigma$ range of SO} and NSI parameters used are taken from Tables~\ref{Tab1} and \ref{Tab2}, respectively.}
  \label{fig1}
\end{figure*}

\begin{figure*}[!htbp]
  \centering
  \begin{subfigure}[b]{0.33\textwidth}
    \centering
    \includegraphics[width=\textwidth]{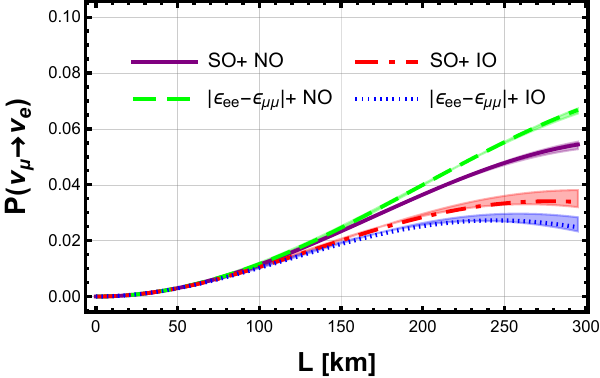}
    \caption{T2K}
    \label{2a_fig:sub1}
  \end{subfigure}
  \hfill
  \begin{subfigure}[b]{0.33\textwidth}
    \centering
    \includegraphics[width=\textwidth]{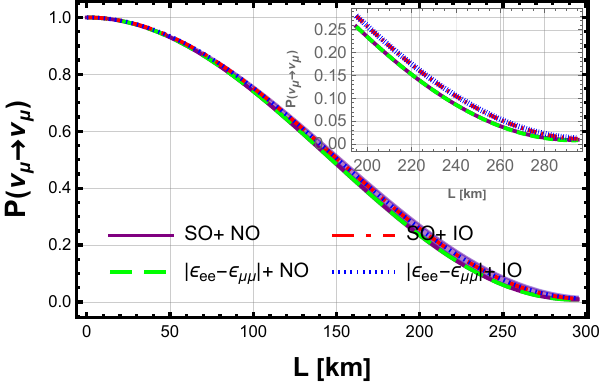}
    \caption{T2K}
    \label{2a_fig:sub2}
  \end{subfigure}
   \begin{subfigure}[b]{0.33\textwidth}
    \centering
    \includegraphics[width=\textwidth]{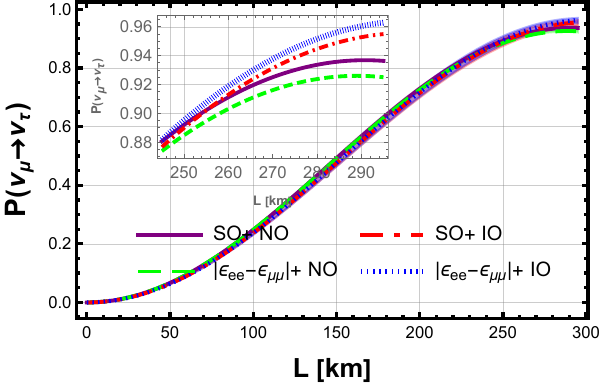}
    \caption{T2K}
    \label{2a_fig:sub3}
  \end{subfigure}
  \hfill
  \\
  \begin{subfigure}[b]{0.33\textwidth}
    \centering
    \includegraphics[width=\textwidth]{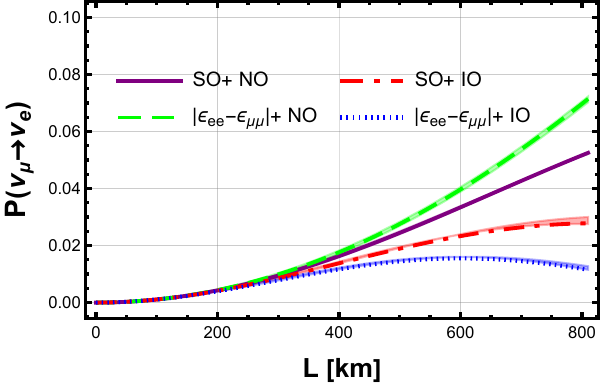}
    \caption{NO$\nu$A}
    \label{2a_fig:sub4}
  \end{subfigure}
  \hfill
  \begin{subfigure}[b]{0.33\textwidth}
    \centering
    \includegraphics[width=1.001\textwidth]{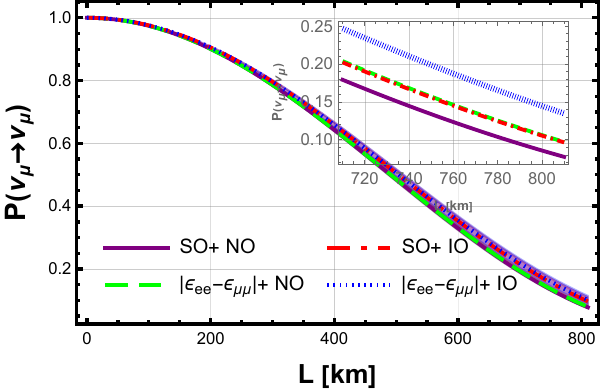}
    \caption{NO$\nu$A}
    \label{2a_fig:sub5}
  \end{subfigure}
  \hfill
  \begin{subfigure}[b]{0.33\textwidth}
    \centering
    \includegraphics[width=\textwidth]{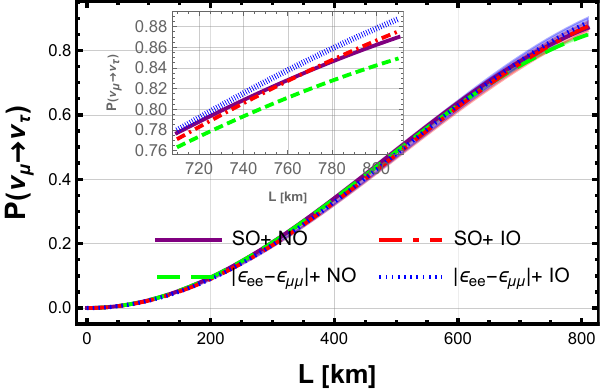}
    \caption{NO$\nu$A}
    \label{2a_fig:sub6}
  \end{subfigure}
\hfill
\\  
\begin{subfigure}[b]{0.33\textwidth}
    \centering
    \includegraphics[width=\textwidth]{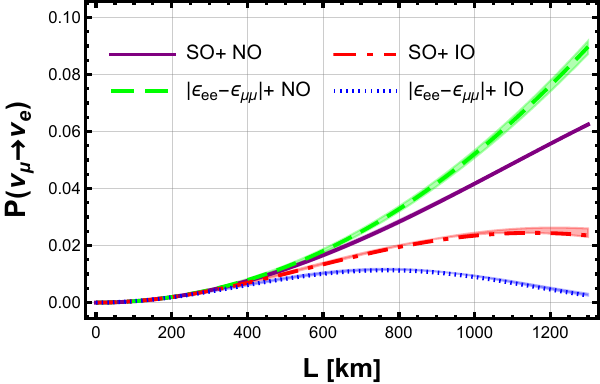}
    \caption{DUNE}
     \label{2a_fig:sub7}
  \end{subfigure}
  \hfill
  \begin{subfigure}[b]{0.33\textwidth}
    \centering
    \includegraphics[width=\textwidth]{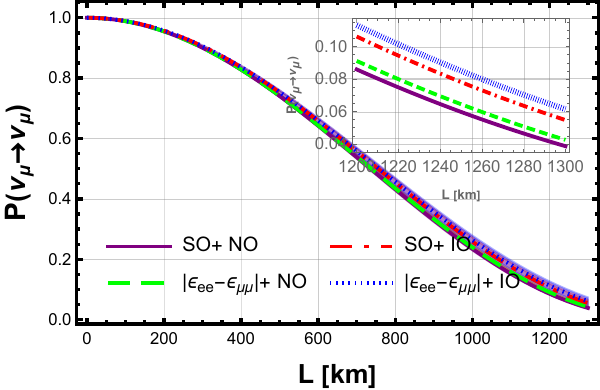}
    \caption{DUNE}
    \label{2a_fig:sub8}
  \end{subfigure}
  \hfill
  \begin{subfigure}[b]{0.33\textwidth}
    \centering
    \includegraphics[width=\textwidth]{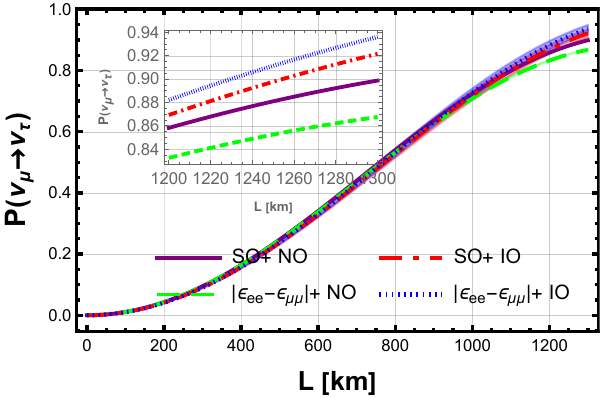}
    \caption{DUNE}
    \label{2a_fig:sub9}
  \end{subfigure}
  \caption{\justifying{From top to bottom, the left, middle, and right panels show the three-flavor-transition probabilities $P({\nu_\mu\rightarrow \nu_e})$, $P({\nu_\mu\rightarrow \nu_\mu})$, and $P({\nu_\mu\rightarrow \nu_\tau})$, respectively, as functions of the baseline length $L\,\text{(km)}$ of the initial muon-flavor neutrino state $\ket{\nu_\mu}$, evolved under four scenarios: SO+NO (purple solid line), SO+IO (red dot-dashed line), NSI+NO (green dashed line), and NSI+IO (blue dotted line), using the best-fit$\pm 1\sigma$ range of $CP$-violating phases $\delta_{CP}$ in NO and IO. The transition probabilities are compared across these scenarios for the diagonal NSI parameter $\left |\epsilon_{ee}-\epsilon_{\mu\mu}\right |$, evaluated at the baselines and energies of the T2K (a)-(c), (d)-(f), and DUNE (g)-(i) experiments. The best-fit$\pm 1\sigma$ range of SO and NSI parameters used are taken from Tables~\ref{Tab1} and \ref{Tab3}, respectively.}}
  \label{fig2}
\end{figure*}

The flavor–transition probability, denoted by  ${P}({\nu_\alpha\rightarrow\nu_\beta })$, represents the likelihood that a neutrino produced in the flavor state $\ket{\nu_\alpha}$ is observed as flavor $\ket{\nu_\beta}$ after traveling the distance $L$ and is given by
\begin{equation}\label{2.15}
   {P}(\nu_\alpha\rightarrow \nu_\beta )(L)=|\mathcal{S}^{f}_{\alpha\beta}(L)|^2.
\end{equation}

The appearance probabilities correspond to flavor transitions with $\alpha \neq \beta$, such as ${P}({\nu_\mu \rightarrow \nu_e })$ and ${P}({\nu_\mu \rightarrow \nu_\tau })$, whereas disappearance probabilities correspond to $\alpha= \beta$, namely ${P}({\nu_\mu \rightarrow \nu_\mu })$. The probabilities ${P}({\nu_\alpha\rightarrow \nu_\beta })$ are determined by the standard neutrino oscillation parameters:  two independent mass-squared splittings, $\Delta m^2_{21}=m^2_2-m^2_1$, $\Delta m^2_{31}=m^2_3-m^2_1=\Delta m_{32}^{2}+\Delta m_{21}^{2}$; three mixing angles, $\theta_{12}$, $\theta_{13}$, $\theta_{23}$; and the $CP$-violating phase $\delta_{CP}$. In addition, the probability of flavor transition is sensitive to the choice of mass ordering: for normal ordering, the spectrum follows 
 $m_3 > m_2 > m_1$, whereas for inverted ordering, it corresponds to $m_3 < m_1 < m_2$ \cite{Giganti:2017fhf}. By convention, the mass splitting is taken as  $\Delta m^2_{3l} = \Delta m^2_{31} > 0$ for NO, and $\Delta m^2_{3l} = \Delta m^2_{32} < 0$ for IO \cite{Esteban:2024eli,NuFIT}. The standard neutrino oscillation parameters with NO and IO scenarios are summarized in Table\,\ref{Tab1}.

Furthermore, ${ P}({\nu_\alpha\rightarrow\nu_ \beta })$, in Eq.\,(\ref{2.15}), is also a function of the baseline distance ($L$) between the neutrino source and the detector, as well as the neutrino energy ($E$). Since both $L$ and $E$ differ for different neutrino experimental configurations, the corresponding flavor–transition probabilities also vary from one neutrino experiment to another. In this work, we perform our analysis using three accelerator-based long-baseline neutrino oscillation experiments: T2K \cite{Abe_2023}, NO$\nu$A \cite{Acero_2022}, and DUNE \cite{Abi:2020wmh}. These facilities operate with different baselines, beam configurations, and detector technologies, thereby probing neutrino oscillations across complementary regions of energy and matter effect sensitivities.
\begin{enumerate}
\item T2K experiment: The Tokai to Kamioka experiment \cite{Abe_2023}  is a long-baseline accelerator neutrino experiment designed to measure neutrino oscillation parameters, particularly the $CP$-violating phase $\delta_{ CP}$. The T2K experiment sends a muon neutrino beam from J-PARC to the Super-Kamiokande detector over a 295 km baseline. Its off-axis beam configuration produces a narrow energy spectrum with a maximum flux around 0.6 GeV, optimized for studying $\nu _{\mu}\rightarrow \nu_{e}$ oscillations. Operating in both neutrino and antineutrino modes, T2K has strong sensitivity to $\delta_{ CP}$ and precise measurements of $\theta_{23}$ and $\Delta m^{2}_{32}$.
\item NO$\nu$A experiment:
The NuMI off-Axis $\nu_e$ appearance experiment \cite{Acero_2022} is a long-baseline neutrino oscillation experiment that sends a neutrino beam from Fermilab to a far detector located in Ash River, Minnesota, covering a baseline of 810 km. It uses the NuMI beam, operated in a 14 mrad off-axis configuration that produces a narrow band neutrino energy spectrum with maximum flux $\sim$ 2 GeV, which closely matches the first oscillation maximum for its baseline. The off-axis technique significantly reduces high-energy backgrounds and enhances sensitivity to the $\nu _{\mu}\rightarrow \nu_{e}$ appearance channel. Because of its long baseline, NO$\nu$A experiences moderate matter effects, giving it strong sensitivity to the neutrino mass ordering, the octant $\theta_{23}$, and the $CP$-violating phase $\delta_{CP}$.
\item DUNE experiment:
The Deep Underground Neutrino Experiment \cite{Abi:2020wmh} is a next-generation long-baseline neutrino oscillation experiment designed to deliver precision measurements of neutrino oscillation parameters and search for BSM. It sends an intense wideband neutrino beam from Fermilab to the Sanford Underground Research Facility, spanning a 1300 km baseline. This long baseline amplifies matter effects, giving DUNE exceptional sensitivity to the neutrino mass ordering. The neutrino beam covers a broad energy range of approximately 0.5 - 8 GeV, with the flux peaking at around 2.5 - 3 GeV, allowing DUNE to probe multiple oscillation maxima rather than relying solely on the first peak. Owing to its long baseline and strong matter effects, DUNE offers exceptional sensitivity to the neutrino mass ordering, the $CP$-violating phase $\delta_{CP}$, and possible signatures of nonstandard interactions in neutrino propagation.
\end{enumerate}

\begin{table}[t!] 
    \centering
    \begin{tabular}{|c|c|c|c|}
        \hline
       Mass ordering  & NSI & $ \left |\epsilon_{\alpha\beta}\right |$ & $\phi_{\alpha\beta}/ \pi$  \\
       \hline
       \hline
        NO & $\epsilon_{e\mu}$ & 0.13& 1.35\\
        IO & $\epsilon_{e\mu}$ &  0.05 & 1.44\\
        \hline
        NO & $\epsilon_{e\tau}$ &0.22 & 1.70 \\
        IO & $\epsilon_{e\tau}$ & 0.23 & 1.54 \\
        \hline
        NO & $\epsilon_{\mu\tau}$ &0.35 & 0.60 \\
        IO & $\epsilon_{\mu\tau}$ & 0.17 & 0.14 \\
        \hline
    \end{tabular}
    \caption{\justifying{Best-fit values of the different off-diagonal NSI parameters $\epsilon_{e\mu}$ and $\epsilon_{e\tau}$ are taken from Ref. \cite{Chatterjee:2024kbn} and $\epsilon_{\mu\tau}$ is from Ref.\,\cite{Denton:2020uda} for both NO and IO scenarios.} }
    \label{Tab2}
\end{table}

\begin{table}[t]
	\centering
    \begin{tabular}{|c|c|c|}
		\hline
        NSI parameter  & Range (99 \% C.L.)\\      
		\hline\hline
        $ \left |\epsilon_{e e}- \epsilon_{\mu \mu} \right |$ & [$-4.8, -1.6$] $\oplus$ [$-0.40, +2.6$]\\
		\hline
		$ \left |\epsilon_{\tau \tau}- \epsilon_{\mu \mu} \right |$ & [$-0.075, + 0.080$]\\
		\hline
	\end{tabular}
    \caption{\justifying{The diagonal NSI parameters as presented here are taken from Ref. \cite{Coloma:2023ixt} and their upper bound values are used in our analysis for both mass ordering scenarios.}}
    \label{Tab3}
\end{table}
Using the transition probability expression given in Eq.\,\eqref{2.15}, we now explore how the presence of NSIs modifies the flavor evolution of an initial muon neutrino state ($\ket{\nu_\mu}$) as it propagates through matter. The probability ${ P}({\nu_\alpha\rightarrow\nu_ \beta })$ encapsulates not only the SO dynamics but also the modifications induced by NSI parameters, which include diagonal NSI parameters as well as off-diagonal NSI parameters one at a time. To illustrate these effects, Fig.~\ref{fig1}, and also Figs.\,\ref{fig6} and \ref{fig7} (see the Appendix) presents the variation of the appearance and disappearance probabilities for the case of the off-diagonal NSI parameter $\left |\epsilon_{\alpha \beta}\right |$ with its associated phase $\phi_{\alpha \beta}$, only one off-diagonal NSI parameter at a time is considered in calculation. Similarly, Figs.\,\ref{fig2} and \ref{fig8} (see the Appendix) show the corresponding results for the diagonal NSI parameters, where each diagonal parameter is also introduced one at a time. Throughout our analysis, we consider neutrino propagation under four scenarios: SO+NO, SO+IO, NSI+NO, and NSI+IO, with their respective best-fit $CP$-violating phase $\delta_{CP}$.

The probabilities in these plots are evaluated at the characteristic neutrino energies corresponding to the maximum flux and length scales of the three long-baseline experiments considered in this work: T2K \cite{Abe_2023} with $L\approx 295\,\rm km$ and $E\approx 0.6\,\rm GeV$, NO$\nu$A \cite{Acero_2022} with $L\approx 810\,\rm km$ and $E\approx 2\,\rm GeV$, and DUNE \cite{Abi:2020wmh} with $L\approx 1300\,\rm km$ and $E\approx 3\,\rm GeV$, allowing us to compare how the impact of NSIs evolves across different experimental setups for both NO and IO. 

In generating these plots, the standard oscillation parameters listed in Table\,\ref{Tab1} are used. For the off-diagonal NSI parameters, we have taken the parameter values given in Table \ref{Tab2}. The chosen values of the off-diagonal NSI parameters $\left |\epsilon_{\alpha \beta}\right |$ and their corresponding phases $\phi_{\alpha \beta}$ are motivated, as they have been shown to play a significant role in addressing the tension between the NO$\nu$A and T2K measurements of the leptonic $CP$-violating phase $\delta_{CP}$ and the atmospheric mixing angle $\theta_{23}$ \cite{Chatterjee:2024kbn,Denton:2020uda}. The diagonal NSI parameters used in this work are taken from Table \ref{Tab3}. In our analysis, we use only the differences between the diagonal NSI parameters, since neutrino oscillation experiments are sensitive not to the absolute values of the diagonal terms, but to combinations such as $\epsilon_{ee}-\epsilon_{\mu\mu}$ and $\epsilon_{\tau \tau}-\epsilon_{\mu\mu}$. This follows from the fact that subtracting any real number to all diagonal elements of the Hamiltonian in Eq.\,\eqref{2.6} leaves the oscillation probabilities unchanged \cite{ESSnuSB:2025vsf, Coloma:2023ixt}. 

Figure\,\ref{fig1} illustrates the effect of the off-diagonal NSI parameter $\epsilon_{\mu\tau}$  on the appearance and disappearance channels $P(\nu_{\mu}\rightarrow \nu_{e})$, $P(\nu_{\mu}\rightarrow \nu_{\tau})$, and $P(\nu_{\mu}\rightarrow \nu_{\mu})$, plotted as a function of the baseline $L$ in kilometers. The results are presented for the ongoing long-baseline accelerator experiments T2K, shown in Figs.~\ref{fig1}(\subref{1a_fig:sub1})-\ref{fig1}(\subref{1a_fig:sub3}), and NO$\nu$A shown in the middle row in Figs.~\ref{fig1}(\subref{1a_fig:sub4})-\ref{fig1}(\subref{1a_fig:sub6}), as well as for the upcoming accelerator experiment DUNE in the bottom row in Figs.~\ref{fig1}(\subref{1a_fig:sub7})-\ref{fig1}(\subref{1a_fig:sub9}). Each panel displays the behavior for both NO and IO, using the corresponding $CP$-violating phases $\delta_{CP}= 177^{\circ}$ and $\delta_{CP}= 285^{\circ}$, respectively. For the T2K experiment, the appearance probability $P(\nu_{\mu}\rightarrow \nu_{e})$, shown in Fig.~\ref{fig1}(\subref{1a_fig:sub1}), clearly separates the two mass orderings for SO scenarios, i.e. SO+NO and SO+IO. The inclusion of the NSI parameter $\epsilon_{\mu\tau}$ introduces a mild overall enhancement in the appearance probability for both mass ordering, with $\epsilon_{\mu\tau}$+NO exhibiting higher probabilities than SO+NO and $\epsilon_{\mu\tau}$+IO compared to the SO+IO. Notably, the ordering-dependent separation even persists in the presence of NSI, where NO and IO remain distinguishable under the NSI framework. In the disappearance channel, shown in Fig.~\ref{fig1}(\subref{1a_fig:sub2}), neither the mass ordering nor the difference between SO and NSI can be discerned, as the corresponding curves largely overlap. Whereas the transition probability $P(\nu_{\mu}\rightarrow \nu_{\tau})$, displayed in Fig.~\ref{fig1}(\subref{1a_fig:sub3}), shows a suppression for both $\epsilon_{\mu\tau}$+NO compared to the SO+NO and $\epsilon_{\mu\tau}$+IO compared to the SO+IO case. However, at the end of the baseline, an overlap is observed between the SO curve and the NSI $\epsilon_{\mu\tau}$ curve for both mass orderings. 

In Figs.~\ref{fig1}(\subref{1a_fig:sub4})-\ref{fig1}(\subref{1a_fig:sub6}), all three oscillation channels are shown for the NO$\nu$A experiment and all four scenarios are clearly distinguishable. In the appearance and disappearance channels shown in Fig.~\ref{fig1}(\subref{1a_fig:sub4}) and Fig.~\ref{fig1}(\subref{1a_fig:sub5}), respectively, NSI leads to an increase in the probabilities for both orderings relative to the SO prediction; this increase is more noticeable for NO in the appearance channel, whereas it becomes more significant for IO in the disappearance channel. For the $\nu_{\tau}$ transition probability present in Fig.~\ref{fig1}(\subref{1a_fig:sub6}), a decrement is observed for both mass orderings relative to the standard case, with the reduction being more pronounced for IO.

In the DUNE experiment, the behavior of the appearance and disappearance probabilities, shown in Figs.~\ref{fig1}(\subref{1a_fig:sub7}) and \ref{fig1}(\subref{1a_fig:sub8}), is qualitatively similar to that observed for NO$\nu$A; however, the impact of the NSI parameter $\epsilon_{\mu\tau}$ is significantly enhanced due to the stronger matter effects associated with DUNE’s longer baseline. In contrast, the $\nu_{\tau}$ appearance probability, displayed in Fig.~\ref{fig1}(\subref{1a_fig:sub9}), exhibits a distinct behavior. The experiment cannot clearly distinguish between SO+NO and SO+IO at baselength approx 1000-1100 km, by contrast, when NSI effects are taken into account, DUNE can discriminate between $\epsilon_{\mu\tau}$+NO and $\epsilon_{\mu\tau}$+IO. At the end of the baseline, however, this behavior is reversed, demonstrating the influence of long-baseline configurations in the presence of new physics effects.

Figure \ref{fig2} presents the impact of the diagonal NSI parameter $\left |\epsilon_{ee}-\epsilon_{\mu\mu}\right |$ on the transition probabilities as a function of the baseline $L$. The upper, middle, and lower rows correspond to the T2K, NO$\nu$A, and DUNE experimental configurations, respectively, with the associated results displayed in Figs.~\ref{fig2}(\subref{2a_fig:sub1})-\ref{fig2}(\subref{2a_fig:sub3}) for T2K, Figs.~\ref{fig2}(\subref{2a_fig:sub4})-\ref{fig2}(\subref{2a_fig:sub6}) for NO$\nu$A, and Figs.~\ref{fig2}(\subref{2a_fig:sub7})-\ref{fig2}(\subref{2a_fig:sub9}) for DUNE. As in Fig.~\ref{fig1}, the results in Fig.~\ref{fig2} are also shown for both NO and IO, using their respective best-fit $CP$-violating phase values, thereby allowing a direct comparison between the SO scenario and the diagonal NSI parameter-induced modifications across different baselines and energies.

For the T2K experiment in Fig.~\ref{fig2}, the appearance probability in Fig.\,\ref{fig2}(\subref{2a_fig:sub1}), $\left |\epsilon_{ee}-\epsilon_{\mu\mu}\right |$+NO exhibits a noticeable enhancement from the SO+NO case, whereas a suppression relative to the SO+IO prediction is observed for $\left |\epsilon_{ee}-\epsilon_{\mu\mu}\right |$+IO. All four plots remain well separated, ensuring clear discrimination between the two mass orderings. 
For the disappearance channel in Fig.\,\ref{fig2}(\subref{2a_fig:sub2}), the NSI behaviour remains closely aligned with the SO curves, whereas a clear separation between NO and IO is retained for both SO and NSI scenarios. For the $\nu_\tau$ appearance channel in Fig.\,\ref{fig2}(\subref{2a_fig:sub3}), all four scenarios are distinguishable at the end of the baseline and enhancement in IO and suppression in NO after inclusion of NSI is observed as compared to the SO.

In the NO$\nu$A experiment, the electron–neutrino appearance channel in Fig.\,\ref{fig2}(\subref{2a_fig:sub4}) shows an increase in the probability for  $\left |\epsilon_{ee}-\epsilon_{\mu\mu}\right |$+NO and a decrease for the $\left |\epsilon_{ee}-\epsilon_{\mu\mu}\right |$+IO case relative to the SO+NO and SO+IO prediction, respectively. A clear separation between the mass ordering is visible in both the SO and NSI scenarios. The disappearance channel in Fig.\,\ref{fig2}(\subref{2a_fig:sub5}), allows a clear distinction between SO+NO and SO+IO, as well as between NSI+NO and NSI+IO. However, the NSI+NO  curve closely follows the SO+NO prediction, and similarly, the NSI+IO curve overlaps with SO+IO. For the $\nu_\tau$ appearance channel in Fig.\,\ref{fig2}(\subref{2a_fig:sub6}), the SO+NO and SO+IO cases are indistinguishable, whereas a clear separation is observed between NSI+NO and NSI+IO.

The behavior in the appearance channel as shown in Fig.\,\ref{fig2}(\subref{2a_fig:sub7}), for the DUNE experiment follows the same pattern observed in NO$\nu$A, but with a more pronounced separation between the SO and NSI predictions. Unlike the other two experiments, the disappearance channel in DUNE, shown in Fig.\,\ref{fig2}(\subref{2a_fig:sub8}) is able to distinguish all four scenarios clearly. In the $\nu_\tau$ appearance channel in Fig.\,\ref{fig2}(\subref{2a_fig:sub9}), at the end of the baseline all four scenarios are clearly distinguishable. The $\lvert \epsilon_{ee}-\epsilon_{\mu\mu}\rvert$+NO case exhibits a suppression in the probability relative to the SO+NO scenario, while an enhancement is observed for $\lvert \epsilon_{ee}-\epsilon_{\mu\mu}\rvert$+IO compared to SO+IO.

 The above discussion is based on the best-fit values of the standard oscillation parameters, whereas we have also examined the transition probabilities within the best-fit$\pm 1\sigma$ allowed ranges of the oscillation parameters as provided in Table \ref{Tab1}. For the off-diagonal NSI parameter $\epsilon_{\mu \tau}$ as shown in Fig. \ref{fig1}, the T2K configuration exhibits an overlap among all four scenarios in the three oscillation channels, making them difficult to distinguish within the 1$\sigma$ region. For the NO$\nu$A experiment, the $\nu_{e}$ appearance channel provides a clear separation between the SO and NSI scenarios as well as between the NO and IO cases. However, in the disappearance channel, only the $\epsilon_{\mu \tau}$+IO scenario can be clearly distinguished from the remaining cases, while the $\nu_{\tau}$ appearance channel exhibits substantial overlap among all scenarios. For the DUNE configuration, the $\nu_{e}$ appearance channel clearly distinguishes all four scenarios. On the other hand, in both the disappearance and $\nu_{\tau}$ appearance channels, only the $\epsilon_{\mu \tau}$+IO case remains clearly distinguishable, whereas the other three scenarios overlap within the 1$\sigma$ allowed parameter space. Similarly, for the diagonal NSI parameter $\left|\epsilon_{ee}-\epsilon_{\mu\mu}\right|$, as shown in Fig. \ref{fig2}, among the three oscillation channels, only the $\nu_{e}$ appearance channel exhibits a clear separation between all four scenarios for the T2K, NO$\nu$A, and DUNE experiments. In contrast, the disappearance and $\nu_{\tau}$ appearance channels show a substantial overlap of the SO and NSI predictions within the 1$\sigma$ allowed parameter space, making it difficult to distinguish between the different scenarios for all three experimental configurations.

For the other two off-diagonal NSI parameters $\epsilon_{e\mu}$ scenario shown in Fig.~\ref{fig6} and the $\epsilon_{e\tau}$ scenario results presented in Fig.~\ref{fig7} of the Appendix, the corresponding probability plots display only minor deviations from the SO curves across all three experimental setups. As a result, the NSI and SO behaviors remain largely indistinguishable in the disappearance channel as well as in the $\nu_{\tau}$ appearance channel. A noticeable separation from the SO is visible only in the $\nu_{e}$ appearance channel. In contrast, the $\epsilon_{\mu\tau}$ contribution introduces visibly larger deviations, particularly in the muon disappearance and $\nu_{\tau}$ transition channels, as compared to the SO, where the NSI and SO predictions become clearly distinguishable. Because the off-diagonal $\epsilon_{\mu\tau}$ NSI parameter generates the most significant impact in our analysis, we therefore discussed this scenario in detail in this section. Similarly, for the diagonal parameter $\left|\epsilon_{ee}-\epsilon_{\tau\tau}\right|$, shown in Fig.~\ref{fig8} in the Appendix, the deviations from the SO prediction are negligible. Therefore, in this section, we focus our discussion on the diagonal parameter $\left|\epsilon_{ee}-\epsilon_{\mu\mu}\right|$, which exhibits a comparatively more noticeable impact on probabilities. For Figs. \ref{fig6}-\ref{fig8}, we have used only best-fit values of SO and NSI parameters since most of the 1$\sigma$ or 3$\sigma$ give overlapping results among the four scenarios across all three experiments.
 
\section{QSL for bipartite entanglement in three-flavor neutrino Oscillations in matter with NSI}
\label{sec4}
Using quantum-information–theoretic techniques, in the ultrarelativistic regime ($t\approx L$), neutrino flavor eigenstates can be treated as independent quantum modes. For example, in the two-flavor neutrino oscillation framework, where only the flavor eigenstates $\ket{\nu_e}$ and $\ket{\nu_\mu}$ are involved, the occupation-number representation assigns $\ket{1}_e$ to the occupied electron neutrino mode, while the muon neutrino mode remains unoccupied, $\ket{0}_\mu$. The tensor product of these two modes, $\ket{1}_{e}\otimes\ket{0}_{\mu}\equiv {\ket{10}}$, defines the corresponding two-qubit mode state for the flavor state $\ket{\nu_e}$. By the same convention, the flavor state $\ket{\nu_\mu}$ can be represented as $\ket{01}$. The nonclassical correlations established between these two-qubit mode states give rise to entanglement, and the particular form generated among flavor modes is referred to as “neutrino mode (flavor) entanglement.” Under this interpretation, the length-evolved flavor state naturally appears as a Bell-like superposition state, a structure that can be analyzed in both two- and three-flavor neutrino oscillation scenarios \cite{Blasone_2009,BLASONE2013320,PhysRevD.77.096002,Banerjee:2015mha,Alok:2014gya,Dixit:2017ron,Dixit:2018kev,Dixit:2019swl,Dixit:2020ize,KumarJha:2020pke,Yadav:2022grk,Li:2022mus,wang2023monogamy,Bittencourt:2023asd,Blasone:2023qqf,Konwar:2024nrd,Konwar:2025ipv,Quinta:2022sgq,Banerjee:2024lih,PhysRevA.98.050302,Ming:2020nyc}.

In the three-flavor neutrino oscillation framework, we map the neutrino flavor eigenstates at distance $L=0$ to a state in three-qubit mode as \cite{KumarJha:2020pke,Bouri:2024kcl}
\begin{eqnarray}
{\ket{\nu_e}\equiv\ket{1}_{e}\otimes\ket{0}_{\mu}\otimes\ket{0}_\tau\equiv\ket{100};}
&\nonumber\\
{\ket{\nu_\mu}\equiv\ket{0}_{e}\otimes\ket{1}_{\mu}\otimes\ket{0}_\tau\equiv\ket{010};}&\nonumber\\
{\ket{\nu_\tau}\equiv\ket{0}_{e}\otimes\ket{0}_{\mu}\otimes\ket{1}_\tau\equiv\ket{001}.}&
\label{4.1}
\end{eqnarray}

By substituting Eq.\,(\ref{4.1}) into Eq.\,(\ref{2.11}), the evolution equation of the initial flavor state $\ket{\nu_\mu}$ in SO with NSI effects, after propagating a distance $L$, can be expressed in a linear superposition of the three-qubit mode (flavor) basis as
\begin{equation} 
    \ket{\nu_\mu (L)}={\mathcal{S}}_{\mu e}^{f}(L)\ket{100}+{\mathcal{S}}_{\mu \mu}^{f}(L)\ket{010} + {\mathcal{S}}_{\mu\tau}^{f}(L)\ket{001}.
    \label{4.2}
    \end{equation}
    
In the standard basis $\ket{klm}$ with $k,l,m\in \{0,1\}$, the density matrix of the state $\ket{\nu_\mu(L)}$ is given by $\rho(L)=\ket{\nu_\mu(L)}\bra{\nu_\mu(L)}=$ 
 \begin{equation}
 \begin{pmatrix}  
 0 & 0 & 0 & 0 & 0 & 0 & 0 & 0\\ 
 0 & 0 & 0 & 0 & 0 & 0 & 0 & 0\\ 
0 & 0 & 0 & 0 & 0 & 0 & 0 & 0\\ 
0 & 0 & 0 & \vert{{\mathcal{S}}_{\mu e}^{f}(L)}\vert^2 & 0 & {\mathcal{S}}_{\mu e}^{f}(L) {\mathcal{S}}_{\mu \mu}^{f*}(L) & {\mathcal{S}}_{\mu e}^{f}(L) {\mathcal{S}}_{\mu\tau}^{f*}(L) & 0 \\
0 & 0 & 0 & 0 & 0 & 0 & 0 & 0\\ 
0 & 0 & 0 & {\mathcal{S}}_{\mu\mu}^{f}(L) {\mathcal{S}}_{\mu e}^{f*}(L) & 0 & \vert{{\mathcal{S}}_{\mu\mu}^{f} (L)}\vert^2 & {\mathcal{S}}_{\mu\mu}^{f}(L){\mathcal{S}}_{\mu\tau}^{f*}(L) & 0 \\
0 & 0 & 0 & {\mathcal{S}}_{\mu\tau}^{f}(L) {\mathcal{S}}_{\mu e}^{f*}(L) & 0 & {\mathcal{S}}_{\mu\tau}^{f}(L) {\mathcal{S}}_{\mu\mu}^{f*} (L) & \vert{{\mathcal{S}}_{\mu\tau}^{f} (L)}\vert^2 & 0\\ 
0 & 0 & 0 & 0 & 0 & 0 & 0 & 0 
\end{pmatrix}.
\label{4.3}
\end{equation} 

Here, the length-evolved muon–flavor neutrino state $\ket{\nu_\mu(L)}$ is a pure quantum state, as evidenced by the properties of its density matrix: $\rho(L)=\rho^2(L)$  and $\rm Tr\,(\rho^2(L))=1$. By taking the partial trace with respect to the other qubit(s), two reduced density matrices can be obtained as  $\rho_{\mu e}(L)= \mathrm{Tr}_{(\tau)}(\rho(L))$ and $\rho_{\tau}(L)=\mathrm{Tr}_{(\mu e)}(\rho(L))$. Both reduced density matrices correspond to mixed states, as reflected by the inequalities $ \mathrm{Tr}\,(\rho^2_{\mu e}(L)<1$ and $ \mathrm{Tr}\,(\rho^2_{\tau}(t))<1$. Consequently, $\rho_{\mu e}(L)$ and $\rho_{\tau}(L)$ represent the two subsystems of a bipartite quantum system $\rho(L)$, constructed under the assumption that, initially, each flavor state ($\ket{\nu_\alpha}$) is mapped onto a three-qubit state. This demonstrates that the state $\ket{\nu_\mu(L)}$ forms a pure bipartite quantum state within the three-qubit system. By considering a bipartition in which, for instance, $\tau$ constitutes one subsystem of $\rho(L)$, while $\mu e$ (treating as a single quantum object) forms the complementary subsystem, the four eigenvalues ($\eta_i$) of the reduced density matrix $\rho_{\mu e}(L)$ can be obtained as follows 
\begin{align} \eta_1(L)&=\eta_2(L)=0, \hspace{0.5em}\eta_3(L)=|\mathcal{S}^f_{\mu\mu}(L)|^2=P(\nu_{\mu}\rightarrow \nu_{\mu}), \hspace{0.5em} \text{and}&\nonumber\\ \eta_4(L)&=|\mathcal{S}^f_{\mu e}(L)|^2+|\mathcal{S}^f_{\mu \tau}(L)|^2=P(\nu_{\mu}\rightarrow \nu_{e}) + P(\nu_{\mu}\rightarrow \nu_{\tau}),
    \label{4.4}
\end{align}
  where $\eta_3(L)+\eta_4(L)=1$.  With these two non-vanishing eigenvalues, standard bipartite entanglement quantifiers, namely the entanglement entropy (\ref{2}) and the capacity of entanglement (\ref{4}), can be directly evaluated for the length–evolved muon neutrino flavor state $\ket{\nu_{\mu}(L)}$ in SO with NSI effects. The entanglement entropy can be expressed as
\begin{align}
    S_{EE}(\rho_{\mu e}(L))&= -\mathrm{Tr}\left(\rho_{\mu e}(L)\log_2\rho_{\mu e}(L)\right)&\nonumber\\
 &  = -\eta_3(L) \log_2\eta_3(L)-\eta_4(L)\log_2\eta_4(L),
    \label{4.5}
\end{align}

while the capacity of entanglement gives
\begin{align}
     C_{E}(\rho_{\mu e}(L))=&-\eta_3(L) \log^2_2\eta_3(L)-\eta_4(L)\log^2_4\eta_4(L) \nonumber \\&
     -(-\eta_3(L) \log_2\eta_3(L)-\eta_4(L)\log_2\eta_4(L))^2.
     \label{4.6}
\end{align}
Thus, the two bipartite entanglement measures $S_{EE}(\rho_{\mu e}(L))$ and $C_{E}(\rho_{\mu e}(L))$ given in Eqs.\,(\ref{4.5}) and (\ref{4.6}) for the three-flavor neutrino oscillation in SO with NSI effects can be quantified in terms of neutrino transition probabilities.  One of the prominent bipartite entanglement measures, negativity, can be related to $\eta_3(L)$ and  $\eta_4(L)$ in three-flavor neutrino oscillations as $N(\rho_{\mu e}(L))=2\sqrt{\eta_3(L) \eta_4(L)}$ \cite{KumarJha:2020pke}. This implies $\eta_3(L)=(1+\sqrt{1-N^2(\rho_{\mu e}(L))})/2$ and $\eta_4(L)=(1-\sqrt{1-N^2(\rho_{\mu e}(L))})/2$. Hence, the entanglement entropy $S_{EE}(\rho_{\mu e}(L)$, Eq.\,(\ref{4.5}), can be reexpressed in terms of negativity as 
\begin{align}
    S_{EE}(N(\rho_{\mu e}(L)))&=\frac{1}{2}(1+\sqrt{1-N^2(\rho_{\mu e}(L))})\nonumber\\
    &\text{log}_2(\frac{1}{2}(1+\sqrt{1-N^2(\rho_{\mu e}(L))}))&\nonumber\\
    &-\frac{1}{2}(1-\sqrt{1-N^2(\rho_{\mu e}(L))})\nonumber\\
    &\text{log}_2(\frac{1}{2}(1-\sqrt{1-N^2(\rho_{\mu e}(L))})).
    \label{4.7}
\end{align}
Since, $\eta_3>0$ implies $\eta_4<1$, which in turm implies  $N(\rho_{\mu e}(L))\neq0$, consequently, $S_{EE}(N(\rho_{\mu e}(L)))\neq0$. Similarly, one can also relate the capacity of entanglement, Eq.\,(\ref{4.6}), as a function of negativity. In this vein, various other bipartite entanglement measures, such as concurrence, tangle, and linear entropy, among others, have previously been investigated and simplified in terms of three-flavor neutrino transition probabilities \cite{KumarJha:2020pke}. 

\begin{figure*}[!htbp]
    \centering
    \begin{subfigure}{0.49\textwidth}
        \centering
        \includegraphics[width=\textwidth]{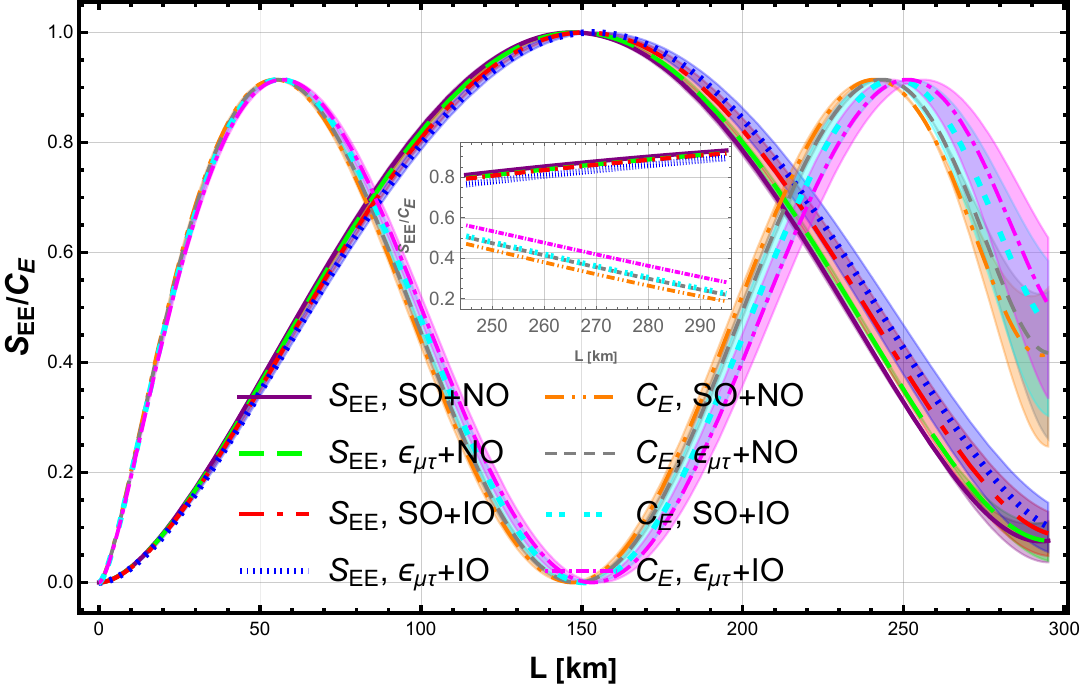}
        \caption{T2K}
        \label{3a_fig:sub1}
    \end{subfigure}
    \hfill
    \begin{subfigure}{0.49\textwidth}
        \centering
        \includegraphics[width=\textwidth]{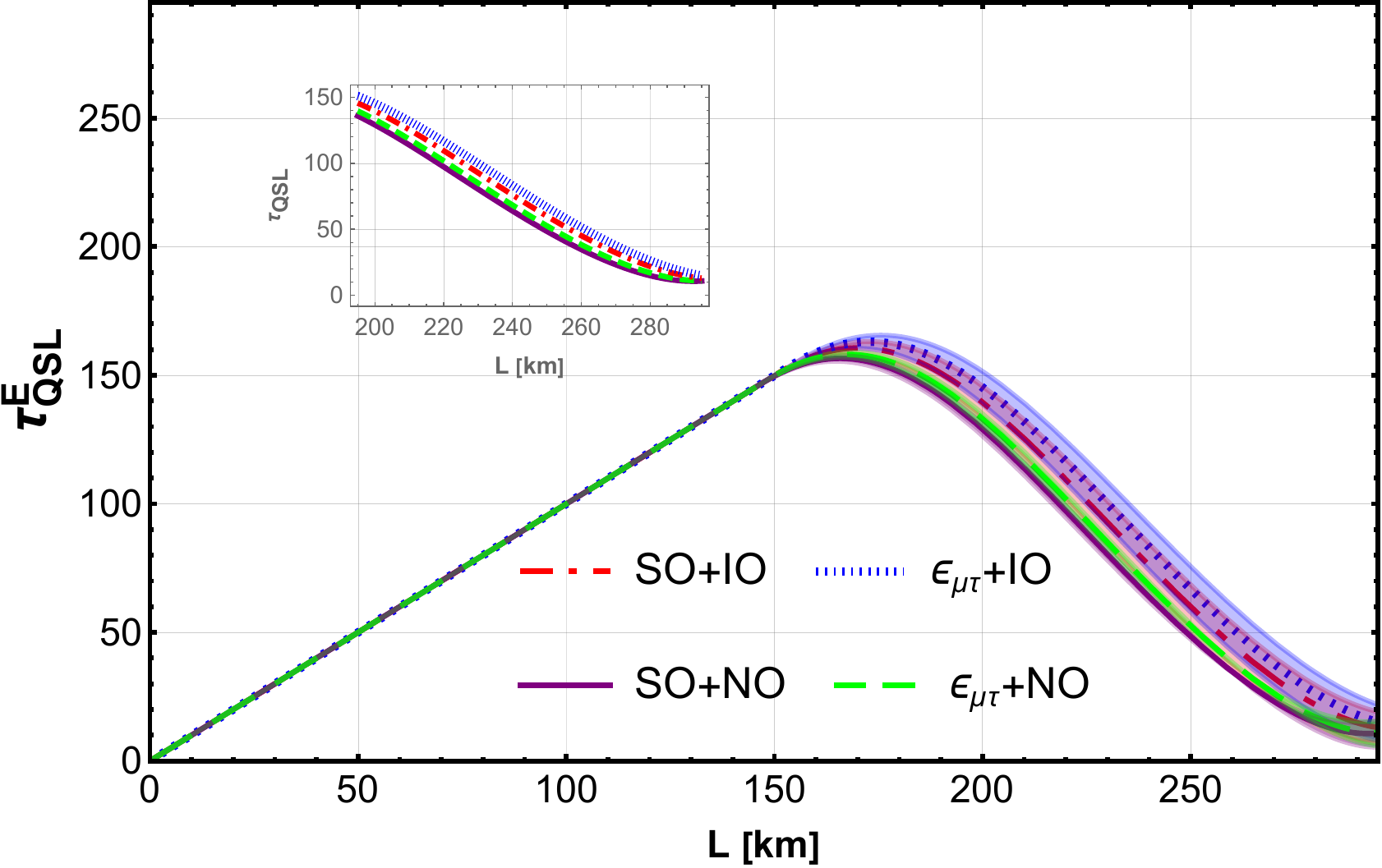}
        \caption{T2K}
        \label{3a_fig:sub2}
    \end{subfigure}
    
    \medskip
    
    \begin{subfigure}{0.49\textwidth}
        \centering
        \includegraphics[width=\textwidth]{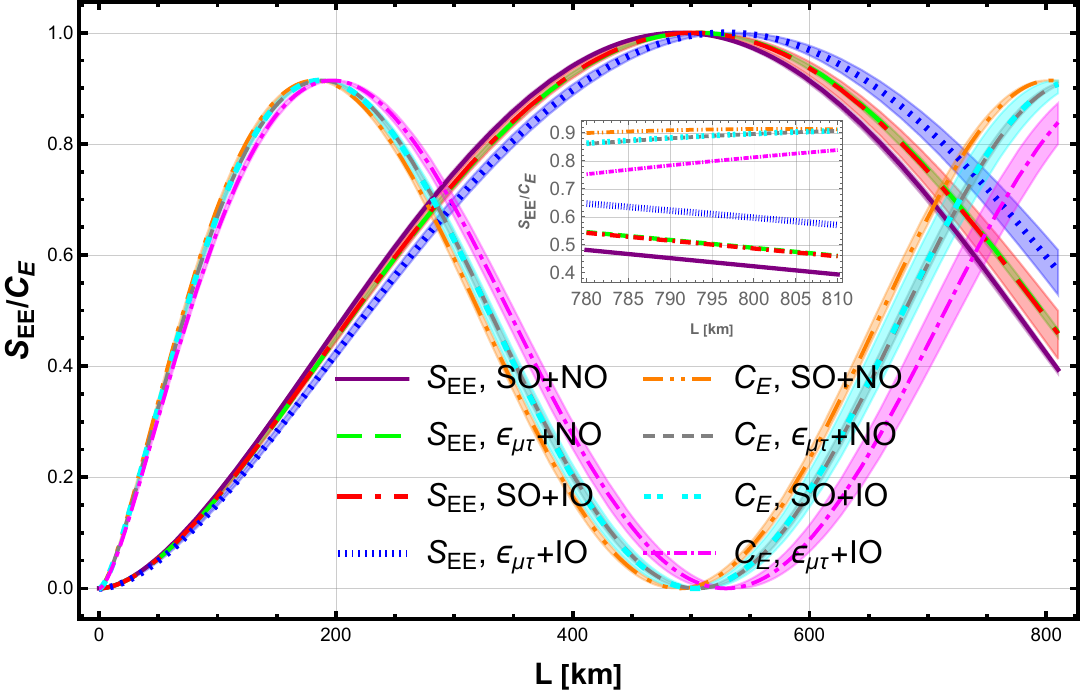}
        \caption{NO$\nu$A}
        \label{3a_fig:sub3}
    \end{subfigure}
    \hfill
    \begin{subfigure}{0.49\textwidth}
        \centering
        \includegraphics[width=\textwidth]{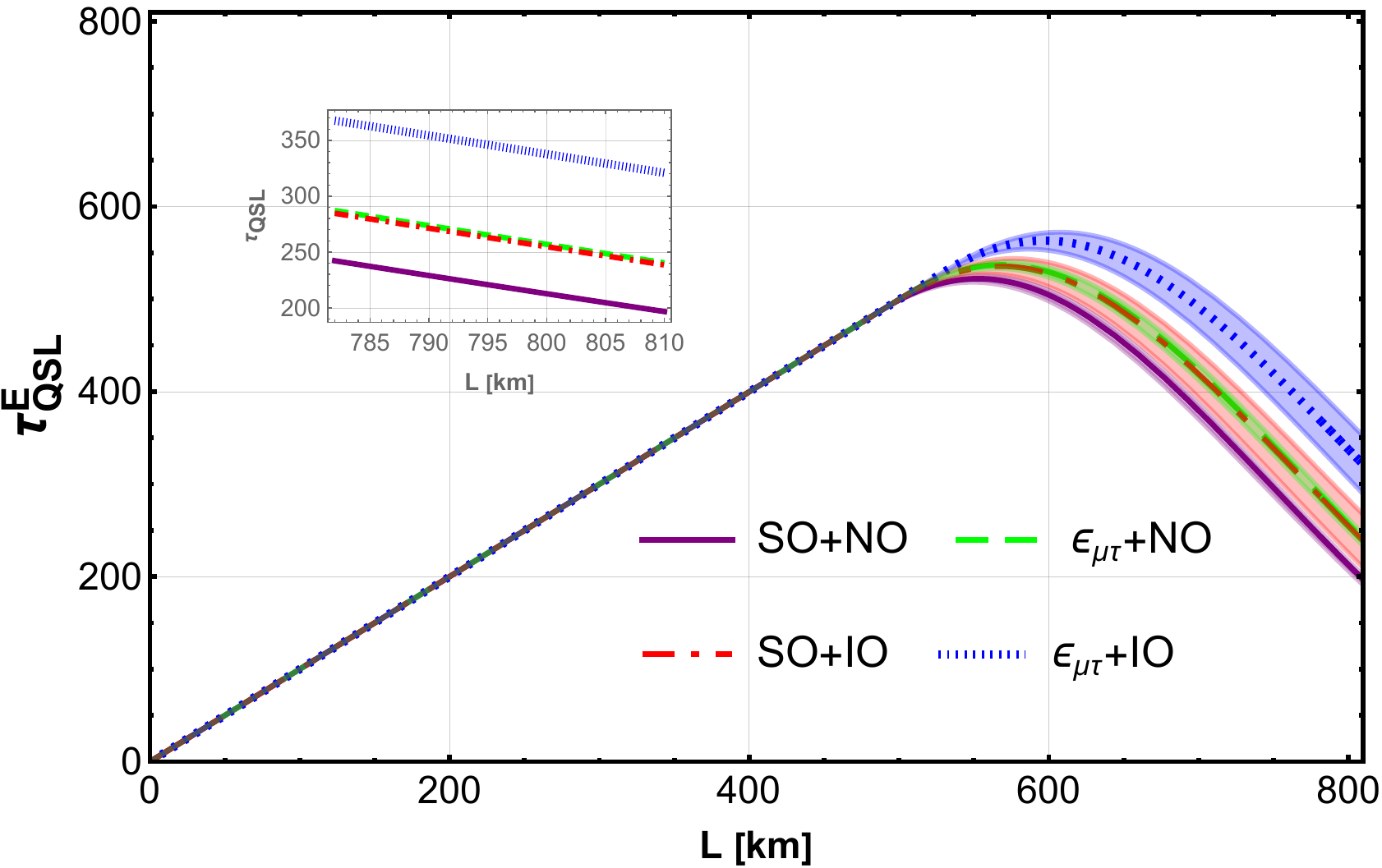}
        \caption{NO$\nu$A}
        \label{3a_fig:sub4}
    \end{subfigure}
    
    \medskip
    
    \begin{subfigure}{0.49\textwidth}
        \centering
        \includegraphics[width=\textwidth]{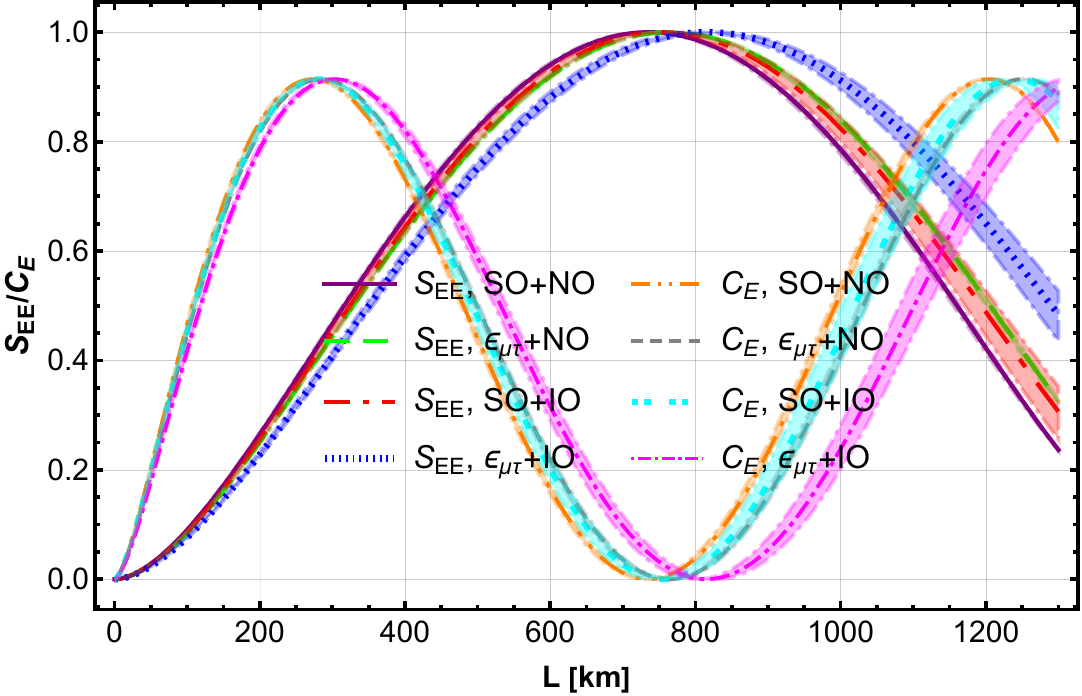}
        \caption{DUNE}
        \label{3a_fig:sub5}
    \end{subfigure}
    \hfill
    \begin{subfigure}{0.49\textwidth}
        \centering
        \includegraphics[width=\textwidth]{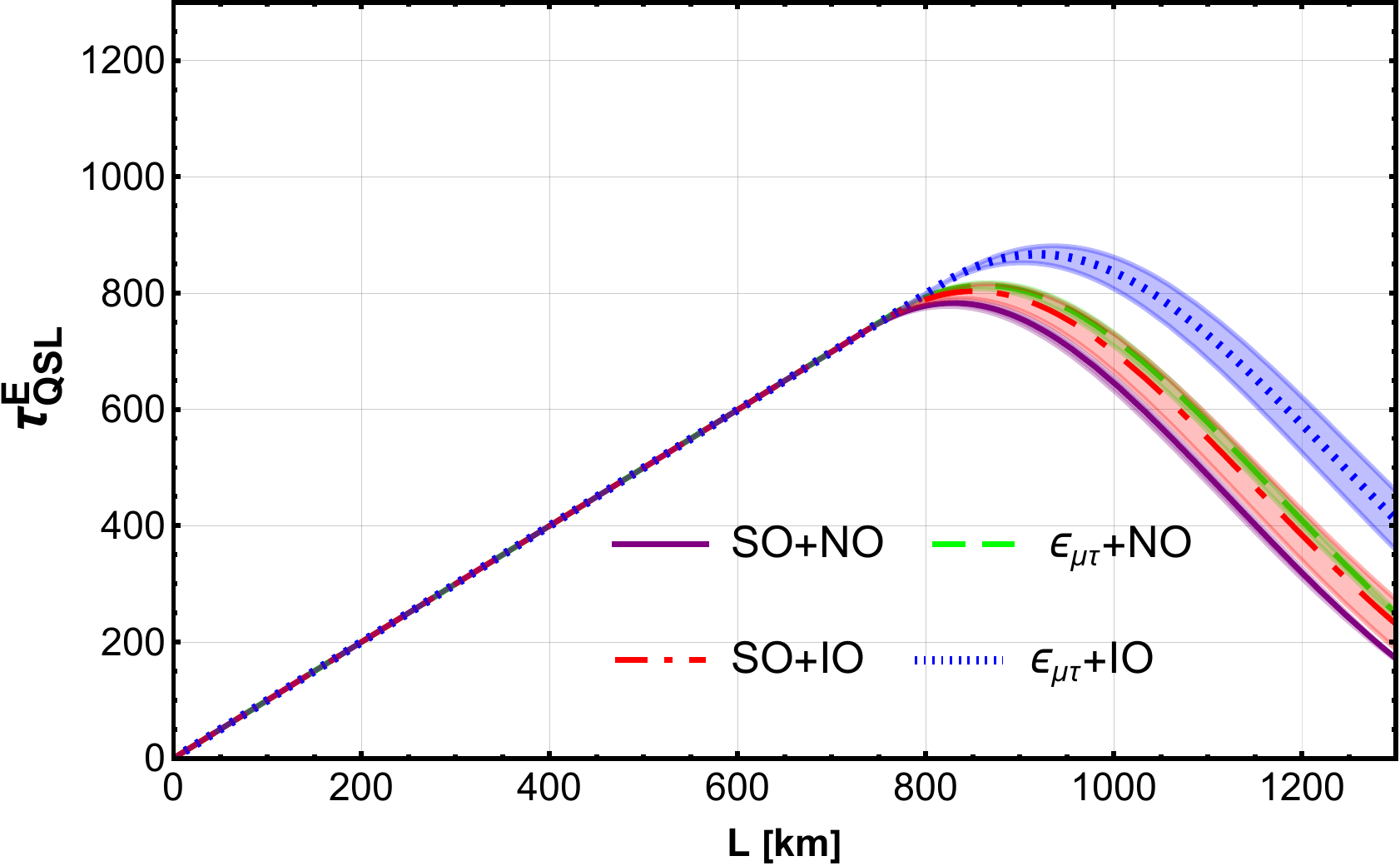}
        \caption{DUNE}
        \label{3a_fig:sub6}
    \end{subfigure}
    \caption{\justifying{From top to bottom, the left panels show the entanglement entropy ($S_{EE}$) and capacity of entanglement ($C_{E}$), while the right panels show the QSL time for bipartite entanglement ($\tau^E_{\rm QSL}$), as functions of the baseline length $L\,\text{(km)}$ of the initial muon-flavor neutrino state $\ket{\nu_\mu}$, evolved under four scenarios: SO+NO (purple solid line), SO+IO (red dot-dashed line), NSI+NO (green dashed line), and NSI+IO (blue dotted line), using the best-fit$\pm 1\sigma$ range of $CP$-violating phases $\delta_{CP}$ in NO  and IO. The $S_{EE}$, $C_{E}$, and $\tau^E_{\rm QSL}$ are compared across these scenarios for the off-diagonal NSI parameter $\left|\epsilon_{\mu\tau}\right|$ with complex phase $\phi_{\mu\tau}$, evaluated at the baselines and energies of the T2K (a)-(b), NO$\nu$A (c)-(d), and DUNE (e)-(f) experiments. The best-fit$\pm 1\sigma$ range of SO and NSI parameters used are taken from Tables~\ref{Tab1} and \ref{Tab2}, respectively.}}
    \label{fig3}
\end{figure*}

\begin{figure*}[!htbp]
    \centering
    \begin{subfigure}{0.44\textwidth}
        \centering
        \includegraphics[width=\textwidth]{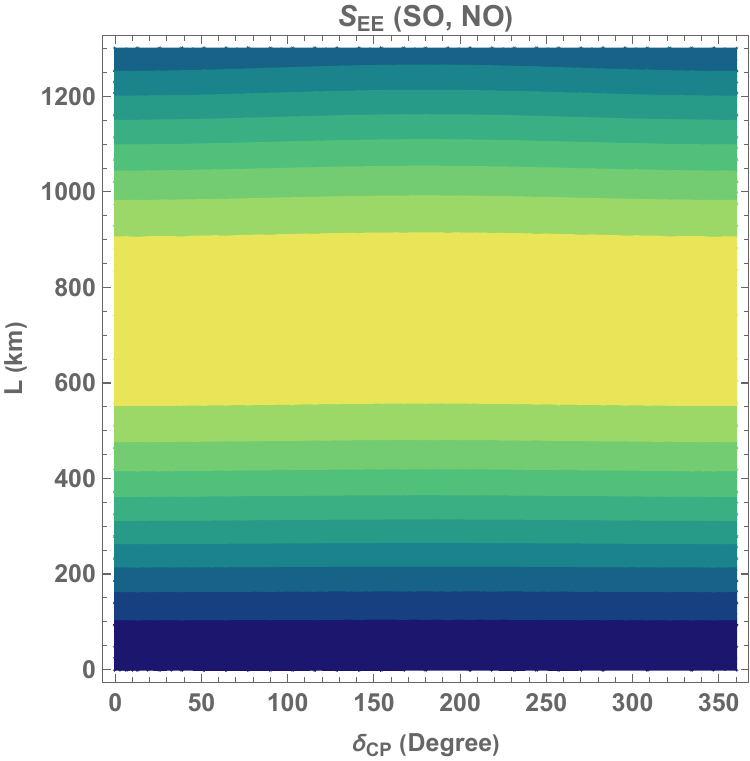}
        \caption{}
        \label{4a_fig:sub1}
    \end{subfigure}
   \hfill
    \begin{subfigure}{0.44\textwidth}
        \centering
        \includegraphics[width=\textwidth]{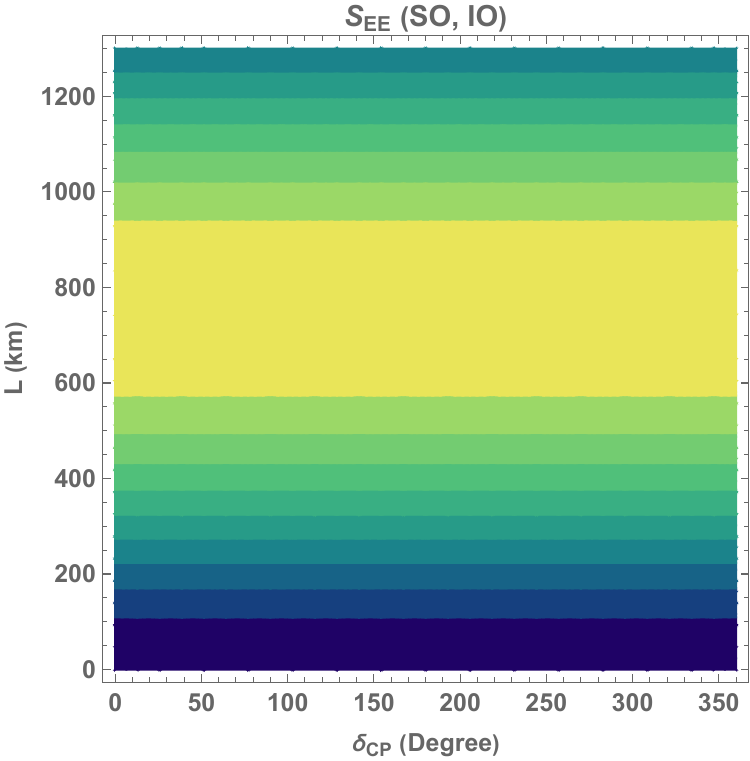}
        \caption{}
        \label{4b_fig:sub2}
    \end{subfigure}
    \hfill
    \begin{subfigure}{0.07\textwidth}
        \centering
        \includegraphics[width=\textwidth]{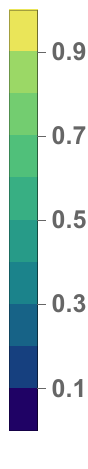}
         \vspace{0.8cm}
     
        \label{4c_fig:sub3}
    \end{subfigure}
    
    \medskip
    
    \begin{subfigure}{0.44\textwidth}
        \centering
        \includegraphics[width=\textwidth]{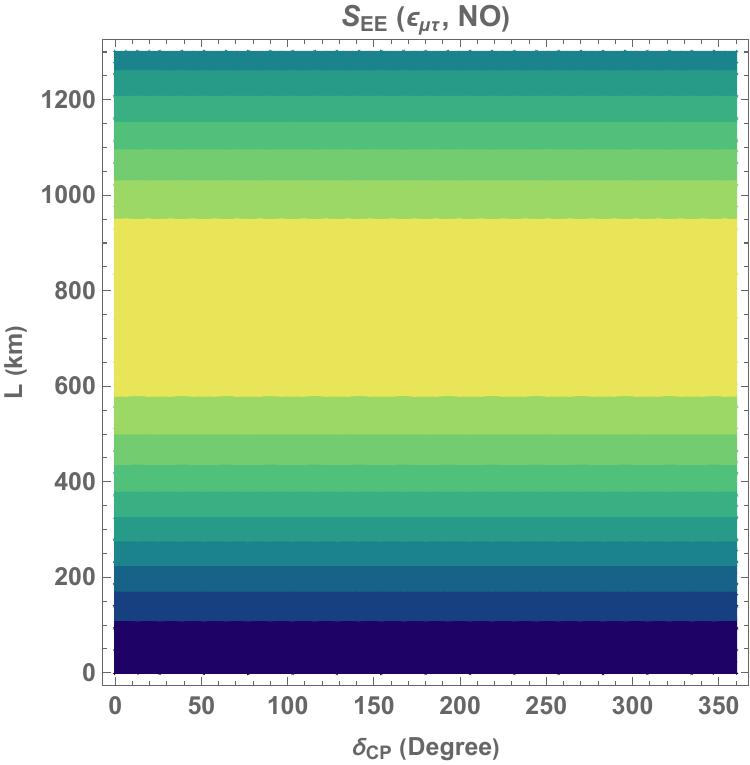}
        \caption{}
        \label{4d_fig:sub4}
    \end{subfigure}
    \hfill
    \begin{subfigure}{0.44\textwidth}
        \centering
        \includegraphics[width=\textwidth]{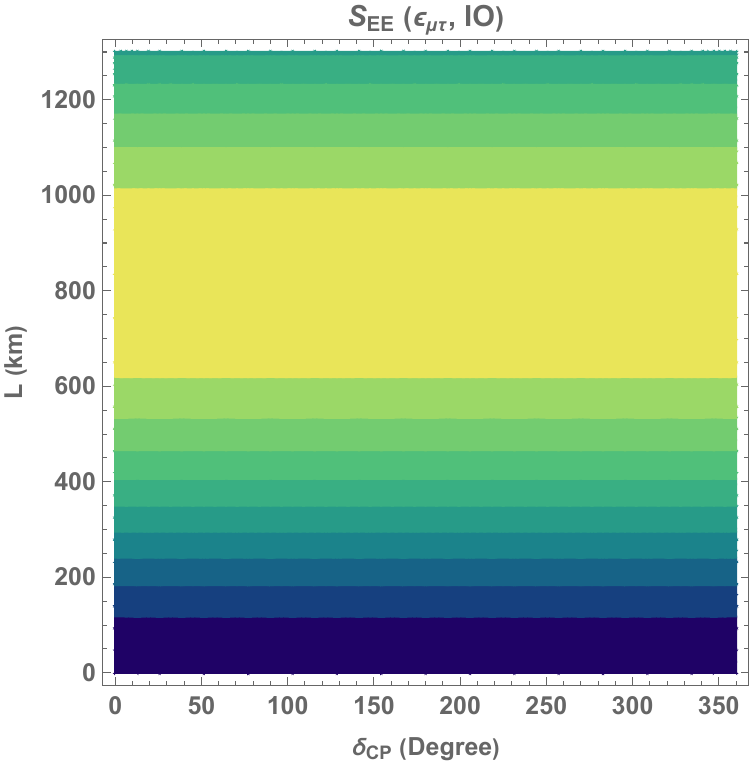}
        \caption{}
        \label{4e_fig:sub5}
    \end{subfigure}
    \hfill
     \begin{subfigure}{0.07\textwidth}
        \centering
        \includegraphics[width=\textwidth]{S_label.pdf}
                \vspace{0.8cm}
    
        \label{4f_fig:sub6}
    \end{subfigure}
    \caption{\justifying{From top to bottom, the panels display the oscillogram plots of the entanglement entropy ($S_{EE}$) in a bipartite system for the initial muon-flavor neutrino state $\ket{\nu_\mu}$, as functions of the baseline length $L\,\text{(km)}$ and the whole range of $CP$-violating phase $\delta_{CP}$ ($0^\circ \leq \delta_{CP}\leq 360^\circ$), under four scenarios: SO+NO (a), SO+IO (b), NSI+NO (c), and NSI+IO (d). The $S_{EE}$ is shown for the off-diagonal NSI parameter $\left|\epsilon_{\mu\tau}\right|$ with complex phase $\phi_{\mu\tau}$, evaluated at the baseline and energy of the DUNE experiment. The best-fit values of SO and NSI parameters used are taken from Tables~\ref{Tab1} and \ref{Tab2}, respectively.}}
    \label{fig4}
\end{figure*}

\begin{figure*}[!htbp]
    \centering
    \begin{subfigure}{0.49\textwidth}
        \centering
        \includegraphics[width=\textwidth]{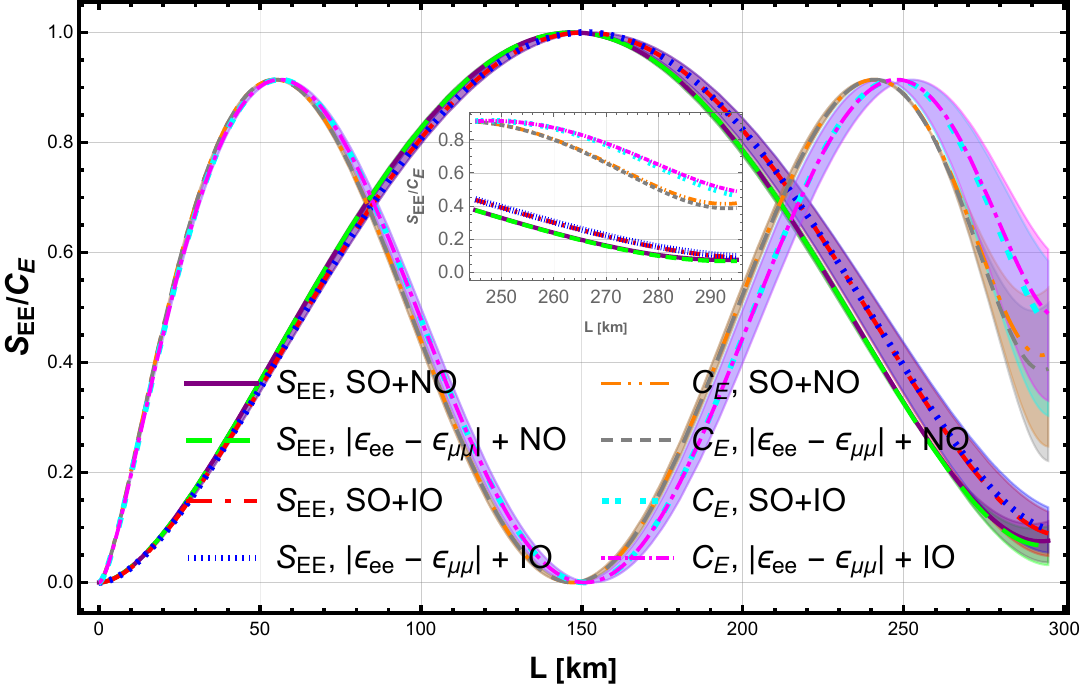}
        \caption{T2K}
        \label{5a_fig:sub1}
    \end{subfigure}
    \hfill
    \begin{subfigure}{0.49\textwidth}
        \centering
        \includegraphics[width=\textwidth]{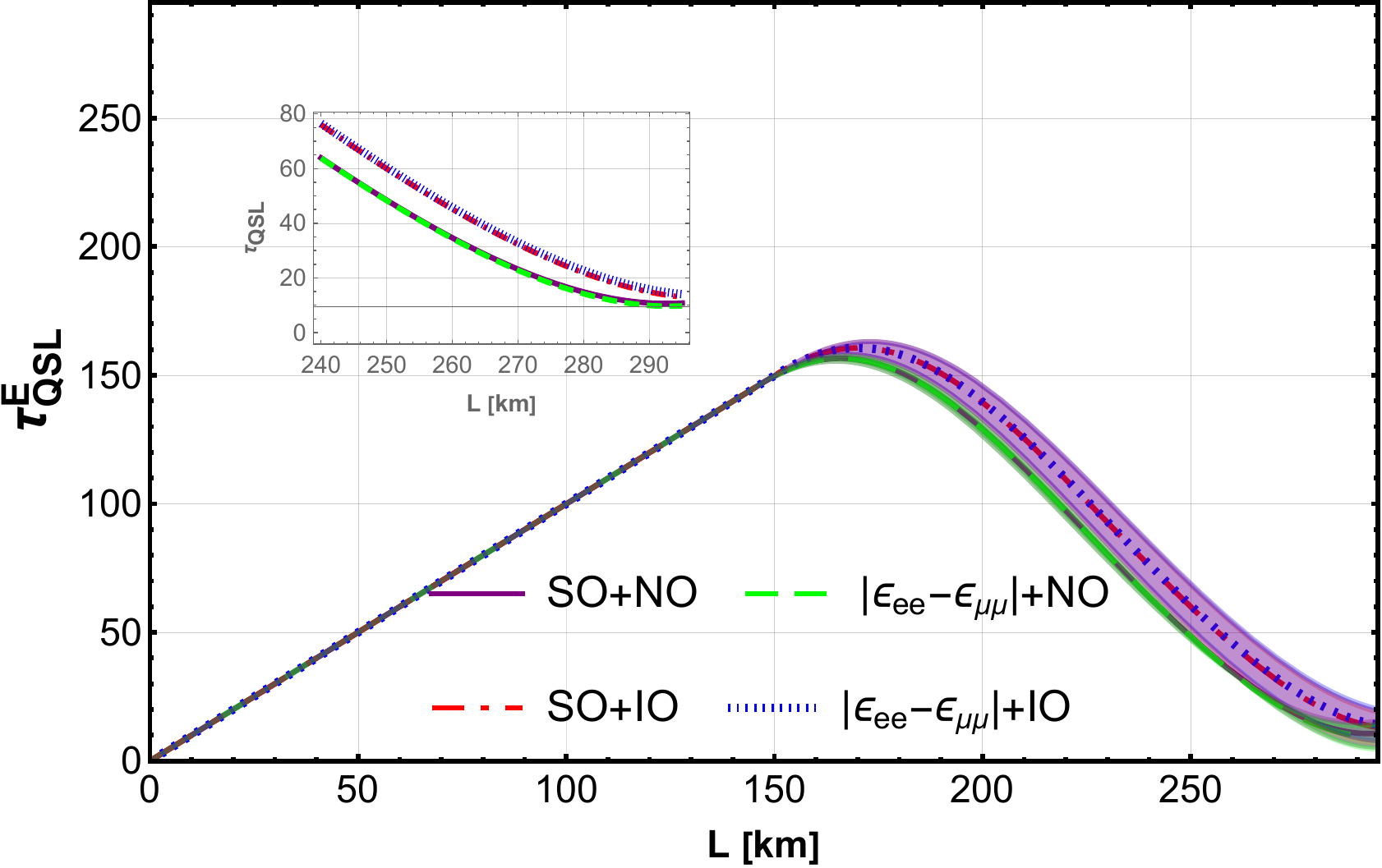}
        \caption{T2K}
        \label{5a_fig:sub2}
    \end{subfigure}
    
    \medskip
    
    \begin{subfigure}{0.49\textwidth}
        \centering
        \includegraphics[width=\textwidth]{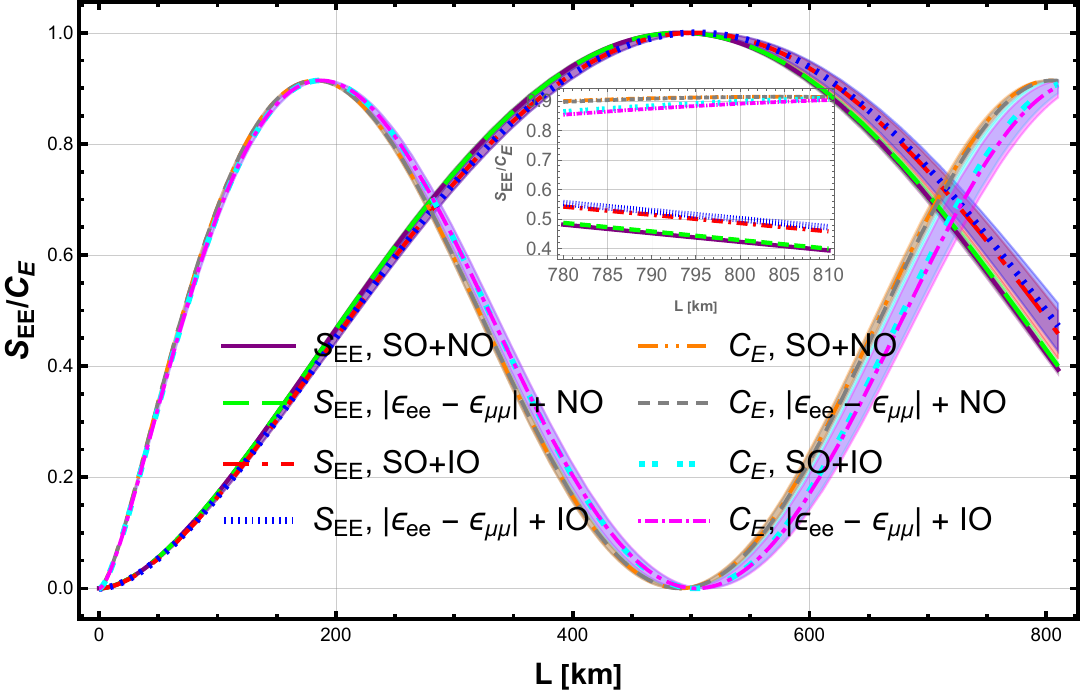}
        \caption{NO$\nu$A}
        \label{5a_fig:sub3}
    \end{subfigure}
    \hfill
    \begin{subfigure}{0.49\textwidth}
        \centering
        \includegraphics[width=\textwidth]{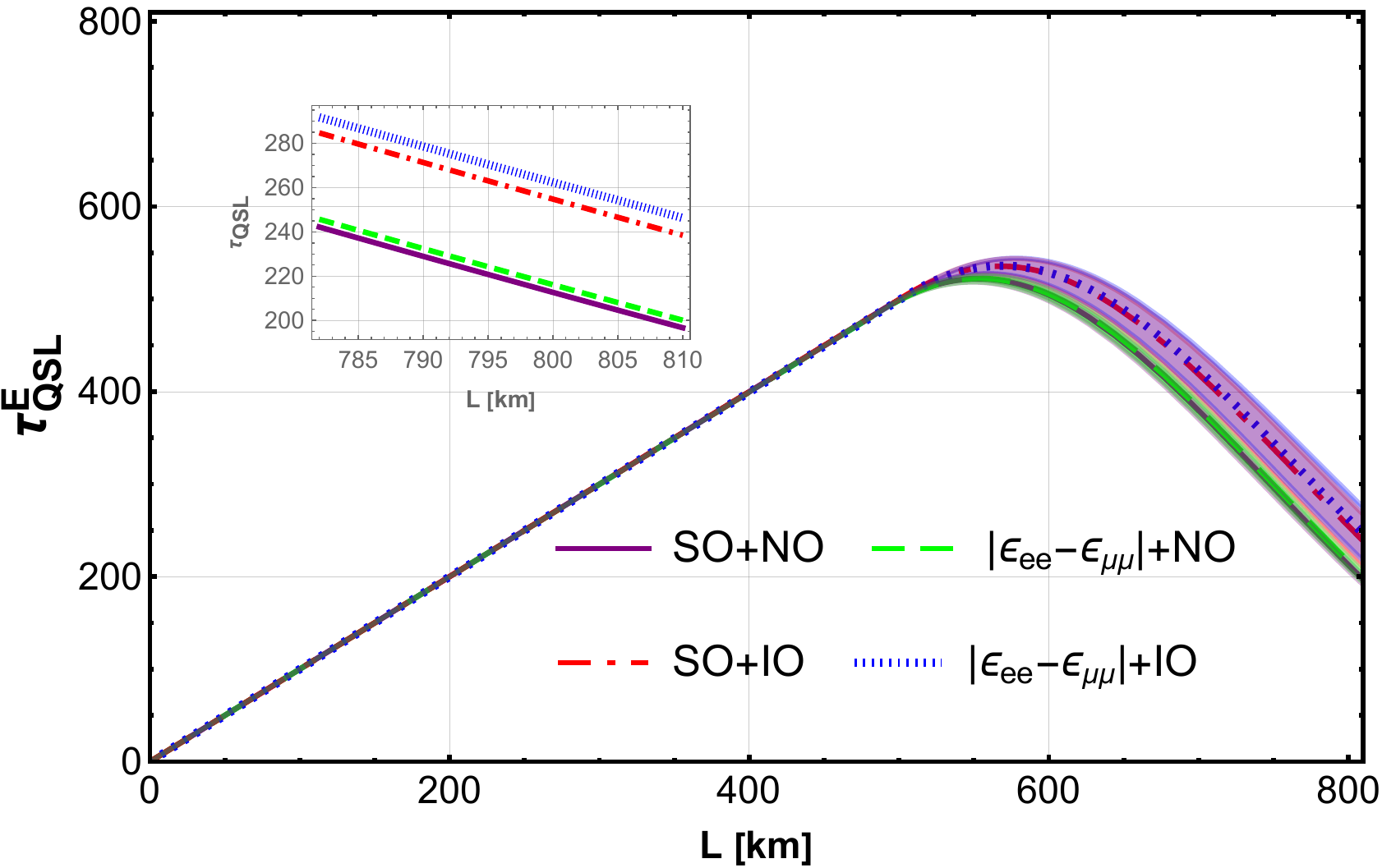}
        \caption{NO$\nu$A}
        \label{5a_fig:sub4}
    \end{subfigure}
    
    \medskip
    
    \begin{subfigure}{0.49\textwidth}
        \centering
        \includegraphics[width=\textwidth]{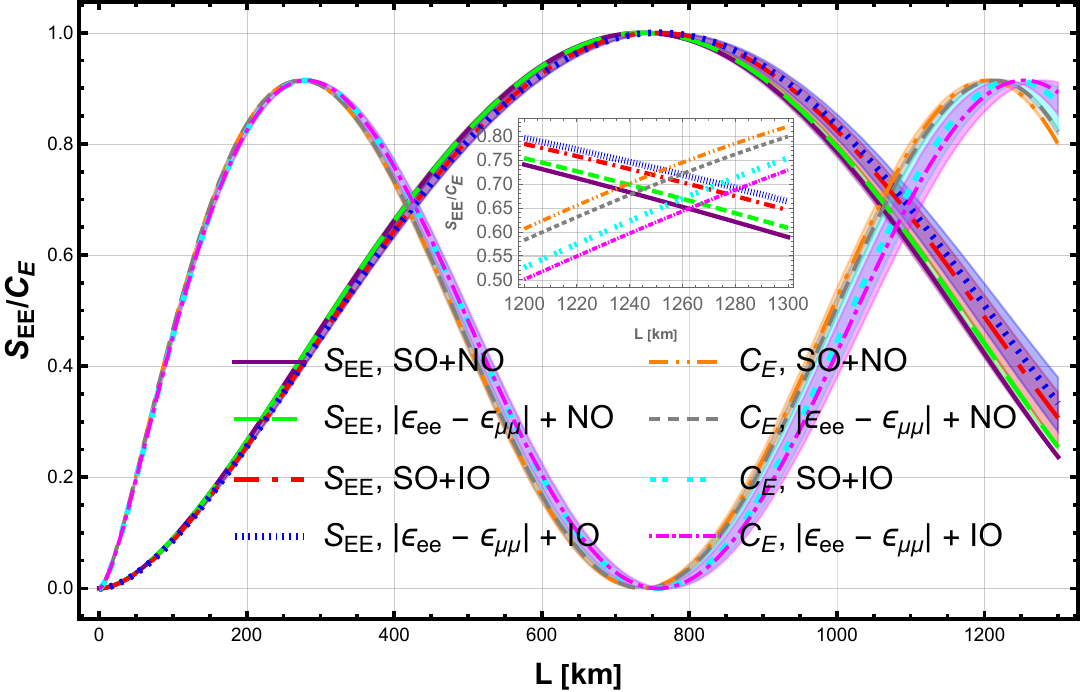}
        \caption{DUNE}
        \label{5a_fig:sub5}
    \end{subfigure}
    \hfill
    \begin{subfigure}{0.49\textwidth}
        \centering
        \includegraphics[width=\textwidth]{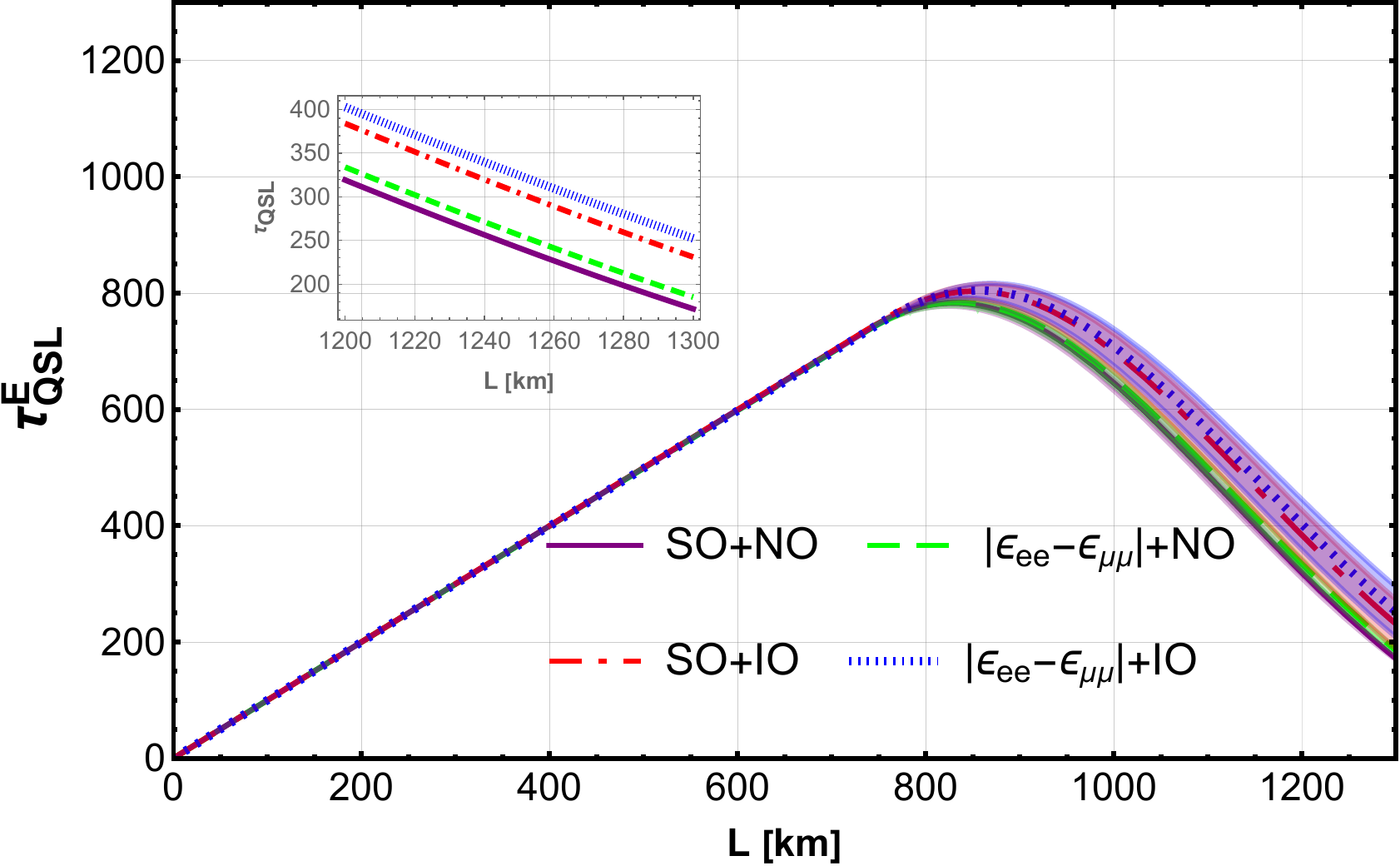}
        \caption{DUNE}
        \label{5a_fig:sub6}
    \end{subfigure}
    \caption{\justifying{From top to bottom, the left panels show the entanglement entropy ($S_{EE}$) and capacity of entanglement ($C_{E}$), while the right panels show the QSL time for bipartite entanglement ($\tau^E_{\rm QSL}$), as functions of the baseline length $L\,\text{(km)}$ of the initial muon-flavor neutrino state $\ket{\nu_\mu}$, evolved under four scenarios: SO+NO (purple solid line), SO+IO (red dot-dashed line), NSI+NO (green dashed line), and NSI+IO (blue dotted line), using the best-fit$\pm 1\sigma$ range of $CP$-violating phases $\delta_{CP}$ in NO and IO. The $S_{EE}$, $C_{E}$, and $\tau^E_{\rm QSL}$ are compared across these scenarios for the diagonal NSI parameter $\left |\epsilon_{ee}-\epsilon_{\mu\mu}\right |$, evaluated at the baselines and energies of the T2K (a),(b), NO$\nu$A (c),(d), and DUNE (e),(f) experiments. The best-fit$\pm 1\sigma$ range of SO and NSI parameters used are taken from Tables~\ref{Tab1} and \ref{Tab3}, respectively.}}
    \label{fig5}
\end{figure*}

Moreover, following the previous section, this section also considers four cases of neutrino propagation: SO+NO, SO+IO, NSI+NO and NSI+IO, with their respective best-fit $CP$-violating phases $\delta_{CP}$. The behavior of the entanglement entropy ($S_{EE}$) and the capacity of entanglement ($C_E$) as functions of the baseline length $L\text{(km)}$ are illustrated in Figs.~\ref{fig3}(\subref{3a_fig:sub1}), \ref{fig3}(\subref{3a_fig:sub3}), and \ref{fig3}(\subref{3a_fig:sub5}) for the initial state $\ket{\nu_\mu}$ propagating under SO, with and without the off-diagonal NSI parameter $\epsilon_{\mu\tau}$, in both NO and IO, using their respective best-fit $CP$-violating phases $\delta_{CP}$, for T2K, NO$\nu$A, and DUNE, respectively. We find that at $L=0$, the bipartite entanglement measures $S_{EE}$ and $C_E$ vanish for all experiments in all four cases, indicating that the initial muon-flavor neutrino state $\ket{\nu_\mu}$ is an unentangled state (i.e., a separable state). For $L>0$, both $S_{EE}$ and $C_E$ become nonzero across all experiments in all four cases, implying that the evolved muon-flavor neutrino state $\ket{\nu_\mu(L)}$ is a pure bipartite entangled state.

Moreover, Fig.~\ref{fig3}(\subref{3a_fig:sub1}), corresponding to the T2K baseline and energy, shows that $S_{EE}$ increases monotonically with baseline length and reaches its maximal value of unity for all cases near $L \approx 150\,\mathrm{km}$. This indicates that the initial muon-flavor neutrino state $\ket{\nu_\mu}$ evolved in T2K exhibits maximally pure bipartite entanglement in all cases. A separation in $S_{EE}$ is observed between SO+NO (purple solid line) and SO+IO (red dash-dotted line) in the baseline range $150<L\,(\mathrm{km})<295$ for T2K, with a similar discrepancy also observed between $\epsilon_{\mu\tau}$+NO (green dashed line) and $\epsilon_{\mu\tau}$+IO (blue dotted line) when the off-diagonal NSI parameter $\epsilon_{\mu\tau}$ is taken into account. We conclude that, in this baseline range, $S_{EE}$ is suppressed for the SO+NO case compared to the other three cases, with observable discrepancies among all four cases. However, at the end of the baseline ($L = 295\,\mathrm{km}$), $S_{EE}$ coincides for all four cases, showing no discrepancies. Similar to $S_{EE}$, discrepancies in $C_{E}$ are also observed among all four cases at the T2K baseline. In all four cases, $C_E$ vanishes for the separable ($S_{EE}=0$) and maximally entangled ($S_{EE}=1$) states, reaching its highest value at the partially entangled state ($S_{EE}=1/2$).

Furthermore, Fig.~\ref{fig3}(\subref{3a_fig:sub3}) indicates that, for the NO$\nu$A baseline and energy, $S_{EE}$ for SO+NO is suppressed relative to all other cases toward the end of the baseline. The inclusion of the off-diagonal NSI parameter $\epsilon_{\mu\tau}$ leads to an enhancement of $S_{EE}$ in both mass orderings: $S_{EE}$ is enhanced for the $\epsilon_{\mu\tau}$+NO case compared to SO+NO, and similarly enhanced for the $\epsilon_{\mu\tau}$+IO case relative to SO+IO, whereas the $C_E$ is suppressed in both NSI scenarios relative to their corresponding SO cases. Unlike T2K, discrepancies among different cases are present in $S_{EE}$ at the end of the NO$\nu$A baseline, except between the SO+IO and $\epsilon_{\mu \tau}$+NO cases, which almost overlap.
 
Moreover, $S_{EE}\rightarrow 1$ near $L\approx 525\,\rm km$ implies that the initial state $\ket{\nu_\mu}$ evolving in NO$\nu$A forms a maximally pure bipartite entangled state. On the other hand, for all four cases, $C_E\rightarrow 0$ in a separable ($S_{EE}=0$) and a maximally entangled state ($S_{EE}=1$), while it reaches a maximum in a partially entangled state ($S_{EE}=1/2$). 

For $S_{EE}$ and $C_{E}$ in all four cases, Fig.\,\ref{fig3}(\subref{3a_fig:sub5}), corresponding to the baseline and energy of DUNE, shows behavior similar to that observed in Fig.\,\ref{fig3}(\subref{3a_fig:sub3}) for NO$\nu$A. However, in Fig.\,\ref{fig3}(\subref{3a_fig:sub5}), a large suppression of entanglement is observed for SO+NO compared to all other cases toward the end of the baseline. Near this region, a small separation in $S_{EE}$ is also observed between SO+IO and $\epsilon_{\mu \tau}$+NO. Unlike NO$\nu$A, where discrepancies in $S_{EE}$ and $C_{E}$ among the four cases are relatively small, DUNE exhibits pronounced discrepancies. This arises because DUNE’s longer baseline induces stronger matter effects during neutrino propagation, leading to enhanced contributions to $S_{EE}$ and $C_{E}$ compared to NO$\nu$A. Moreover, $S_{EE}$ tends to 0 and 1 near $L=0$ and $L\approx 800\,\rm km$, respectively; consequently, $C_{E}$ also tends to zero at these distances. However, $C_E$ attains a maximum in a partially entangled state ($S_{EE}=1/2$). These results indicate that, similar to T2K and NO$\nu$A, the evolved state $\ket{\nu_\mu(L)}$ in DUNE also forms a maximally pure bipartite entangled state.
Notably, $S_{EE}$ and $C_{E}$ exhibit distinct behavior among all four cases at the end of the DUNE baseline, in contrast to T2K and NO$\nu$A.

It is worth noting that Figs.\,\ref{fig3}(\subref{3a_fig:sub1}), \ref{fig3}(\subref{3a_fig:sub3}) and \ref{fig3}(\subref{3a_fig:sub5}), corresponding to T2K, NO$\nu$A and DUNE, show that a maximally pure bipartite entanglement is achieved for the state $\ket{\nu_\mu(L)}$ in all four cases at baselines and energies $L \approx 150\,\rm{km}$ and $E\approx 0.6\,\rm GeV$, $L \approx 525\,\rm{km}$ and $E\approx 2\,\rm GeV$, and $L \approx 800\,\rm{km}$ and $E\approx 3\,\rm GeV$, respectively. This occurs because, at these baselines and energies, the transition probabilities $P(\nu_{\mu}\rightarrow \nu_{\mu})$ and $P(\nu_{\mu}\rightarrow \nu_{e}) + P(\nu_{\mu}\rightarrow \nu_{\tau})$ [or $\eta_3(L)$ and $\eta_4(L)$ in Eq.\,(\ref{4.4})] both equal $1/2$, as evident from the corresponding probability plots for each experiment: Figs.\,\ref{fig1}(\subref{1a_fig:sub1})-\ref{fig1}(\subref{1a_fig:sub3}) for T2K, Figs.\,\ref{fig1}(\subref{1a_fig:sub4})-\ref{fig1}(\subref{1a_fig:sub6}) for NO$\nu$A, and Figs.\,\ref{fig1}(\subref{1a_fig:sub7})-\ref{fig1}(\subref{1a_fig:sub9}) for DUNE. Thus, at these baselines and energies of T2K, NO$\nu$A and DUNE, the initial muon-flavor neutrino state $\ket{\nu_\mu}$ propagating in all four cases exhibits a Bell-like state, which corresponds to a maximally pure bipartite entangled state in quantum information theory. 

For the analysis of $S_{EE}$ in the bipartite system, we used a fixed best-fit value of $\delta_{CP}$ for the four cases. However, the general behavior of $S_{EE}$ as a function of baseline length remains consistent for other values of $\delta_{CP}$ in the four cases. To examine such behavior, an oscillogram is shown in Fig.\,\ref{fig4}, where we focus on the dependency of the baseline length and energy of the DUNE experiment, along with the full range of $\delta_{CP}$ ($0^\circ \leq \delta_{CP} \leq 360^\circ$), on $S_{EE}$ in the four cases using the complex off-diagonal NSI parameter $\epsilon_{\mu \tau}$. In all four panels of Fig.\,\ref{fig4}, dark blue represents the minimum value of $S_{EE}$, while dark yellow indicates the maximum value of $S_{EE}$. It is observed that in the length range of $[550 - 900]\,\rm km$, for the SO+NO case [Fig.\,\ref{fig4}(\subref{4a_fig:sub1})], $S_{EE}$ achieves the maximum value ($\sim 1$) for all values of $\delta_{CP}$. Similarly, SO+IO [Fig.\,\ref{fig4}(\subref{4b_fig:sub2})] nearly mirrors the behavior of SO+NO, attaining a maximum value of about $1$ at the length scale of $[580 - 950]\,\rm km$. The length scale for maximum $S_{EE}$ in the case of $\epsilon_{\mu \tau}$+NO [Fig.\,\ref{fig4}(\subref{4d_fig:sub4})] is slightly broader, around $590 \leq L \,\rm (km) \leq 950$, as compared to the SO+NO case, for all values of $\delta_{CP}$. For the $\epsilon_{\mu \tau}$+IO case [Fig.\,\ref{fig4}(\subref{4e_fig:sub5})], the length scale for which maximum $S_{EE}$ is achieved is around $[610 - 1010]\,\rm km$. These results emphasize that, among all four cases, distinct behavior of maximally pure bipartite entanglement for the initial state $\ket{\nu_\mu}$ is observed across different ranges of the DUNE baseline and over the entire $\delta_{CP}$ range.

The weak dependence of $S_{EE}$ on $\delta_{CP}$ observed in Fig.\,\ref{fig4} can be understood from the definition of the entanglement entropy. In our analysis, $S_{EE}$ depends only on the survival probability $P(\nu_\mu \rightarrow \nu_\mu)$ and its complement, $1-P(\nu_\mu \rightarrow \nu_\mu)$. Since the $\nu_\mu$ disappearance channel exhibits only a weak sensitivity to the $CP$-violating phase $\delta_{CP}$, both in the standard three-flavor framework and in the NSI scenarios considered here, variations in $\delta_{CP}$ produce only minor changes in the survival probability. Consequently, these small variations are directly reflected in the nearly $\delta_{CP}$-independent behavior of $S_{EE}$ across all four panels of Fig.\,\ref{fig4}.

Furthermore, Figs.\,\ref{fig5}(\subref{5a_fig:sub1}), \ref{fig5}(\subref{5a_fig:sub3}), and \ref{fig5}(\subref{5a_fig:sub5}) illustrate $S_{EE}$ and $C_E$ as functions of $L$ for T2K, NO$\nu$A, and DUNE, respectively, when the initial state $\ket{\nu_\mu}$ propagates under SO and in the presence of the diagonal NSI parameter $|\epsilon_{ee}-\epsilon_{\mu\mu}|$, for both NO and IO scenarios, with the best-fit value of $\delta_{CP}$. 
Fig.\,\ref{fig5}(\subref{5a_fig:sub1}) shows that, while $S_{EE}$ merges for all four cases at the end of the T2K baseline, $C_E$ exhibits separation between SO+NO and SO+IO, as well as between $|\epsilon_{ee}-\epsilon_{\mu\mu}|$+NO and $|\epsilon_{ee}-\epsilon_{\mu\mu}|$+IO.  However, Figs.\,\ref{fig5}(\subref{5a_fig:sub3}) and \ref{fig5}(\subref{5a_fig:sub5}), corresponding to NO$\nu$A and DUNE, respectively, show mild discrepancies in $S_{EE}$ and $C_E$ among the four cases near the end of the baselines. Moreover, for the state $\ket{\nu_\mu(L)}$, a maximally pure bipartite entangled state is realized in T2K, NO$\nu$A, and DUNE for all four cases. These results arise from the behavior of $P(\nu_{\mu}\rightarrow \nu_{e})$, $P(\nu_{\mu}\rightarrow \nu_{\mu})$, and $P(\nu_{\mu}\rightarrow \nu_{\tau})$, as observed in Fig.\,\ref{fig2} for T2K, NO$\nu$A, and DUNE.

Similar to the above analysis, in the Appendix we also present the behavior of $S_{EE}$ and $C_E$ as functions of $L$ in Figs.\,\ref{fig9}-\ref{fig11} for the initial state $\ket{\nu_\mu}$ propagating under SO with and without complex off-diagonal NSI parameters $\epsilon_{e\mu}$ and $\epsilon_{e\tau}$, as well as the diagonal NSI parameter $|\epsilon_{\tau\tau}-\epsilon_{\mu\mu}|$, respectively. The results are shown for both NO and IO, using the best-fit value of $\delta_{CP}$, and considering the baseline and energy configurations of T2K, NO$\nu$A, and DUNE in Figs.\,\ref{fig9}(\subref{9a_fig:sub1}), \ref{fig9}(\subref{9a_fig:sub3}), \ref{fig9}(\subref{9a_fig:sub5}); Figs.\,\ref{fig10}(\subref{10a_fig:sub1}), \ref{fig10}(\subref{10a_fig:sub3}), \ref{fig10}(\subref{10a_fig:sub5}); and Figs.\,\ref{fig11}(\subref{11a_fig:sub1}), \ref{fig11}(\subref{11a_fig:sub3}), \ref{fig11}(\subref{11a_fig:sub5}), respectively. Across all experiments, no pronounced discrepancies in $S_{EE}$ and $C_E$ are observed among the four cases. However, the qualitative conclusion remains unchanged: T2K, NO$\nu$A, and DUNE exhibit maximally pure bipartite entanglement for all four cases of neutrino propagation. These results can be attributed to the behavior of $P(\nu_{\mu}\rightarrow \nu_{e})$, $P(\nu_{\mu}\rightarrow \nu_{\mu})$, and $P(\nu_{\mu}\rightarrow \nu_{\tau})$, as shown in Figs.\,\ref{fig6}-\ref{fig8} in the Appendix.

Moreover, for the state $\ket{\nu_\mu(L)}$, following Eq.\,(\ref{2.7}), the effective Hamiltonian in flavor basis is
\begin{equation}
    \mathcal{H}^f_{tot}=U \mathcal{H}_{tot} U^{-1},
    \label{4.8}
\end{equation} 
where $\mathcal{H}^f_{tot}$ is Hermitian, ${\mathcal{H}^f}^\dagger_{tot}=\mathcal{H}^f_{tot}$ and time independent. 
Using Eqs.\,(\ref{4.8}) and (\ref{5}), we determine variance of the time independent driving Hamiltonian ($ \mathcal{H}^f_{tot}$) as
\begin{equation}
   \Delta{\mathcal{H}}^f_{tot}= \sqrt{\langle   (\mathcal{H}^f_{tot})^2\rangle - \langle 
\mathcal{H}^f_{tot}\rangle ^2}.
\label{4.9}
\end{equation}
 Thus, substituting Eqs.\,(\ref{4.5}), (\ref{4.6}) and (\ref{4.9}) into Eq.\,(\ref{2}), the QSL time for bipartite entanglement of the length-evolved state $\ket{\nu_\mu(L)}$ in SO with NSI effects can be quantified as
\begin{equation}
 \tau^E_{\mathrm{QSL}}(L) =
\frac{\left| S_{EE}(\rho_{\mu e}(L))\right|}
{2\Delta{\mathcal{H}}^f_{tot}\frac{1}{L}\int_0^L \sqrt{ C_{E}(\rho_{\mu e}(L))}\,dl} \hspace{0.2cm},
\label{4.10}
\end{equation}
where $S_{EE}(\rho_{\mu e}(0))=0$.

Furthermore, as follows from Eq.\,(\ref{2.7}), the total Hamiltonian ($\mathcal{H}_{tot}$), 
is expressed in the mass basis. Let $\mathcal{H}_{vac}=\text{diag}(E_1, E_2, E_3)$, and $\mathcal{H}^{'}_{NSI}=\mathcal{H}_{mat}+ \mathcal{H}_{NSI}=U^{\dagger }V_{\rm CC}\epsilon_{NSI}U$ denote the vacuum and NSI with matter Hamiltonian terms appearing in Eq.\,(\ref{2.7}), respectively. Then,
$ [\mathcal{H}_{vac},\mathcal{H}^{'}_{NSI}]_{ij}=(E_i-E_j)(\mathcal{H}^{'}_{NSI})_{ij}.$
So the commutator vanishes only if 
$(E_i-E_j)(\mathcal{H}^{'}_{NSI})_{ij}=0 $
for all $i$, $j$. Since generally $E_{i}\neq E_j$, the two terms commute only if $\mathcal{H}^{'}_{NSI}$ is also diagonal in the mass basis. But because of the mixing matrix $U$, the NSI with matter term is generally not diagonal in the mass basis. Therefore, 
\begin{equation}
    [\mathcal{H}_{vac}, \mathcal{H}^{'}_{NSI}]\neq 0
\end{equation}
in general. They commute in special cases, for example $U^\dagger\epsilon_{NSI} U$ is diagonal, or all relevant mass energies are degenerate, $E_{i}=E_{j}$. Otherwise, the vacuum oscillation part and the NSI with matter part do not commute. This statement holds not only in the mass basis but also in the flavor basis, as commutators are, in general, basis-independent quantities. Moreover, since the total Hamiltonian is time independent and Hermitian ($\mathcal{H}_{tot}^\dagger = \mathcal{H}_{tot}$), the corresponding evolution is unitary. Therefore, the neutrino system may be regarded as a closed bipartite system driven by the external Hamiltonian ($\mathcal{H}^{'}_{NSI}$). As a result, the QSL time expression in Eq.\,(\ref{4.10}) remains applicable to the total Hamiltonian governing neutrino evolution in the presence of NSI with matter term.

Furthermore, besides depending on the neutrino energy and matter-density parameters, the total Hamiltonian ($\mathcal{H}_{tot}$) also depends on the mass-squared differences ($\Delta m^2_{ij}$). Consequently, $\mathcal{H}_{tot}$ differs between the NO and IO scenarios and is therefore not unique across different mass ordering schemes. In this sense, $\mathcal{H}_{tot}$ (or $\mathcal{H}^f_{tot}$) is not fixed in the present analysis.

In Figs.\,\ref{fig3}(\subref{3a_fig:sub2}), \ref{fig3}(\subref{3a_fig:sub4}), and \ref{fig3}(\subref{3a_fig:sub6}), the QSL time for bipartite entanglement, $\tau^E_{\rm QSL}$, is illustrated as a function of the baseline length $L\text{(km)}$, for the initial state $\ket{\nu_\mu}$ propagating under SO, with and without the off-diagonal NSI parameter $\epsilon_{\mu\tau}$, in both NO and IO, using their respective best-fit $CP$-violating phases $\delta_{CP}$, for T2K, No$\nu$A, and DUNE, respectively. In Fig.\,\ref{fig3}(\subref{3a_fig:sub2}), the behavior of $\tau^E_{\rm QSL}$ as a function of $L$ leads to the time-bound condition $\tau^E_{\rm QSL}/L=1$ in the baseline range $0<L\,(\rm km)<150$. This result implies that, in all four cases, bipartite entanglement during the evolution of the initial state $\ket{\nu_\mu}$ has already reached its maximum possible speed. However, once the initial state $\ket{\nu_\mu}$ reaches the maximum entanglement entropy ($S_{EE}=1$) at $L \approx 150\,\mathrm{km}$, as shown in Fig.\,\ref{fig3}(\subref{3a_fig:sub1}), the evolved state $\ket{\nu_\mu(L)}$ begins to satisfy the time-bound condition $\tau^E_{\rm QSL}/L < 1$, indicating that the bipartite entanglement in the evolved state can be further speed up.
Thus, Fig.\,\ref{fig3}(\subref{3a_fig:sub2}) indicates that, for T2K in the range $150 < L\,(\mathrm{km}) < 295$, the following empirical relation holds:
\begin{align}
    \left(\frac{\tau^E_{\rm QSL}}{L}\right)^{\rm T2K}_{\rm SO+NO}&<\left(\frac{\tau^E_{\rm QSL}}{L}\right)^{\rm T2K}_{\rm \epsilon_{\mu\tau}+NO}<\left(\frac{\tau^E_{\rm QSL}}{L}\right)^{\rm T2K}_{\rm SO+IO}&\nonumber\\
    &<\left(\frac{\tau^E_{\rm QSL}}{L}\right)^{\rm T2K}_{\rm \epsilon_{\mu\tau}+IO}<1.
    \label{4.11}
\end{align}
Since T2K at the end of the baseline ($L = 295\,\mathrm{km}$) exhibits no discrepancy in $S_{EE}$ among all four cases [see Fig.\,\ref{fig3}(\subref{3a_fig:sub1})], no differences in the evolution sped of the entangled muon-flavor neutrino state are observed among the four cases.

However, unlike T2K, in Fig.\,\ref{fig3}(\subref{3a_fig:sub4}), at the end of the NO$\nu$A baseline, the behavior of $\tau^E_{\rm QSL}$ as a function of $L\,(\rm km)$ leads to the time-bound condition $\tau^E_{\rm QSL}/L<1$. The results demonstrate a slight discrepancy in $\tau^E_{\rm QSL}$ among the four cases. 
The following empirical relations are observed from Fig.\,\ref{fig3}(\subref{3a_fig:sub4}) at the end of the baseline of NO$\nu$A
\begin{align}
    \left(\frac{\tau^E_{\rm QSL}}{L}\right)^{\rm NO\nu A}_{\rm SO+NO}&<\left(\frac{\tau^E_{\rm QSL}}{L}\right)^{\rm NO\nu A}_{\rm \rm SO+IO}<\left(\frac{\tau^E_{\rm QSL}}{L}\right)^{\rm NO\nu A}_{\rm \epsilon_{\mu\tau}+NO}&\nonumber\\
    &<\left(\frac{\tau^E_{\rm QSL}}{L}\right)^{\rm NO\nu A}_{\rm \epsilon_{\mu\tau}+IO}<1.
    \label{4.12}
\end{align}
From Fig.\,\ref{fig3}(\subref{3a_fig:sub3}), for the initial state $\ket{\nu_\mu}$, bipartite entanglement is highest in $\epsilon_{\mu\tau}$+IO and lowest in SO+NO among all four cases at the end of the NO$\nu$A baseline. Consequently, from Fig.\,\ref{fig3}(\subref{3a_fig:sub4}) and Eq.\,(\ref{4.12}), it is evident that bipartite entanglement grows fastest in $\epsilon_{\mu\tau}$+IO, while it is rapidly suppressed in SO+NO, compared to the other cases at the end of the NO$\nu$A baseline. 

Moreover, Fig.\,\ref{fig3}(\subref{3a_fig:sub6}) depicts $\tau^E_{\rm QSL}$ as a function of $L$ for the initial state $\ket{\nu_\mu}$ under four cases using the DUNE baseline and energy. Compared to T2K and NO$\nu$A, DUNE exhibits a greater discrepancy in $\tau^E_{\rm QSL}$ among the four cases. Unlike NO$\nu$A, a large deviation in $\tau^E_{\rm QSL}$ between SO+IO and $\epsilon_{\mu\tau}$+NO is also observed. Furthermore, under the time-bound constraint $\tau^E_{\rm QSL}/L < 1$, Fig.\,\ref{fig3}(\subref{3a_fig:sub6}) yields the following empirical relation at the end of the baseline of DUNE:
\begin{align}
    \left(\frac{\tau^E_{\rm QSL}}{L}\right)^{\rm DUNE}_{\rm SO+NO}&<\left(\frac{\tau^E_{\rm QSL}}{L}\right)^{\rm DUNE}_{\rm SO+IO}<\left(\frac{\tau^E_{\rm QSL}}{L}\right)^{\rm DUNE}_{\rm \epsilon_{\mu\tau}+NO}&\nonumber\\
    &<\left(\frac{\tau^E_{\rm QSL}}{L}\right)^{\rm DUNE}_{\rm \epsilon_{\mu\tau}+IO}<1.
    \label{4.13}
\end{align} 

Equation (\ref{4.13}) is similar to Eq.\,(\ref{4.12}), but the disparity in $\tau^E_{\rm QSL}/L$ among the four cases in DUNE is greater than in NO$\nu$A near the end of their respective baselines. Furthermore, bipartite entanglement grows fastest in $\epsilon_{\mu\tau}$+IO, while it is rapidly suppressed in SO+NO compared to the other cases, a behavior that also persists at the end of the DUNE baseline, like NO$\nu$A. This result can be attributed to the behavior of entanglement entropy, as shown in Fig.\,\ref{fig3}(\subref{3a_fig:sub5}). Since both NO$\nu$A and DUNE exhibit nonzero bipartite pure entanglement at the end of their respective baselines, as shown in Figs.\,\ref{fig3}(\subref{3a_fig:sub3}) and \ref{fig3}(\subref{3a_fig:sub5}), respectively, they are able to capture differences in the evolution speed of the entangled muon-flavor neutrino state among the four cases, as depicted in Figs.\,\ref{fig3}(\subref{3a_fig:sub4}) and \ref{fig3}(\subref{3a_fig:sub6}).

Furthermore, Figs.\,\ref{fig5}(\subref{5a_fig:sub2}), \ref{fig5}(\subref{5a_fig:sub4}), and \ref{fig5}(\subref{5a_fig:sub6}) illustrate $\tau^E_{\rm QSL}$ as a function of $L$ for the initial state $\ket{\nu_\mu}$ propagating under SO, with and without the diagonal NSI parameter $|\epsilon_{ee}-\epsilon_{\mu\mu}|$, in both NO and IO scenarios, using their respective best-fit $CP$-violating values $\delta_{CP}$. We observe no discrepancies in $\tau^E_{\rm QSL}$ among the four cases at the end of the T2K baseline, as depicted in Fig.\,\ref{fig5}(\subref{5a_fig:sub2}). However, slight discrepancies in $\tau^E_{\rm QSL}$ among the four cases for NO$\nu$A and larger discrepancies for DUNE near the end of their respective baselines are found, as shown in Figs.\,\ref{fig5}(\subref{5a_fig:sub3}) and \ref{fig5}(\subref{5a_fig:sub5}), respectively. Under the time-bound constraint $\tau^E_{\rm QSL}/L < 1$, the empirical relations satisfied by NO$\nu$A, and DUNE at the end of their respective baselines are given by, respectively,
\begin{align}
    \left(\frac{\tau^E_{\rm QSL}}{L}\right)^{\rm NO\nu A}_{\rm SO+NO}&<\left(\frac{\tau^E_{\rm QSL}}{L}\right)^{\rm  NO\nu A}_{\rm |\epsilon_{ee}-\epsilon_{\mu\mu}|+NO}<\left(\frac{\tau^E_{\rm QSL}}{L}\right)^{\rm NO\nu A}_{\rm SO+IO}&\nonumber\\
    &<\left(\frac{\tau^E_{\rm QSL}}{L}\right)^{\rm  NO\nu A}_{\rm |\epsilon_{ee}-\epsilon_{\mu\mu}|+IO}<1;
    \label{4.14a}
\end{align}

\begin{align}
    \left(\frac{\tau^E_{\rm QSL}}{L}\right)^{\rm DUNE}_{\rm SO+NO}&<\left(\frac{\tau^E_{\rm QSL}}{L}\right)^{\rm  DUNE}_{\rm |\epsilon_{ee}-\epsilon_{\mu\mu}|+NO}<\left(\frac{\tau^E_{\rm QSL}}{L}\right)^{\rm DUNE}_{\rm SO+IO}&\nonumber\\
    &<\left(\frac{\tau^E_{\rm QSL}}{L}\right)^{\rm  DUNE}_{\rm |\epsilon_{ee}-\epsilon_{\mu\mu}|+IO}<1.
    \label{4.14}
\end{align}
The results in Eqs.\,(\ref{4.14a}) and (\ref{4.14}) indicate that, in both NO$\nu$A and DUNE, respectively, at the end of their baselines, bipartite entanglement grows fastest in the $|\epsilon_{ee}-\epsilon_{\mu\mu}|$+IO case, while it is rapidly suppressed in SO+NO compared to the other cases. These results can be attributed to the $S_{EE}$ behaviors shown in Figs.\,\ref{fig5}(\subref{5a_fig:sub3}) and \ref{fig5}(\subref{5a_fig:sub5}).

Moreover, in Appendix, we also illustrate $\tau^E_{\rm QSL}$ as a function of $L$ in Figs.\,\ref{fig9}-\ref{fig11} for the initial state $\ket{\nu_\mu}$ propagating under SO, with and without the complex off-diagonal NSI parameters $\epsilon_{e\mu}$ and $\epsilon_{e\tau}$, as well as the diagonal NSI parameter $|\epsilon_{\tau\tau}-\epsilon_{\mu\mu}|$, respectively. The results are presented for both NO and IO, using their respective best-fit $CP$-violating phases $\delta_{CP}$, and considering the baseline lengths and neutrino energies corresponding to the T2K, NO$\nu$A, and DUNE experiments, as shown in Figs.\,\ref{fig9}(\subref{9a_fig:sub2}), \ref{fig9}(\subref{9a_fig:sub4}), \ref{fig9}(\subref{9a_fig:sub6}); Figs.\,\ref{fig10}(\subref{10a_fig:sub2}), \ref{fig10}(\subref{10a_fig:sub4}), \ref{fig10}(\subref{10a_fig:sub6}); and Figs.\,\ref{fig11}(\subref{11a_fig:sub2}), \ref{fig11}(\subref{11a_fig:sub4}), \ref{fig11}(\subref{11a_fig:sub6}), respectively. Across all experiments, at the end of their baselines, no pronounced discrepancies in $\tau^E_{\rm QSL}$ are observed among the four cases, either in the presence of complex off-diagonal or diagonal NSI parameters. The result that T2K is unable to capture differences in the evolution speed of the entangled muon-flavor neutrino state among the four cases, whereas NO$\nu$A and DUNE can, remains valid in this context as well.
However, focusing on the zoomed-in plots for DUNE at its end of baseline and imposing the time-bound condition $\tau^E_{\rm QSL}/L < 1$, the following empirical relations can be inferred from Figs.\,\ref{fig9}(\subref{9a_fig:sub6}), \ref{fig10}(\subref{10a_fig:sub6}), and \ref{fig11}(\subref{11a_fig:sub6}), respectively:
\begin{align}
    \left(\frac{\tau^E_{\rm QSL}}{L}\right)^{\rm DUNE}_{\rm SO+NO}&<\left(\frac{\tau^E_{\rm QSL}}{L}\right)^{\rm DUNE}_{\rm \epsilon_{e\mu}+NO}<\left(\frac{\tau^E_{\rm QSL}}{L}\right)^{\rm DUNE}_{\rm \epsilon_{e\mu}+IO}&\nonumber\\
    &<\left(\frac{\tau^E_{\rm QSL}}{L}\right)^{\rm DUNE}_{\rm SO+IO}<1;&\label{4.15}
\end{align}
\begin{align}
     \left(\frac{\tau^E_{\rm QSL}}{L}\right)^{\rm DUNE}_{\rm SO+NO}&<\left(\frac{\tau^E_{\rm QSL}}{L}\right)^{\rm  DUNE}_{\rm \epsilon_{e\tau}+NO}<\left(\frac{\tau^E_{\rm QSL}}{L}\right)^{\rm DUNE}_{\rm \epsilon_{e\tau}+IO}&\nonumber\\
    &<\left(\frac{\tau^E_{\rm QSL}}{L}\right)^{\rm DUNE}_{\rm SO+IO}<1;&\label{4.16}
    \end{align}
    \begin{align}
    \left(\frac{\tau^E_{\rm QSL}}{L}\right)^{\rm DUNE}_{\rm |\epsilon_{\tau\tau}-\epsilon_{\mu\mu}|+NO}&<\left(\frac{\tau^E_{\rm QSL}}{L}\right)^{\rm  DUNE}_{\rm SO+NO}<\left(\frac{\tau^E_{\rm QSL}}{L}\right)^{\rm DUNE}_{\rm SO+IO}&\nonumber\\
    &<\left(\frac{\tau^E_{\rm QSL}}{L}\right)^{\rm  DUNE}_{\rm |\epsilon_{\tau\tau}-\epsilon_{\mu\mu}|+IO}<1.
    \label{4.17}
\end{align}
From Eqs.\,(\ref{4.15}) and (\ref{4.16}), it is evident that in the presence of the complex off-diagonal NSI parameters $\epsilon_{e\mu}$ and $\epsilon_{e\tau}$, respectively, bipartite entanglement grows fastest in the SO+IO case, while it is rapidly suppressed in the SO+NO case compared to the other cases. However, between the $\epsilon_{e\mu}$+NO and $\epsilon_{e\mu}$+IO cases, a more rapid suppression of bipartite entanglement is observed in the $\epsilon_{e\mu}$+NO scenario. Similarly, between the $\epsilon_{e\tau}$+NO and $\epsilon_{e\tau}$+IO cases, a more rapid suppression of bipartite entanglement is observed in the $\epsilon_{e\tau}$+NO scenario. This behavior can be attributed to the corresponding $S_{EE}$ profiles shown in Figs.\,\ref{fig9}(\subref{9a_fig:sub5}) and \ref{fig10}(\subref{10a_fig:sub5}). This observation differs from the behavior found in Eq.\,(\ref{4.13}) for DUNE in the presence of the complex off-diagonal NSI parameter $\epsilon_{\mu\tau}$.
Furthermore, from Eq.\,(\ref{4.17}), in the presence of the diagonal NSI parameter $|\epsilon_{\tau\tau}-\epsilon_{\mu\mu}|$, bipartite entanglement grows fastest in the $|\epsilon_{\tau\tau}-\epsilon_{\mu\mu}|$+IO case, while it is rapidly suppressed in $|\epsilon_{\tau\tau}-\epsilon_{\mu\mu}|$+NO compared to the other cases at the end of the baseline of DUNE. These results can be attributed to the $S_{EE}$ behavior shown in Fig.\,\ref{fig11}(\subref{11a_fig:sub5}). This trend is distinct from that observed in Eq.\,(\ref{4.14}) for DUNE in the presence of the diagonal NSI parameter $|\epsilon_{ee}-\epsilon_{\mu\mu}|$.

It is worth noting that the empirical relations observed from Eqs.\,(\ref{4.11})-(\ref{4.17}) are strictly based on the single best-fit values of both the SO parameters and the NSI parameters. However, these empirical relations may change when the $1\sigma$ or $3\sigma$ ranges of the parameters are used. 
For the off-diagonal NSI parameter $\left|\epsilon_{\mu\tau}\right|$ with complex phase $\phi_{\mu\tau}$, within the $1\sigma$ range of SO parameters (see Table\,\ref{Tab1}), we observe from Fig.~\ref{fig3} that the $\tau^E_{QSL}$ versus $L$ results overlap among the four scenarios, namely SO+NO, SO+IO, NSI+NO, and NSI+IO, for T2K. However, the corresponding results for NO$\nu$A and DUNE differ. For example, within the $1\sigma$ allowed regions of the SO parameters, Fig.~\ref{fig3}(\subref{3a_fig:sub2}) shows that, for T2K in the range $150 < L\,(\mathrm{km}) < 295$, the empirical relation previously given by Eq.~(\ref{4.11}) becomes
\begin{align}
    \left(\frac{\tau^E_{\rm QSL}}{L}\right)^{\rm T2K}_{\rm SO+NO}&=\left(\frac{\tau^E_{\rm QSL}}{L}\right)^{\rm T2K}_{\rm \epsilon_{\mu\tau}+NO}=\left(\frac{\tau^E_{\rm QSL}}{L}\right)^{\rm T2K}_{\rm SO+IO}&\nonumber\\
   & =\left(\frac{\tau^E_{\rm QSL}}{L}\right)^{\rm T2K}_{\rm \epsilon_{\mu\tau}+IO}<1.&
    \label{4.18}
\end{align} 
These results indicate that, for T2K in the range $150 < L\,(\mathrm{km}) < 295$, the bipartite entanglement in the evolved state $\ket{\nu_\mu(t)}$ could be sped up in all four scenarios; however, the scenarios cannot be distinguished within the $1\sigma$ range of the SO parameters. Furthermore, the results that no differences in the evolution speed of the entangled muon-flavor neutrino state found among the four cases at the end of the T2K baseline ($L = 295\,\mathrm{km}$) remain consistent for both the best-fit values and the $1\sigma$ range of the SO parameters.

Similarly, for NO$\nu$A, Fig.~\ref{fig3}(\subref{3a_fig:sub4}) shows that the empirical relation previously given by Eq.~(\ref{4.12}) becomes
\begin{align}
    \left(\frac{\tau^E_{\rm QSL}}{L}\right)^{\rm NO\nu A}_{\rm SO+NO}&=\left(\frac{\tau^E_{\rm QSL}}{L}\right)^{\rm NO\nu A}_{\rm \rm SO+IO}=\left(\frac{\tau^E_{\rm QSL}}{L}\right)^{\rm NO\nu A}_{\rm \epsilon_{\mu\tau}+NO}&\nonumber\\&<\left(\frac{\tau^E_{\rm QSL}}{L}\right)^{\rm NO\nu A}_{\rm \epsilon_{\mu\tau}+IO}<1,&
    \label{4.19}
\end{align}
and, for DUNE, Fig.~\ref{fig3}(\subref{3a_fig:sub6}) shows that Eq.~(\ref{4.13}) becomes
\begin{align}
    \left(\frac{\tau^E_{\rm QSL}}{L}\right)^{\rm DUNE}_{\rm SO+NO}&=\left(\frac{\tau^E_{\rm QSL}}{L}\right)^{\rm DUNE}_{\rm SO+IO}=\left(\frac{\tau^E_{\rm QSL}}{L}\right)^{\rm DUNE}_{\rm \epsilon_{\mu\tau}+NO}&\nonumber\\
    &<\left(\frac{\tau^E_{\rm QSL}}{L}\right)^{\rm DUNE}_{\rm \epsilon_{\mu\tau}+IO}<1.&
    \label{4.20}
\end{align}
Thus, within the $1\sigma$ range of the SO parameters, the bipartite entanglement grows fastest in the $\epsilon_{\mu\tau}$+IO scenario, whereas the results for SO+NO, SO+IO, and $\epsilon_{\mu\tau}$+NO overlap and are suppressed at the ends of both the NO$\nu$A and DUNE baselines. Therefore, the results obtained using the $1\sigma$ range of the SO parameters differ from those previously obtained using the best-fit values.

Furthermore, within the best-fit $\pm 1\sigma$ range of the SO parameters, for the diagonal NSI parameter $\left|\epsilon_{ee}-\epsilon_{\mu\mu}\right|$, Fig.~\ref{fig5} shows no discrepancies in $\tau^E_{\rm QSL}$ among the four scenarios at the end of the T2K baseline. However, under the time-bound constraint $\tau^E_{\rm QSL}/L < 1$ and within the $1\sigma$ range of the SO parameters, the empirical relations previously given by Eqs.~(\ref{4.14a}) and (\ref{4.14}) for NO$\nu$A and DUNE, respectively, become
\begin{align}
    \left(\frac{\tau^E_{\rm QSL}}{L}\right)^{\rm NO\nu A}_{\rm SO+NO}&=\left(\frac{\tau^E_{\rm QSL}}{L}\right)^{\rm  NO\nu A}_{\rm |\epsilon_{ee}-\epsilon_{\mu\mu}|+NO}=\left(\frac{\tau^E_{\rm QSL}}{L}\right)^{\rm NO\nu A}_{\rm SO+IO}&\nonumber\\
    &=\left(\frac{\tau^E_{\rm QSL}}{L}\right)^{\rm  NO\nu A}_{\rm |\epsilon_{ee}-\epsilon_{\mu\mu}|+IO}<1;&
    \label{4.21}
\end{align}
\begin{align}
    \left(\frac{\tau^E_{\rm QSL}}{L}\right)^{\rm DUNE}_{\rm SO+NO}&=\left(\frac{\tau^E_{\rm QSL}}{L}\right)^{\rm  DUNE}_{\rm |\epsilon_{ee}-\epsilon_{\mu\mu}|+NO}=\left(\frac{\tau^E_{\rm QSL}}{L}\right)^{\rm DUNE}_{\rm SO+IO}&\nonumber\\
    &=\left(\frac{\tau^E_{\rm QSL}}{L}\right)^{\rm  DUNE}_{\rm |\epsilon_{ee}-\epsilon_{\mu\mu}|+IO}<1.&
    \label{4.22}
\end{align}
Thus, the ratio $\tau^E_{\rm QSL}/L$ overlaps for all four scenarios in both the NO$\nu$A and DUNE cases at the end of their baselines. Consequently, the bipartite entanglement in the evolved state $\ket{\nu_\mu(L)}$ could be sped up in all four scenarios at the end of NO$\nu$A and DUNE baselines. Moreover, unlike the results obtained using the best-fit values, no distinction among the four scenarios can be made within the $1\sigma$ range of the SO parameters.

We further emphasize that, within the $1\sigma$ allowed region of the SO parameters, the conclusions drawn above [i.e, from Eqs.\,(\ref{4.21}) and (\ref{4.22})] remain valid for the empirical relations presented in Eqs.~(\ref{4.15})-(\ref{4.17}). In addition to this, we comment that no significant discrepancies in  $S_{EE}$, $C_{E}$ and $\tau^E_{QSL}$ are observed among the four scenarios within the $3\sigma$ range of SO parameters and this behavior remains valid for all the empirical relations [Eqs.\,(\ref{4.11})-(\ref{4.17})] across all experiments, including T2K, NO$\nu$A, and DUNE.

Moreover, in this analysis, the representative neutrino energies adopted in this work correspond to the flux-peak energies of the T2K, NO$\nu$A, and DUNE experiments. Although realistic experiments involve a broad neutrino energy spectrum, our analysis is performed at these representative energies to compare the effects of SO and NSIs on the oscillation probabilities, as well as on the QSL time for bipartite entanglement, across the three experimental configurations. We also examined the behavior of the entanglement and QSL time in the low- and high-energy limits. At low energies, no universal trend in the enhancement or suppression of bipartite entanglement is observed across the three experiments, whereas at high energies all scenarios become nearly indistinguishable, with the QSL bound ratio approaching $\tau_{\rm QSL}^{E}/L = 1$, indicating that the maximum evolution speed of bipartite entanglement has already been reached. These observations justify the use of the flux-peak energies in the present analysis.

\section{Discussion and Conclusion} \label{sec5}
In this work, we have investigated the behavior of bipartite entanglement and its associated QSL time in three-flavor neutrino oscillations, taking into account both SO and NSI effects for four different cases: SO+NO, SO+IO, NSI+NO, and NSI+IO, each with its corresponding best-fit $CP$-violating phases. Both complex off-diagonal ($\epsilon_{\alpha\beta}$) and diagonal ($|\epsilon_{\alpha\alpha}-\epsilon_{\beta\beta}|$) NSI parameters were considered, with each NSI parameter introduced individually in the analysis. We have focused on the evolution of the initial muon-flavor neutrino state by considering the baseline lengths and neutrino energies corresponding to the maximum flux of ongoing long-baseline neutrino experiments, including T2K and NO$\nu$A, as well as the upcoming DUNE experiment. Our analysis is carried out within the mode-entanglement framework, where neutrino flavor states are treated as bipartite quantum systems.

We first examined how NSI parameters modify the flavor-transition probabilities for an initial muon neutrino state propagating through matter in long-baseline accelerator experiments such as T2K, NO$\nu$A, and DUNE. Among the off-diagonal NSI parameters considered, the $\epsilon_{\mu\tau}$ NSI parameter has observed the most pronounced deviations from the SO scenarios for all the probability channels. These effects become increasingly significant with longer baselines, highlighting the enhanced sensitivity of experiments such as DUNE to NSI-induced matter effects. In the $\nu_\tau$ appearance channels of DUNE,
the SO scenario does not provide a clear distinction between the NO and IO for baseline lengths of approximately 1000–1100 km. However, when NSI effects are included, the $\epsilon_{\mu\tau}$+NO and $\epsilon_{\mu\tau}$+IO cases can be distinguished. This behavior is reversed toward the end of the baseline, underscoring the capability of long-baseline experiments to possible signatures of physics BSM. The $\epsilon_{e\mu}$ and $\epsilon_{e\tau}$ parameters lead to comparatively mild deviations, primarily affecting the $\nu_e$ appearance channel. For diagonal NSI parameters, we focused on the $|\epsilon_{ee}-\epsilon_{\mu\mu}|$ diagonal NSI scenario, and observed that their impact is more prominent in appearance channels and in experiments with strong matter effects.

We have further investigated two bipartite entanglement measures, namely the entanglement entropy and the capacity of entanglement. These entanglement measures were quantified in terms of the three-flavor neutrino transition probabilities. Across all experiments, the nonzero values of these entanglement measures indicated that three-flavor neutrino oscillations give rise to a bipartite pure entangled system in all four cases. In fact, T2K, NO$\nu$A, and DUNE have revealed that maximally pure bipartite entanglement can exist during the evolution of the initial muon-flavor neutrino state in all four cases. These results can be primarily ascribed to the neutrino flavor-transition probabilities $P({\nu_{\alpha}\rightarrow \nu_{\beta}})$. Thus, at specific baselines of T2K, NO$\nu$A and DUNE, with best-fit $CP$-violating phases, the initial muon-flavor neutrino state $\ket{\nu_\mu}$ propagating in all four cases exhibits a Bell-like state, which corresponds to a maximally bipartite pure entangled state in quantum information theory. However, we also emphasize here that, according to the four given cases, distinct behavior of maximally pure bipartite entanglement for the initial state $\ket{\nu_\mu}$ can be observed in the oscillogram plots across different length ranges of the long-baseline accelerator experiment and over the entire range of the $CP$-violating phase.

Moreover, using these two entanglement measures along with the variance of the driving Hamiltonian, we investigated the QSL time for bipartite entanglement during the evolution of the initial muon-flavor neutrino state under all four cases. Except for T2K, slight discrepancies in the QSL time among the four cases were observed near the end of the NO$\nu$A baseline, while considerably larger discrepancies appeared for DUNE. These differences are attributed to the behavior of the entanglement entropy in the four cases.

In the presence of the complex off-diagonal NSI parameter $\epsilon_{\mu\tau}$, we found that bipartite entanglement grows fastest in the $\epsilon_{\mu\tau}$+IO case, whereas it is rapidly suppressed in the SO+NO case compared to the other scenarios at the end of the NO$\nu$A and DUNE baselines. In contrast, for other complex off-diagonal NSI parameters, namely $\epsilon_{e\mu}$ and $\epsilon_{e\tau}$, bipartite entanglement grows fastest in the SO+IO case and is rapidly suppressed in the SO+NO case relative to the remaining scenarios. These behaviors are more prominently visible near the end of the DUNE baseline than in NO$\nu$A and T2K. Moreover, when we have used the diagonal NSI parameter $|\epsilon_{ee}-\epsilon_{\mu\mu}|$, bipartite entanglement grows fastest in the $|\epsilon_{ee}-\epsilon_{\mu\mu}|$+IO case, while it is rapidly suppressed in the SO+NO case compared to the other scenarios at the end of the NO$\nu$A and DUNE baselines. On the other hand, for the diagonal NSI parameter $|\epsilon_{\tau\tau}-\epsilon_{\mu\mu}|$, bipartite entanglement exhibits the fastest growth in the $|\epsilon_{\tau\tau}-\epsilon_{\mu\mu}|$+IO case and the rapid suppression in the $|\epsilon_{\tau\tau}-\epsilon_{\mu\mu}|$+NO case relative to the remaining scenarios. These features are most clearly observed at the end of the DUNE baseline than in NO$\nu$A and T2K. Thus, the outcome shows a possible imprint of new physics in neutrino oscillations.

Furthermore, all the results discussed so far are based on the single best-fit values of both the SO and NSI parameters. However, these results may change when the $1\sigma$ or $3\sigma$ ranges of the SO parameters are considered in the analysis.

In conclusion, at the end of the baseline T2K is unable to capture differences in the evolution speed of the entangled muon-flavor neutrino state among the different scenarios considered, whereas such distinctions are clearly resolved in NO$\nu$A and DUNE, specifically when the best-fit values of the SO and NSI parameters are used. These findings could aid in understanding the impact of CP violation and in identifying the neutrino mass ordering in matter when NSIs are present.

\section*{Acknowledgment}
A. K. J. would like to acknowledge the project funded by SERB, India, with Ref. No. CRG/2022/003460, for supporting this research. L. K. acknowledges the Ministry of Education (MoE) for financial support and the Indian Institute of Technology Jodhpur for access to research facilities.

\bibliography{main}
\appendix
\section{Results for the NSI Parameters \(\epsilon_{e\mu}\), \(\epsilon_{e\tau}\), and \(|\epsilon_{\tau\tau}-\epsilon_{\mu\mu}|\)}

\label{Appendix:A}
\begin{figure*}[!htbp]
  \centering
  \begin{subfigure}[b]{0.33\textwidth}
    \centering
    \includegraphics[width=\textwidth]{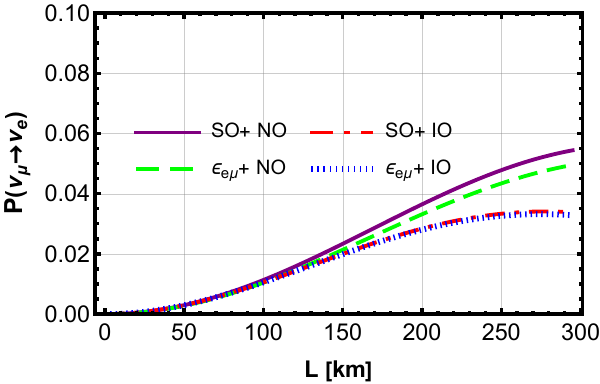}
    \caption{T2K}
    \label{6a_fig:sub1}
  \end{subfigure}
  \hfill
  \begin{subfigure}[b]{0.33\textwidth}
    \centering
    \includegraphics[width=\textwidth]{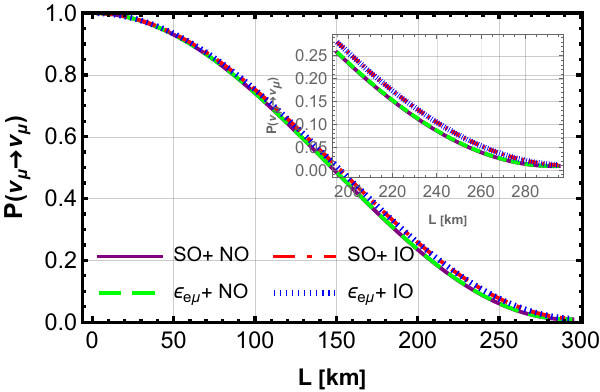}
    \caption{T2K}
    \label{6a_fig:sub2}
  \end{subfigure}
   \begin{subfigure}[b]{0.33\textwidth}
    \centering
    \includegraphics[width=\textwidth]{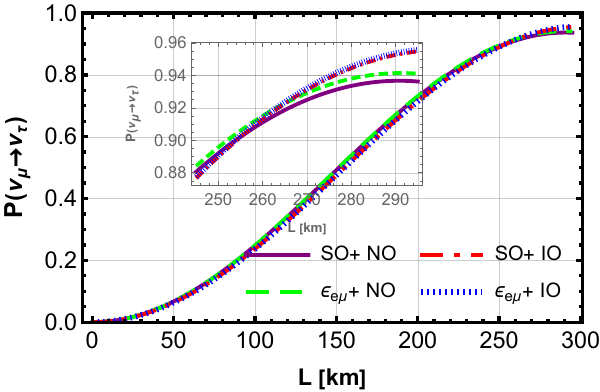}
    \caption{T2K}
    \label{6a_fig:sub3}
  \end{subfigure}
  \hfill
  \\
  \begin{subfigure}[b]{0.33\textwidth}
    \centering
    \includegraphics[width=\textwidth]{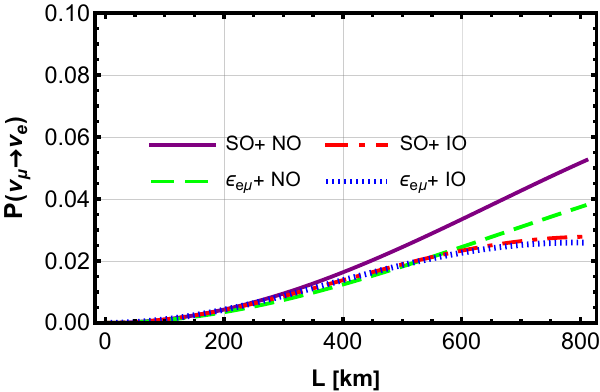}
    \caption{NO$\nu$A}
    \label{6a_fig:sub4}
  \end{subfigure}
  \hfill
  \begin{subfigure}[b]{0.33\textwidth}
    \centering
    \includegraphics[width=1.001\textwidth]{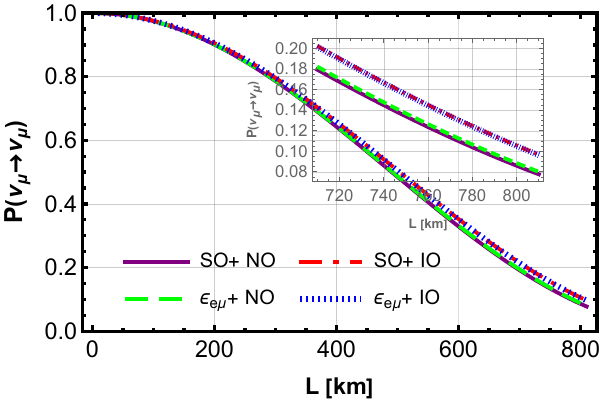}
    \caption{NO$\nu$A}
    \label{6a_fig:sub5}
  \end{subfigure}
  \hfill
  \begin{subfigure}[b]{0.33\textwidth}
    \centering
    \includegraphics[width=\textwidth]{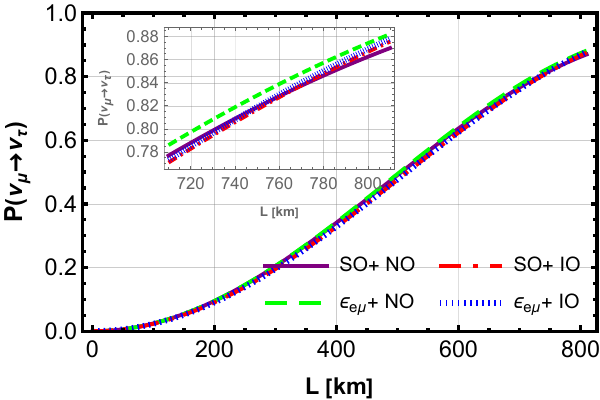}
    \caption{NO$\nu$A}
    \label{6a_fig:sub6}
  \end{subfigure}
\hfill
\\  
\begin{subfigure}[b]{0.33\textwidth}
    \centering
    \includegraphics[width=\textwidth]{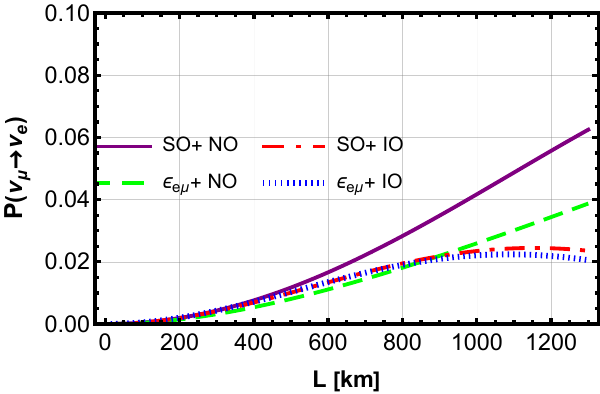}
    \caption{DUNE}
    \label{6a_fig:sub7}
  \end{subfigure}
  \hfill
  \begin{subfigure}[b]{0.33\textwidth}
    \centering
    \includegraphics[width=\textwidth]{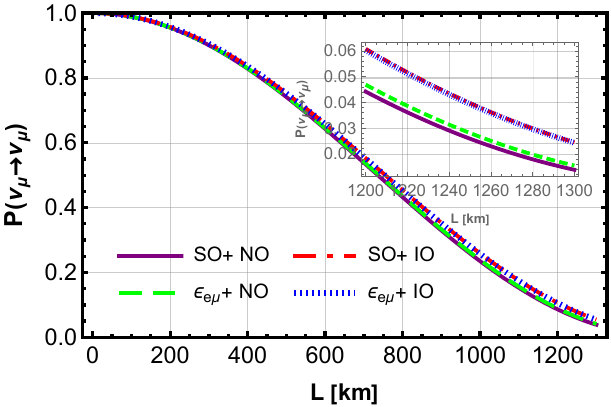}
    \caption{DUNE}
    \label{6a_fig:sub8}
  \end{subfigure}
  \hfill
  \begin{subfigure}[b]{0.33\textwidth}
    \centering
    \includegraphics[width=\textwidth]{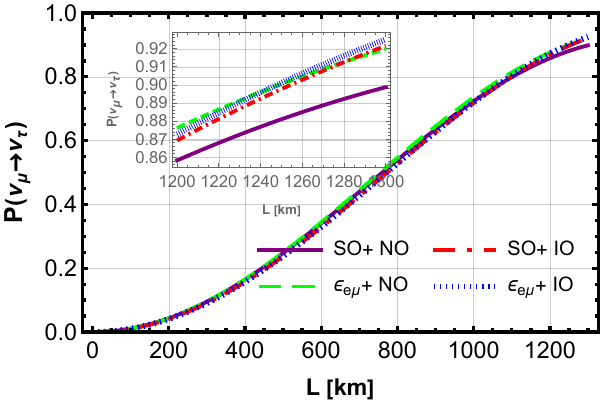}
    \caption{DUNE}
    \label{6a_fig:sub9}
  \end{subfigure}
  \caption{\justifying{Same as Fig\,\ref{fig1}, except that the off-diagonal NSI parameter $\left |\epsilon_{e\mu}\right |$ with complex phase $\phi_{e\mu}$ is used in the analysis, with both the SO and NSI parameters fixed at their best-fit values. T2K (a)-(c), NO$\nu$A (d)-(f), and DUNE (g)-(i)}}
  \label{fig6}
\end{figure*}

\begin{figure*}[!htbp]
  \centering
  \begin{subfigure}[b]{0.33\textwidth}
    \centering
    \includegraphics[width=\textwidth]{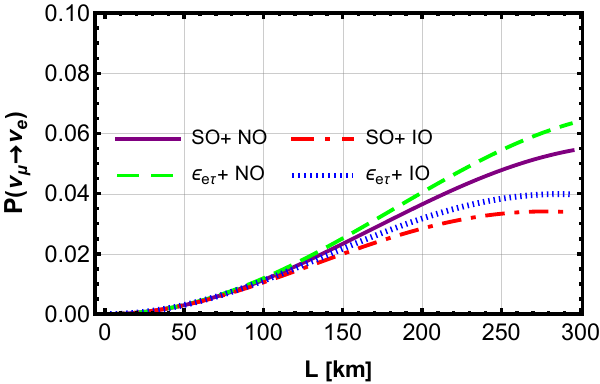}
    \caption{T2K}
    \label{7a_fig:sub1}
  \end{subfigure}
  \hfill
  \begin{subfigure}[b]{0.33\textwidth}
    \centering
    \includegraphics[width=\textwidth]{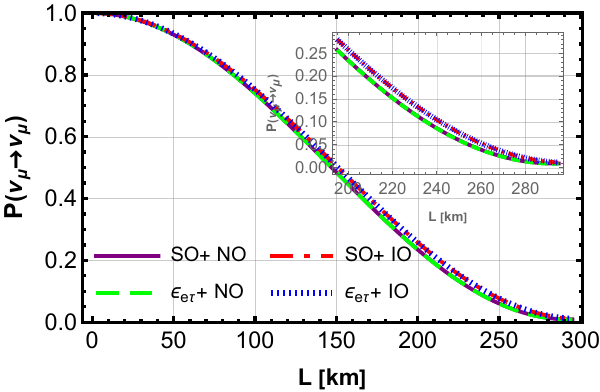}
    \caption{T2K}
    \label{7a_fig:sub2}
  \end{subfigure}
   \begin{subfigure}[b]{0.33\textwidth}
    \centering
    \includegraphics[width=\textwidth]{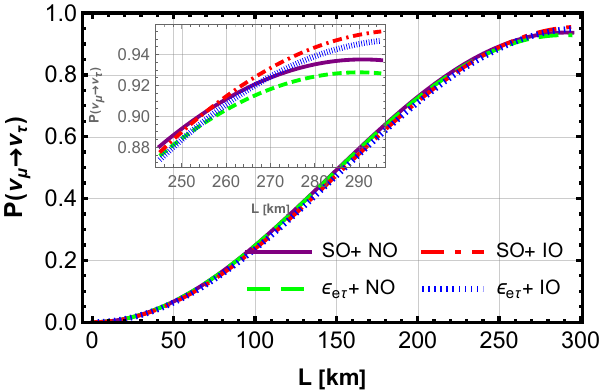}
    \caption{T2K}
    \label{7a_fig:sub3}
  \end{subfigure}
  \hfill
  \\
  \begin{subfigure}[b]{0.33\textwidth}
    \centering
    \includegraphics[width=\textwidth]{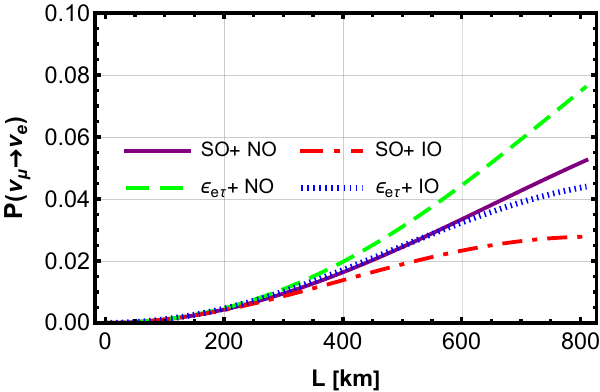}
    \caption{NO$\nu$A}
    \label{7a_fig:sub4}
  \end{subfigure}
  \hfill
  \begin{subfigure}[b]{0.33\textwidth}
    \centering
    \includegraphics[width=1.001\textwidth]{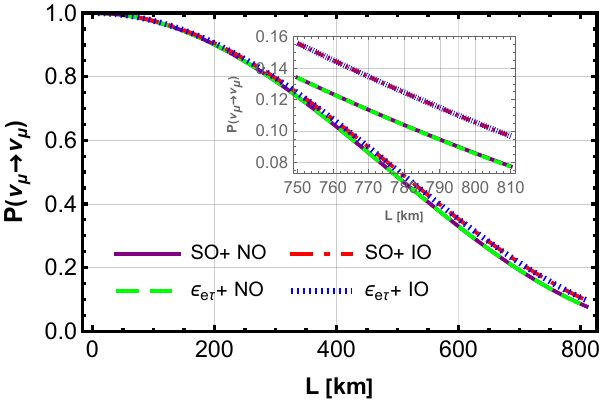}
    \caption{NO$\nu$A}
    \label{7a_fig:sub5}
  \end{subfigure}
  \hfill
  \begin{subfigure}[b]{0.33\textwidth}
    \centering
    \includegraphics[width=\textwidth]{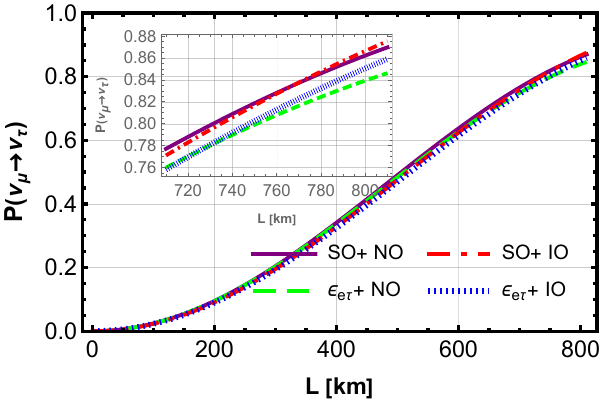}
    \caption{NO$\nu$A}
    \label{7a_fig:sub6}
  \end{subfigure}
\hfill
\\  
\begin{subfigure}[b]{0.33\textwidth}
    \centering
    \includegraphics[width=\textwidth]{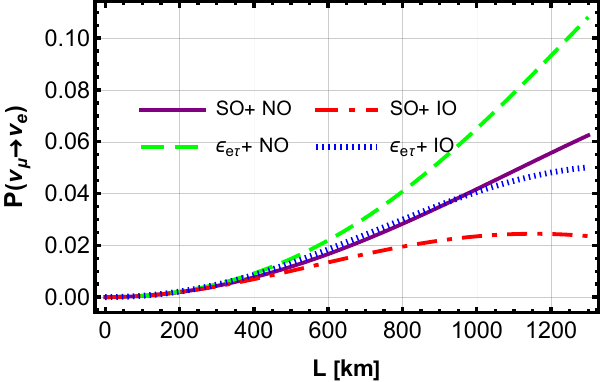}
    \caption{DUNE}
     \label{7a_fig:sub7}
  \end{subfigure}
  \hfill
  \begin{subfigure}[b]{0.33\textwidth}
    \centering
    \includegraphics[width=\textwidth]{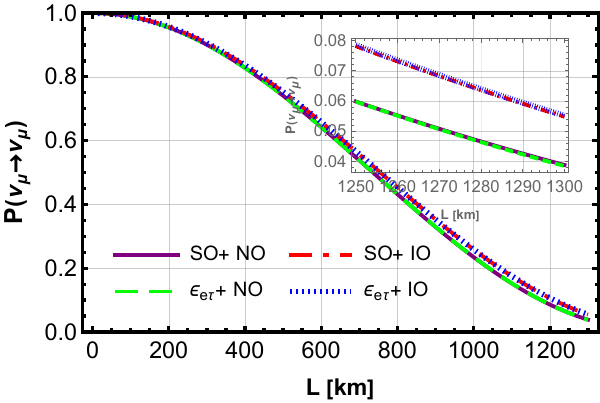}
    \caption{DUNE}
    \label{7a_fig:sub8}
  \end{subfigure}
  \hfill
  \begin{subfigure}[b]{0.33\textwidth}
    \centering
    \includegraphics[width=\textwidth]{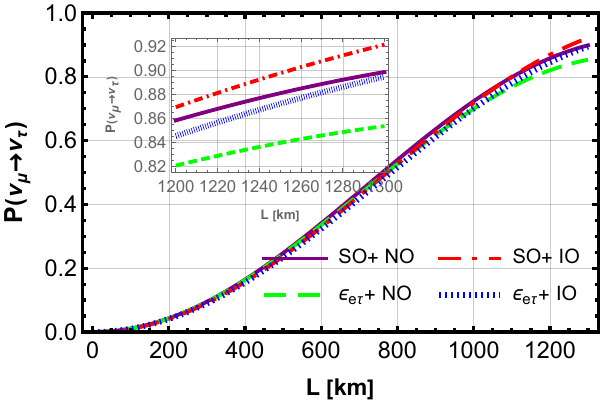}
    \caption{DUNE}
    \label{7a_fig:sub9}
  \end{subfigure}
  \caption{\justifying{Same as Fig\,\ref{fig1}, except that the off- diagonal NSI parameter $\left |\epsilon_{e\tau}\right |$ with complex phase $\phi_{e\tau}$ is used in the analysis, with both the SO and NSI parameters fixed at their best-fit values. T2K (a)-(c), NO$\nu$A (d)-(f), and DUNE (g)-(i)}}
  \label{fig7}
\end{figure*}

\begin{figure*}[!htbp]
  \centering
  \begin{subfigure}[b]{0.33\textwidth}
    \centering
    \includegraphics[width=\textwidth]{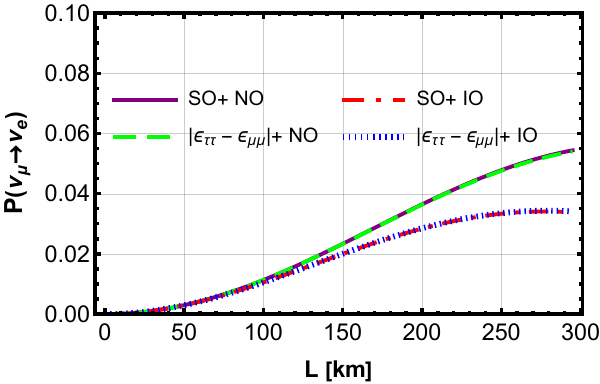}
    \caption{T2K}
    \label{8a_fig:sub1}
  \end{subfigure}
  \hfill
  \begin{subfigure}[b]{0.33\textwidth}
    \centering
    \includegraphics[width=\textwidth]{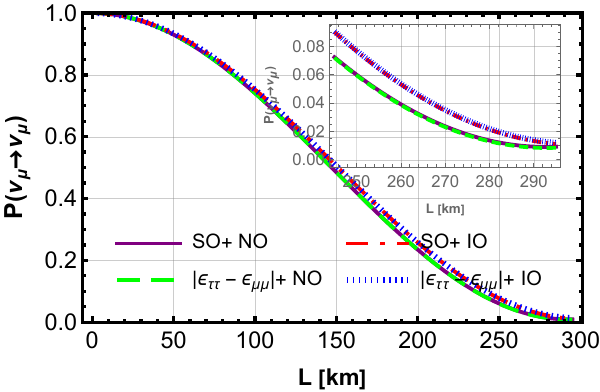}
    \caption{T2K}
    \label{8a_fig:sub2}
  \end{subfigure}
   \begin{subfigure}[b]{0.33\textwidth}
    \centering
    \includegraphics[width=\textwidth]{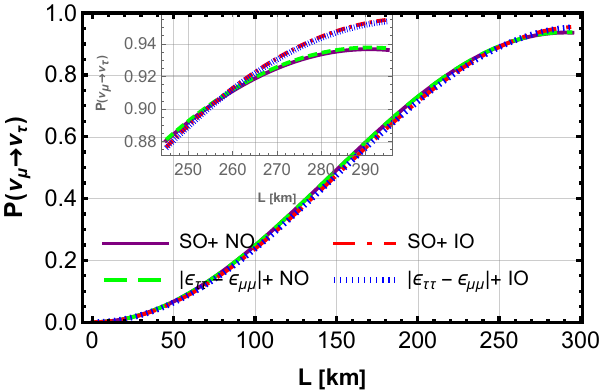}
    \caption{T2K}
    \label{8a_fig:sub3}
  \end{subfigure}
  \hfill
  \\
  \begin{subfigure}[b]{0.33\textwidth}
    \centering
    \includegraphics[width=\textwidth]{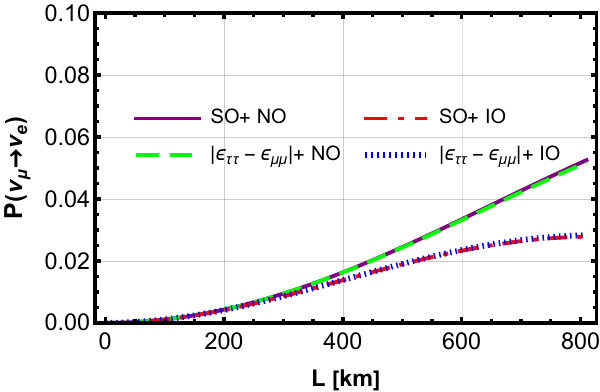}
    \caption{NO$\nu$A}
    \label{8a_fig:sub4}
  \end{subfigure}
  \hfill
  \begin{subfigure}[b]{0.33\textwidth}
    \centering
    \includegraphics[width=1.001\textwidth]{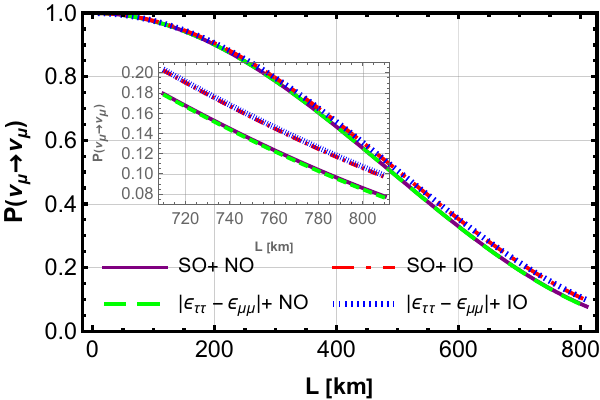}
    \caption{NO$\nu$A}
    \label{8a_fig:sub5}
  \end{subfigure}
  \hfill
  \begin{subfigure}[b]{0.33\textwidth}
    \centering
    \includegraphics[width=\textwidth]{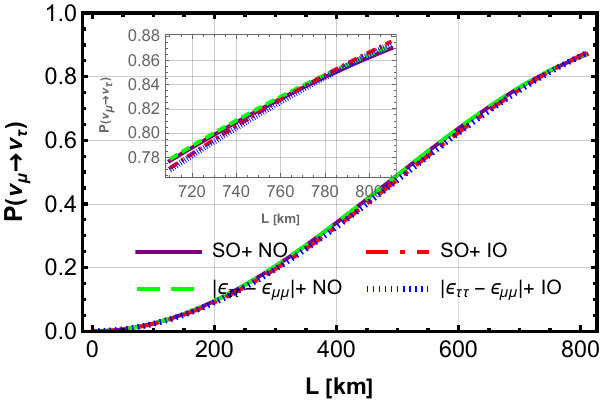}
    \caption{NO$\nu$A}
    \label{8a_fig:sub6}
  \end{subfigure}
\hfill
\\  
\begin{subfigure}[b]{0.33\textwidth}
    \centering
    \includegraphics[width=\textwidth]{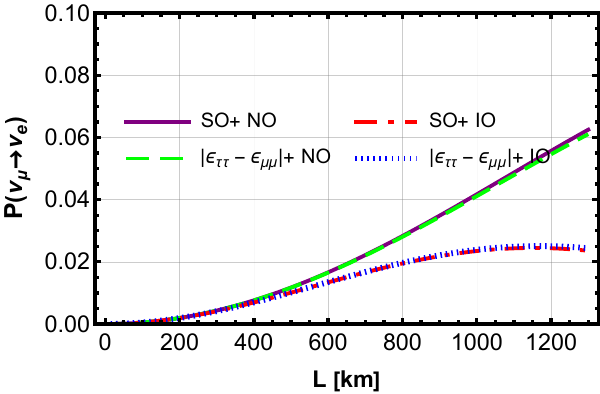}
    \caption{DUNE}
     \label{8a_fig:sub7}
  \end{subfigure}
  \hfill
  \begin{subfigure}[b]{0.33\textwidth}
    \centering
    \includegraphics[width=\textwidth]{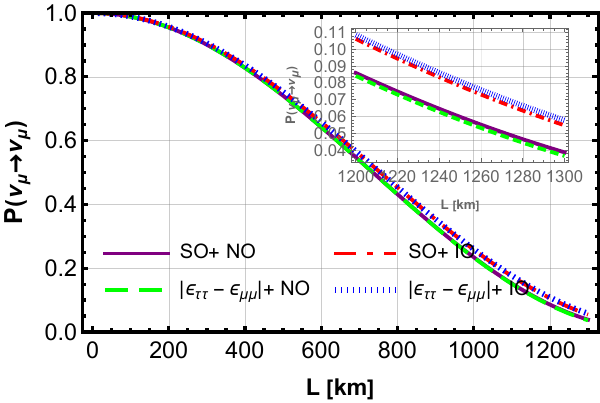}
    \caption{DUNE}
    \label{8a_fig:sub8}
  \end{subfigure}
  \hfill
  \begin{subfigure}[b]{0.33\textwidth}
    \centering
    \includegraphics[width=\textwidth]{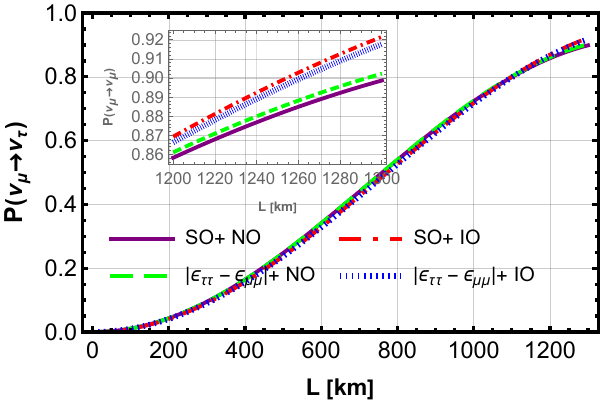}
    \caption{DUNE}
    \label{8a_fig:sub9}
  \end{subfigure}
  \caption{\justifying{Same as Fig\,\ref{fig2}, except that the diagonal NSI parameter $\left |\epsilon_{\tau \tau}-\epsilon_{\mu\mu}\right |$ is used in the analysis, with both the SO and NSI parameters fixed at their best-fit values. T2K (a)-(c), NO$\nu$A (d)-(f), and DUNE (g)-(i)}}
  \label{fig8}
\end{figure*}

\begin{figure*}[!htbp]
    \centering
    \begin{subfigure}{0.49\textwidth}
        \centering
        \includegraphics[width=\textwidth]{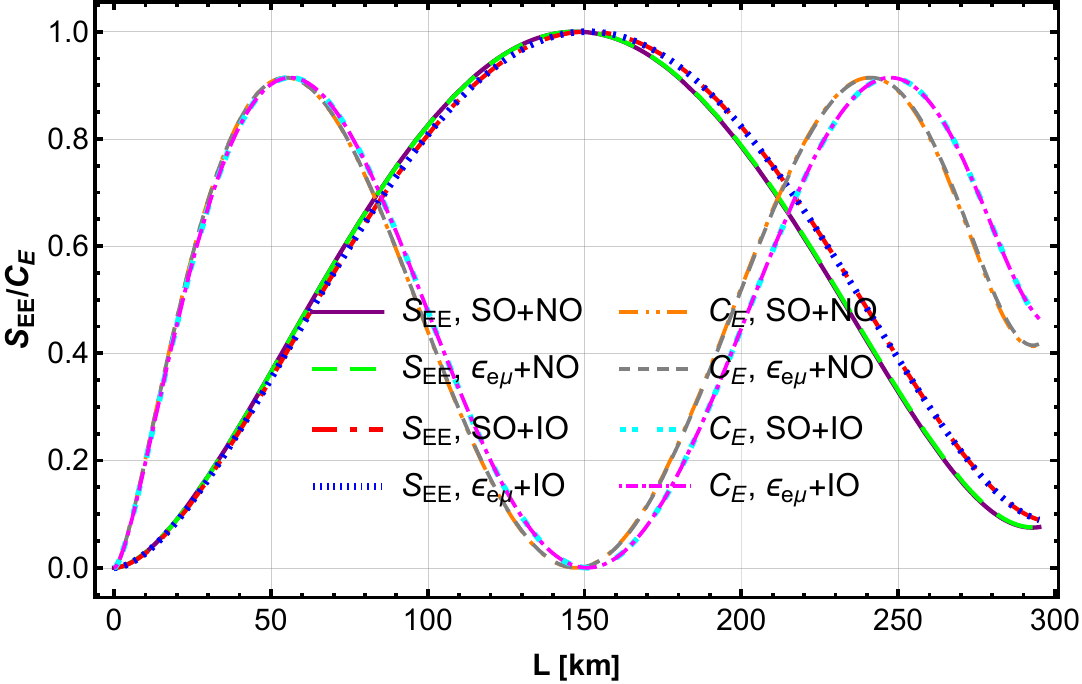}
        \caption{T2K}
        \label{9a_fig:sub1}
    \end{subfigure}
    \hfill
    \begin{subfigure}{0.49\textwidth}
        \centering
        \includegraphics[width=\textwidth]{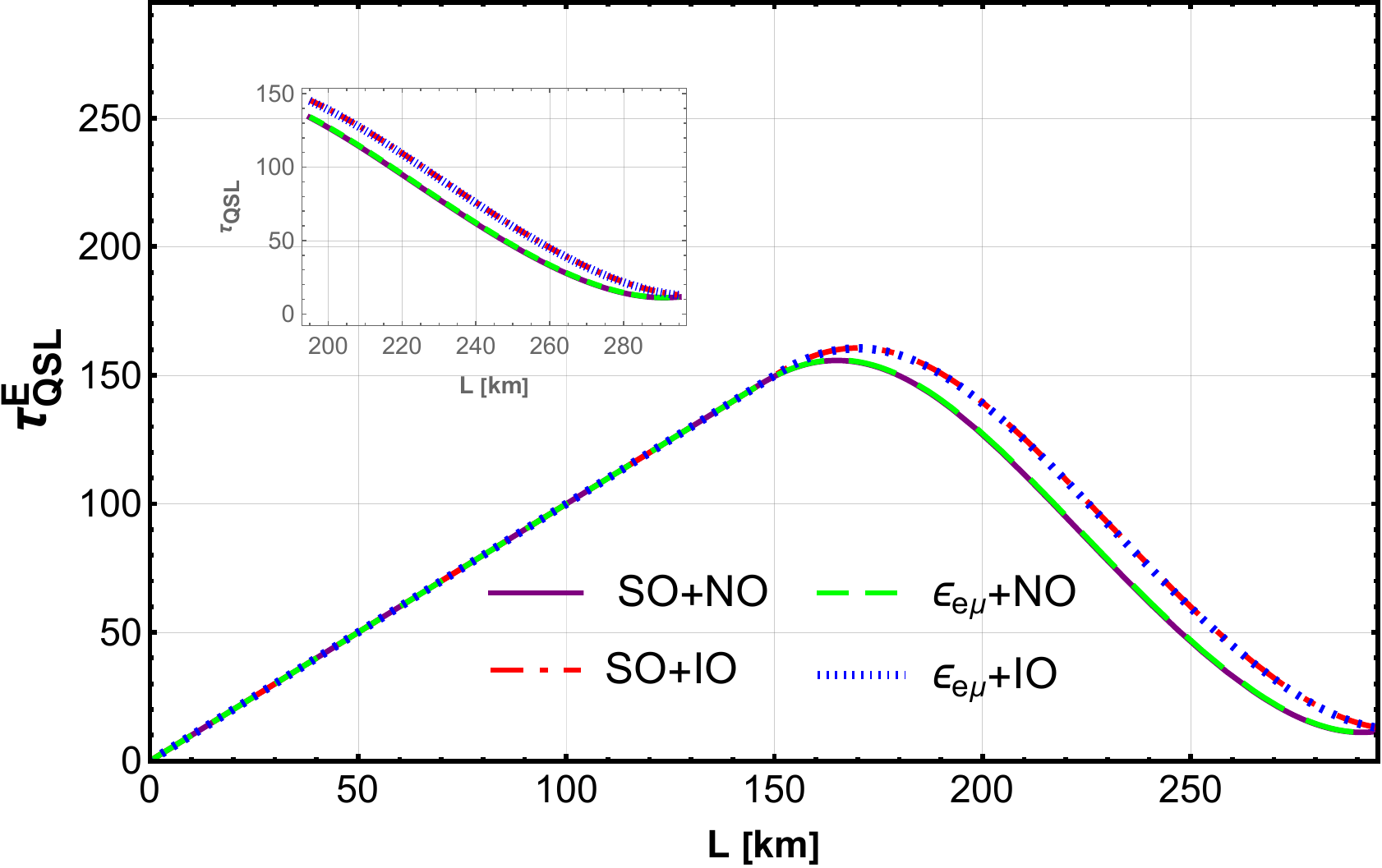}
        \caption{T2K}
        \label{9a_fig:sub2}
    \end{subfigure}
    
    \medskip
    
    \begin{subfigure}{0.49\textwidth}
        \centering
        \includegraphics[width=\textwidth]{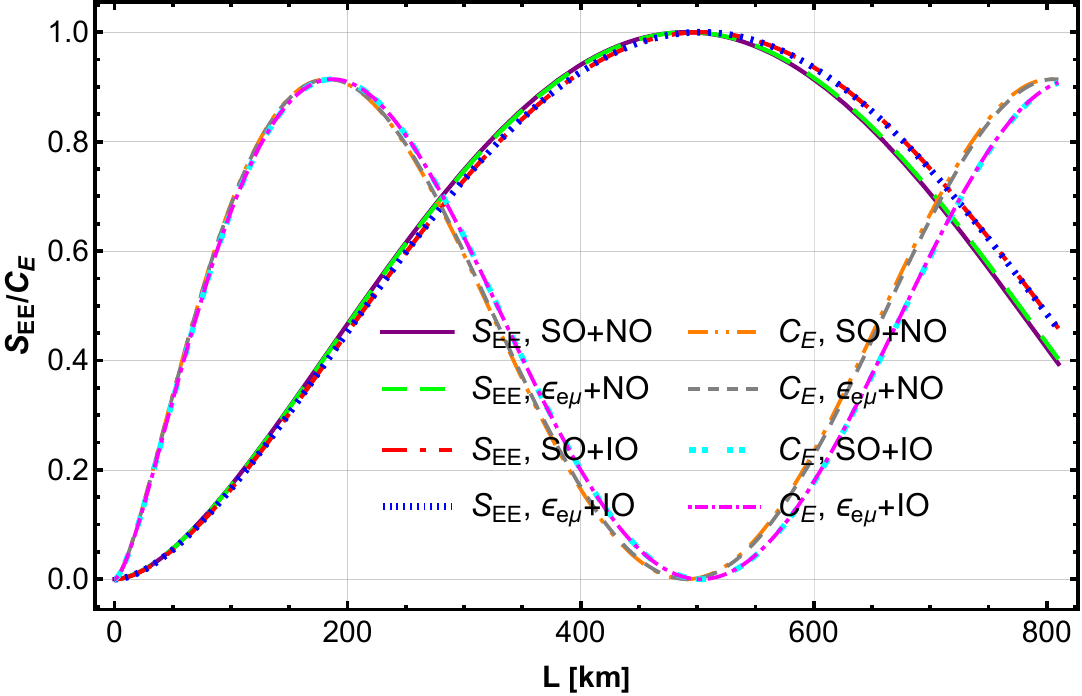}
        \caption{NO$\nu$A}
        \label{9a_fig:sub3}
    \end{subfigure}
    \hfill
    \begin{subfigure}{0.49\textwidth}
        \centering
        \includegraphics[width=\textwidth]{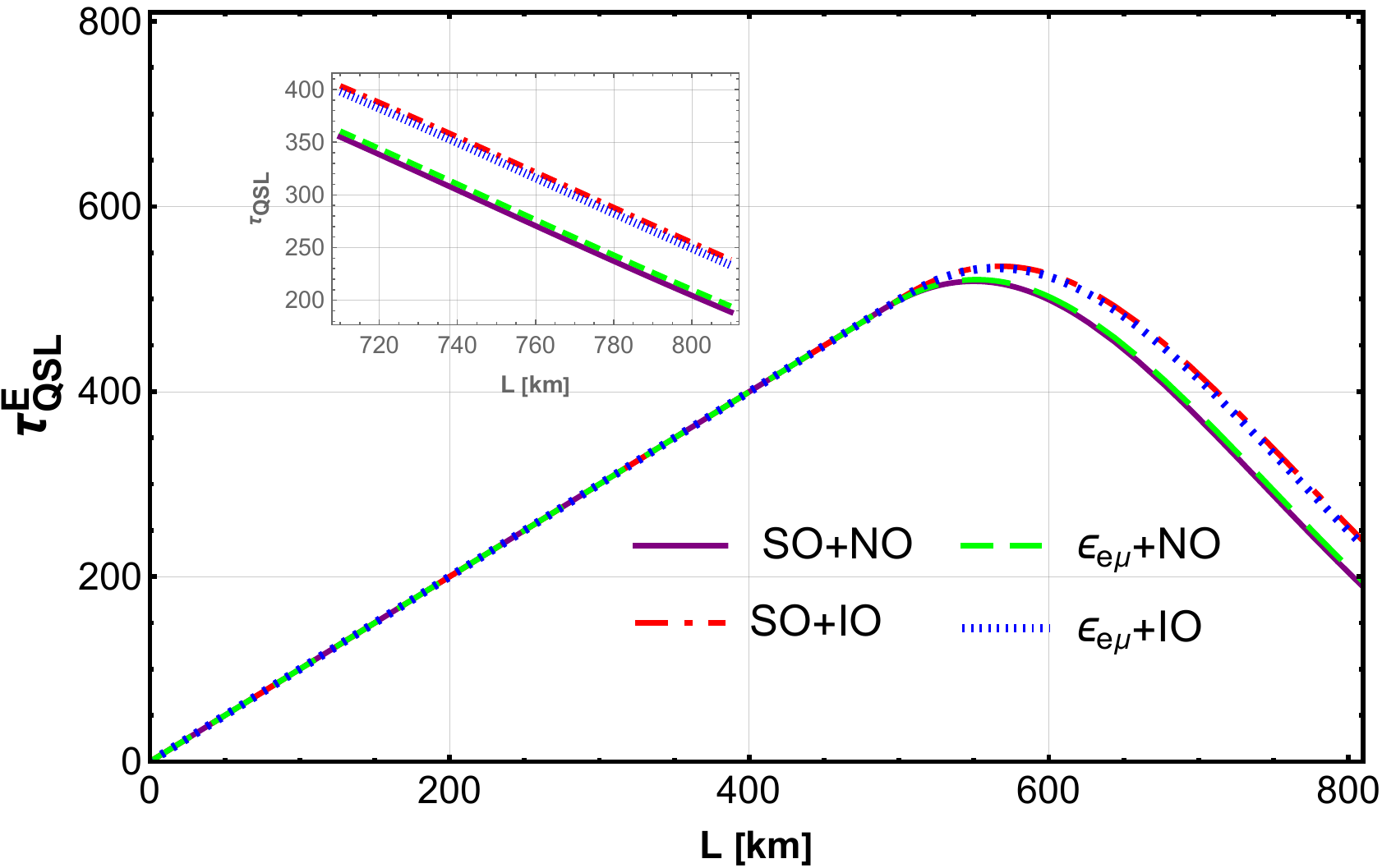}
        \caption{NO$\nu$A}
        \label{9a_fig:sub4}
    \end{subfigure}
    
    \medskip
    
    \begin{subfigure}{0.49\textwidth}
        \centering
        \includegraphics[width=\textwidth]{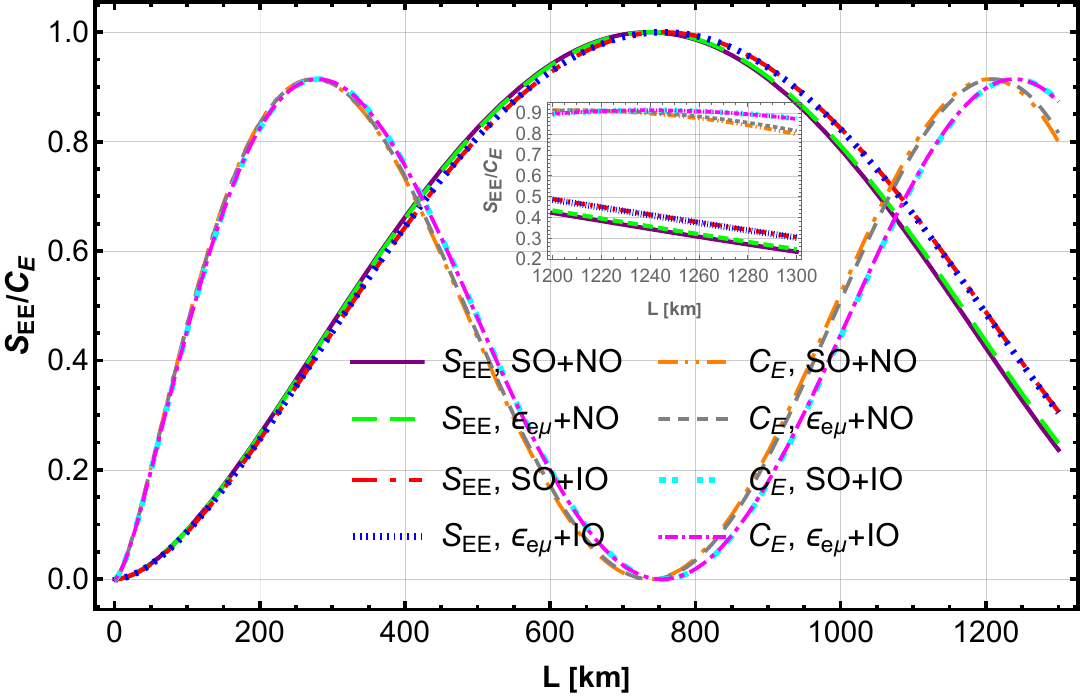}
        \caption{DUNE}
        \label{9a_fig:sub5}
    \end{subfigure}
    \hfill
    \begin{subfigure}{0.49\textwidth}
        \centering
        \includegraphics[width=\textwidth]{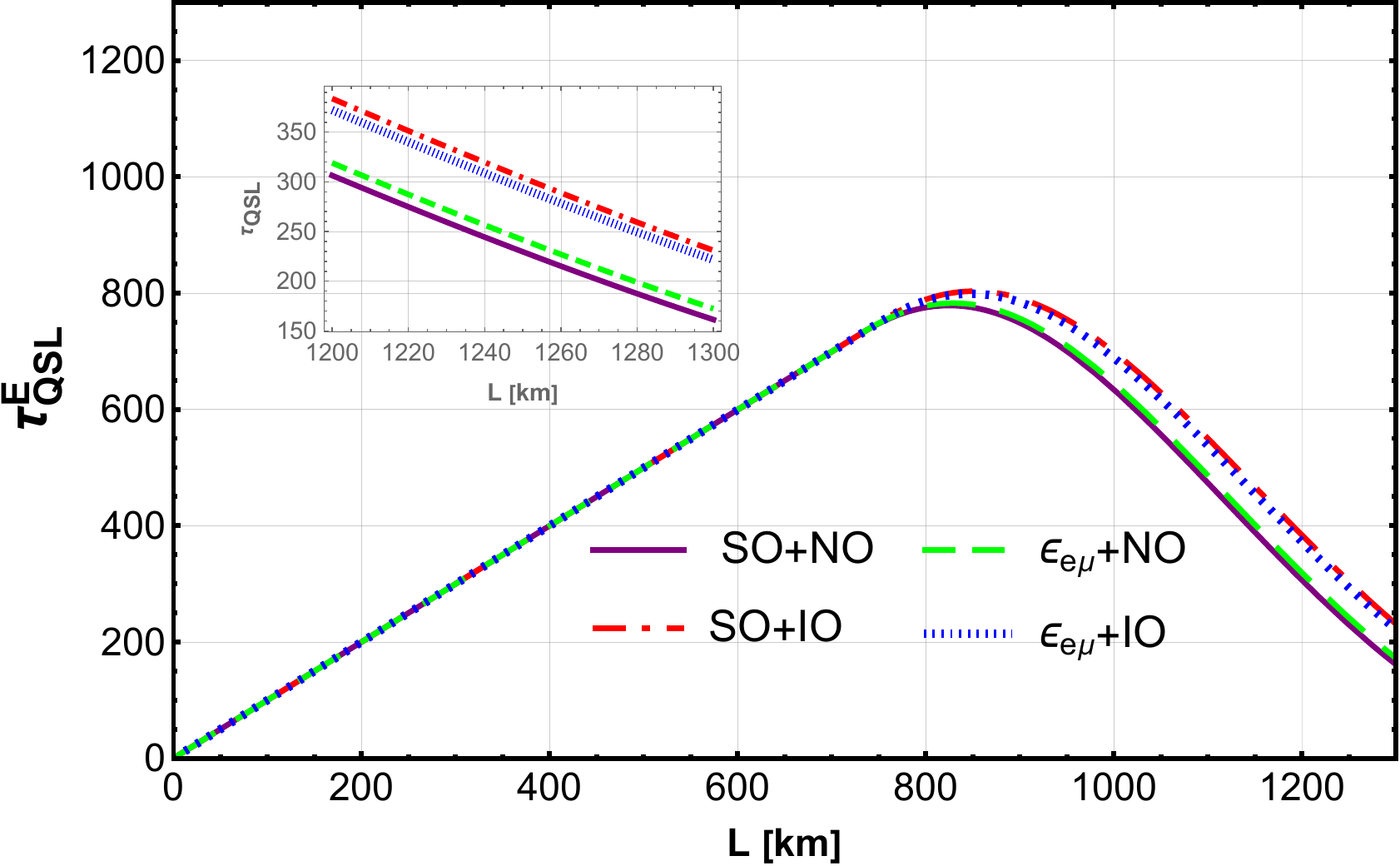}
        \caption{DUNE}
        \label{9a_fig:sub6}
    \end{subfigure}
    \caption{\justifying{Same as Fig\,\ref{fig3}, except that the off-diagonal NSI parameter $\left |\epsilon_{e\mu}\right |$ with complex phase $\phi_{e\mu}$ is used in the analysis, with both the SO and NSI parameters fixed at their best-fit values. (a),(b) T2K, (c),(d) NO$\nu$A, and (e),(f) DUNE }}
    \label{fig9}
\end{figure*}

\begin{figure*}[!htbp]
    \centering
    \begin{subfigure}{0.49\textwidth}
        \centering
        \includegraphics[width=\textwidth]{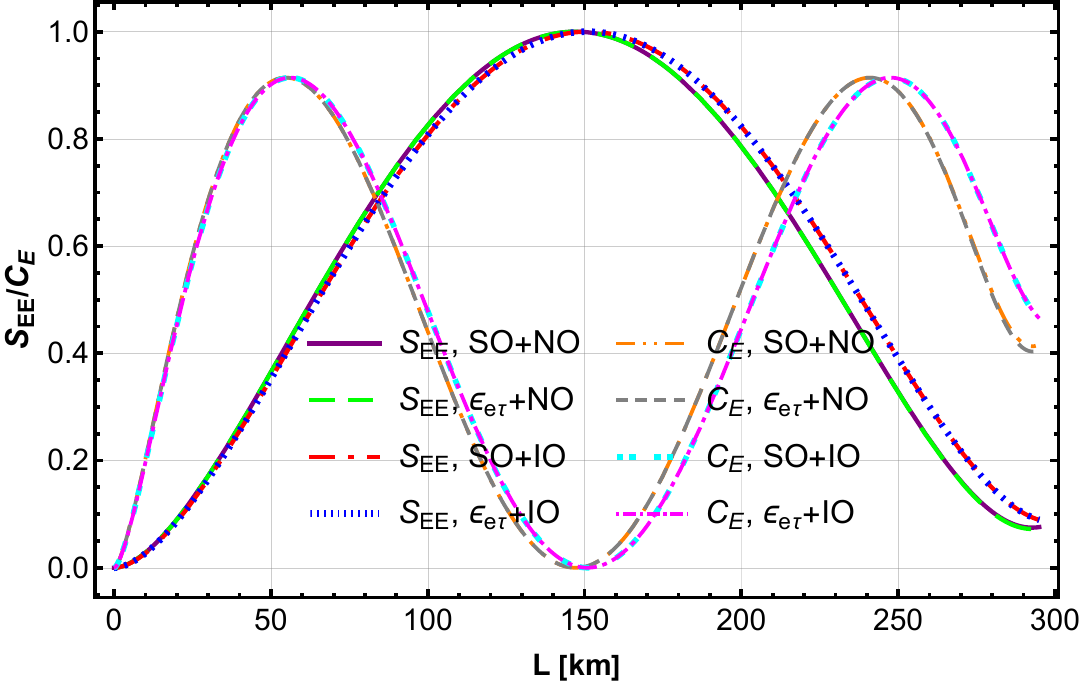}
        \caption{T2K}
        \label{10a_fig:sub1}
    \end{subfigure}
    \hfill
    \begin{subfigure}{0.49\textwidth}
        \centering
        \includegraphics[width=\textwidth]{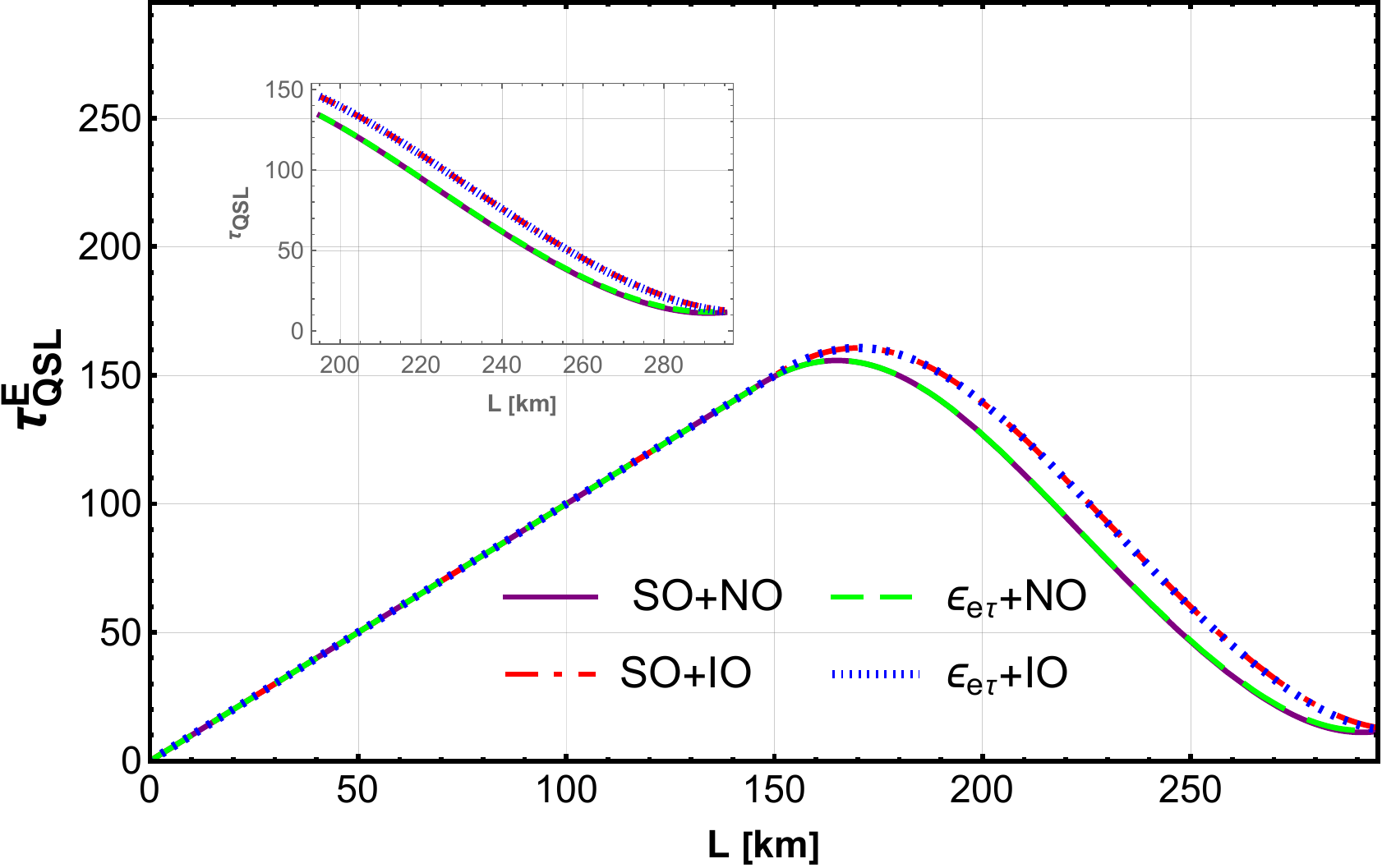}
        \caption{T2K}
        \label{10a_fig:sub2}
    \end{subfigure}
    
    \medskip
    
    \begin{subfigure}{0.49\textwidth}
        \centering
        \includegraphics[width=\textwidth]{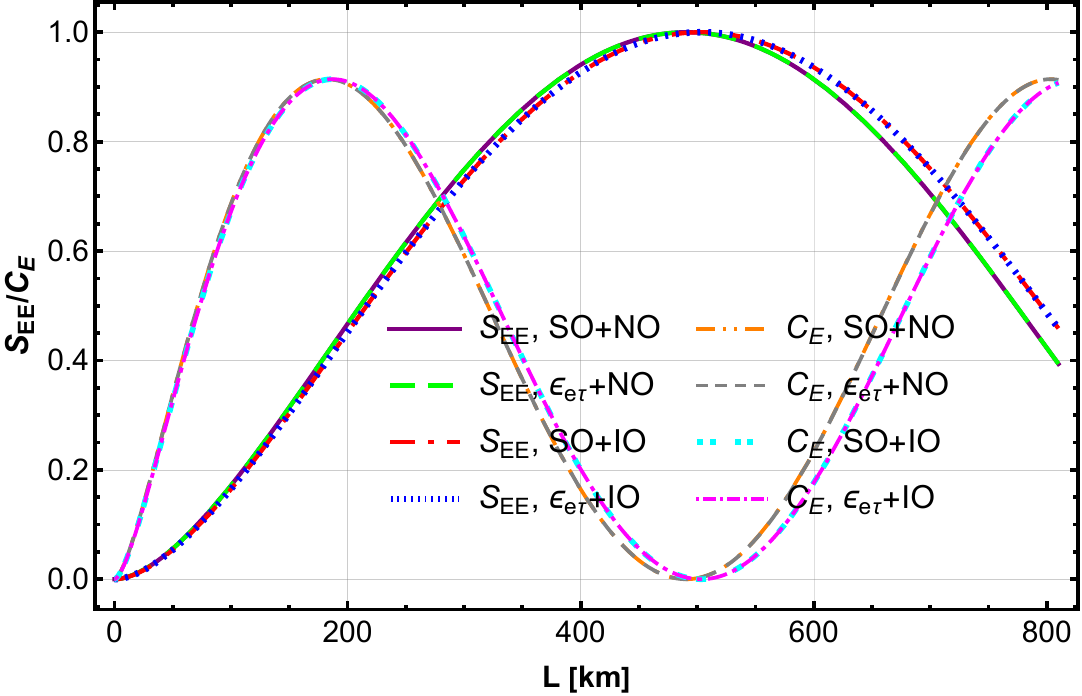}
        \caption{NO$\nu$A}
        \label{10a_fig:sub3}
    \end{subfigure}
    \hfill
    \begin{subfigure}{0.49\textwidth}
        \centering
        \includegraphics[width=\textwidth]{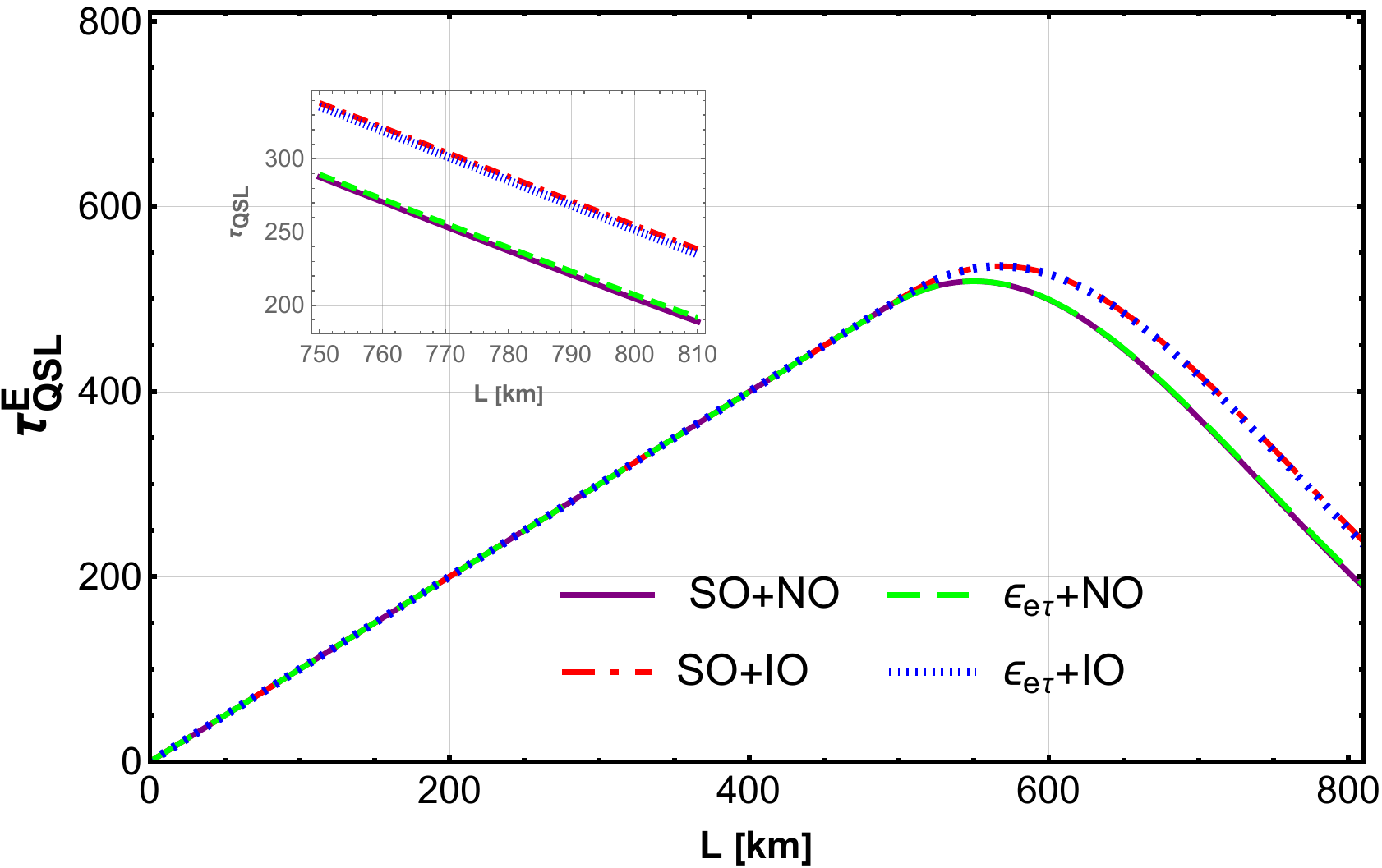}
        \caption{NO$\nu$A}
        \label{10a_fig:sub4}
    \end{subfigure}
    
    \medskip
    
    \begin{subfigure}{0.49\textwidth}
        \centering
        \includegraphics[width=\textwidth]{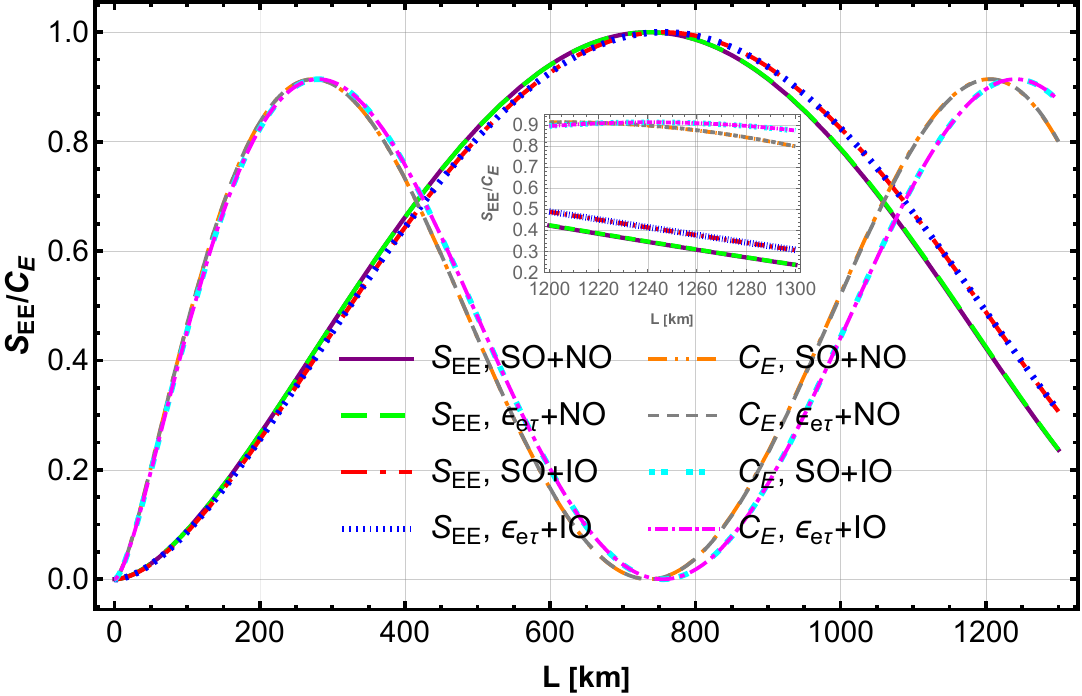}
        \caption{DUNE}
        \label{10a_fig:sub5}
    \end{subfigure}
    \hfill
    \begin{subfigure}{0.49\textwidth}
        \centering
        \includegraphics[width=\textwidth]{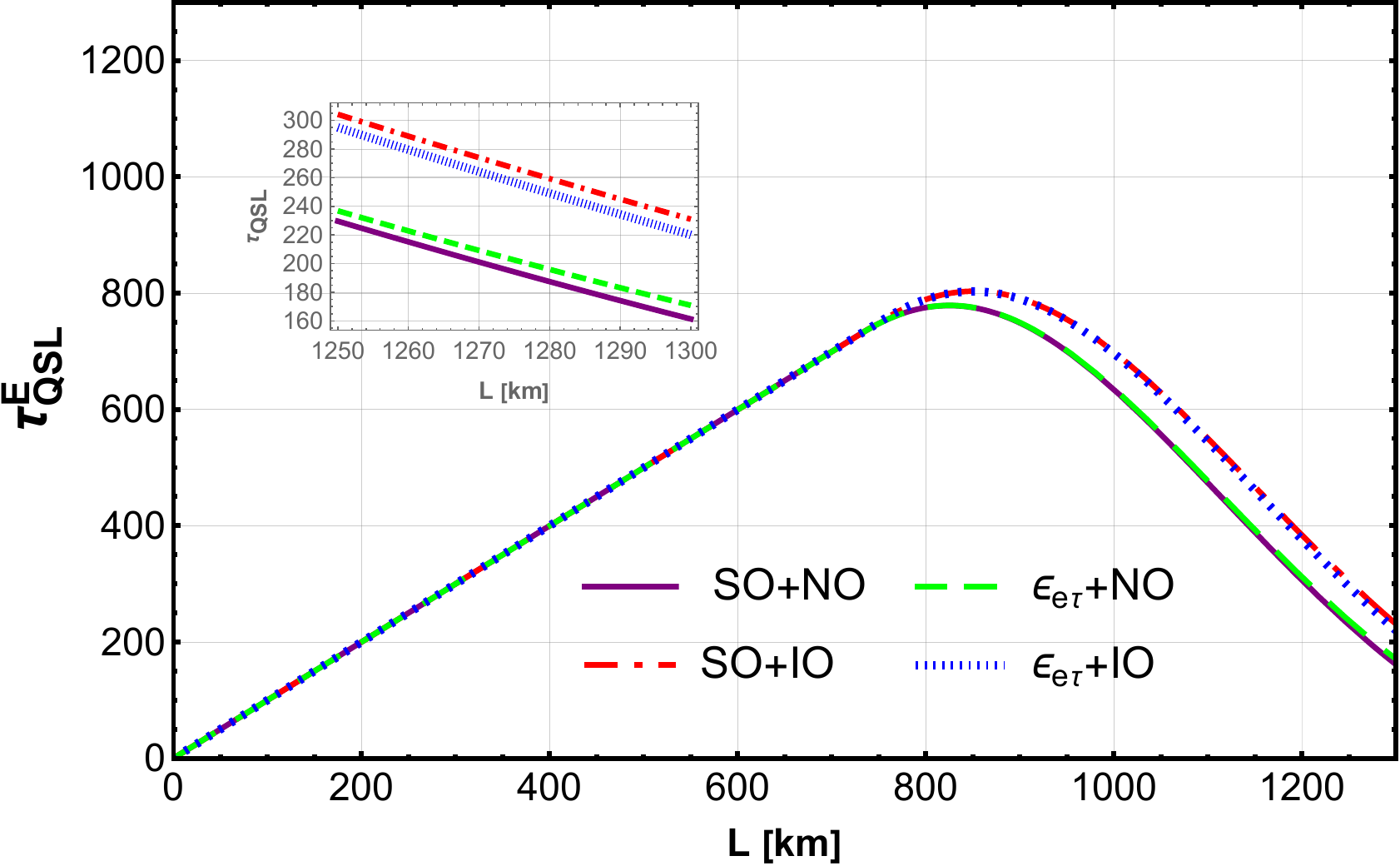}
        \caption{DUNE}
        \label{10a_fig:sub6}
    \end{subfigure}
    \caption{\justifying{Same as Fig\,\ref{fig3}, except that the off-diagonal NSI parameter $\left |\epsilon_{e\tau}\right |$ with complex phase $\phi_{e\tau}$ is used in the analysis, with both the SO and NSI parameters fixed at their best-fit values. (a),(b) T2K, (c),(d) NO$\nu$A, and (e),(f) DUNE }}
    \label{fig10}
\end{figure*}

\begin{figure*}[!htbp]
    \centering
    \begin{subfigure}{0.49\textwidth}
        \centering
        \includegraphics[width=\textwidth]{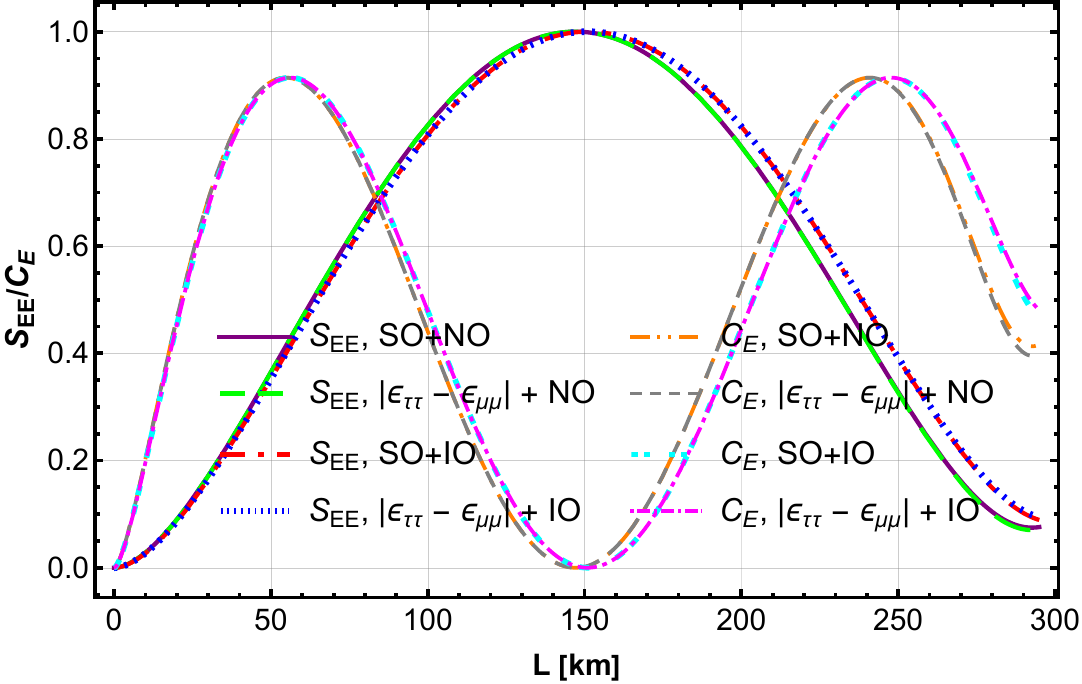}
        \caption{T2K}
        \label{11a_fig:sub1}
    \end{subfigure}
    \hfill
    \begin{subfigure}{0.49\textwidth}
        \centering
        \includegraphics[width=\textwidth]{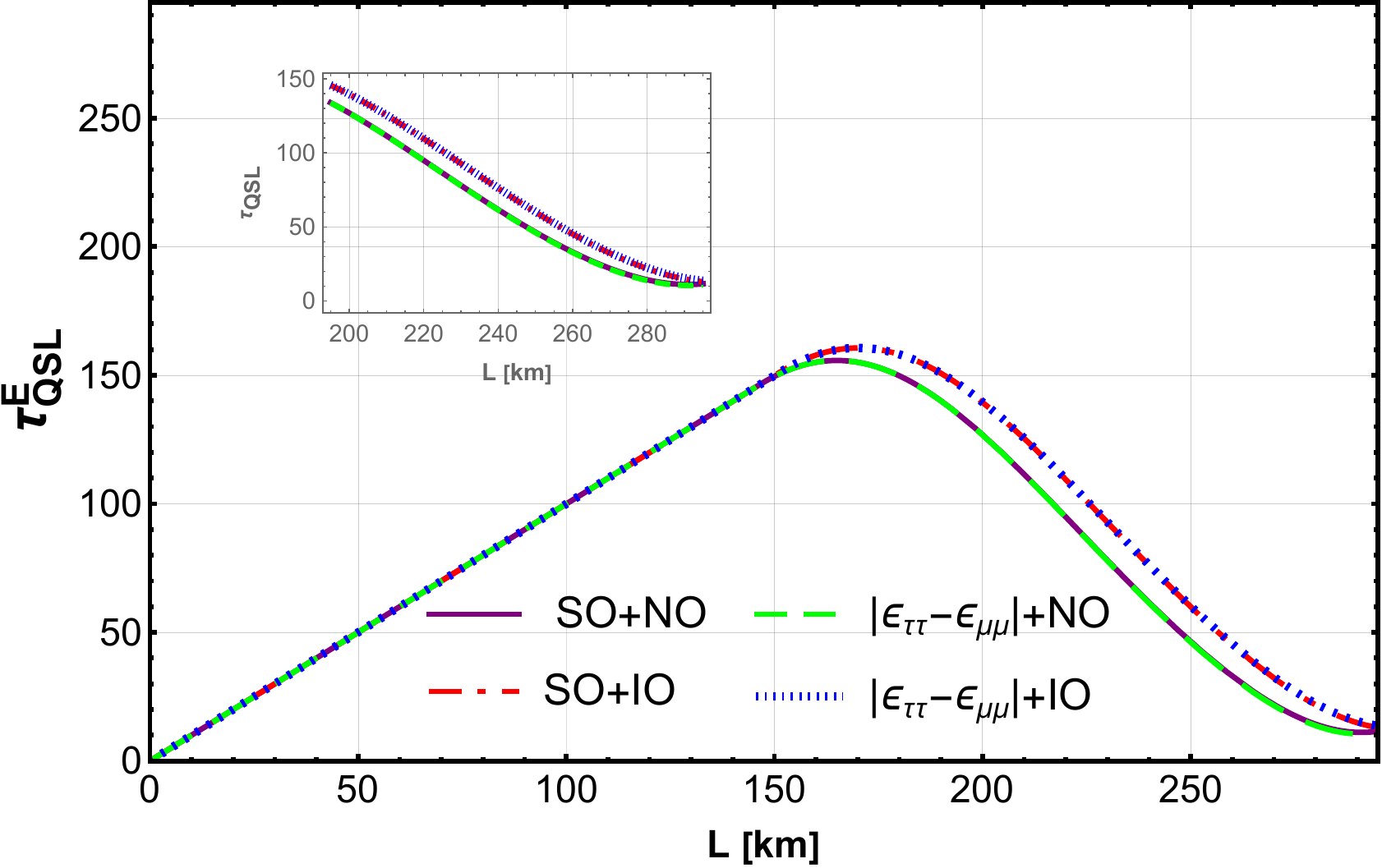}
        \caption{T2K}
        \label{11a_fig:sub2}
    \end{subfigure}
    
    \medskip
    
    \begin{subfigure}{0.49\textwidth}
        \centering
        \includegraphics[width=\textwidth]{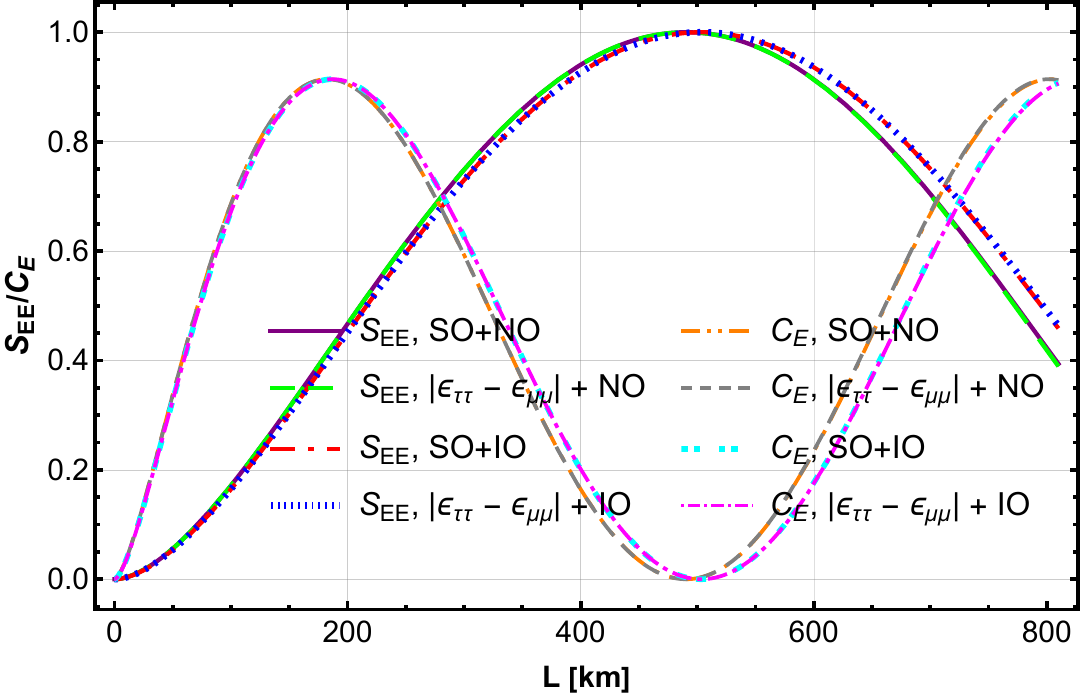}
        \caption{NO$\nu$A}
        \label{11a_fig:sub3}
    \end{subfigure}
    \hfill
    \begin{subfigure}{0.49\textwidth}
        \centering
        \includegraphics[width=\textwidth]{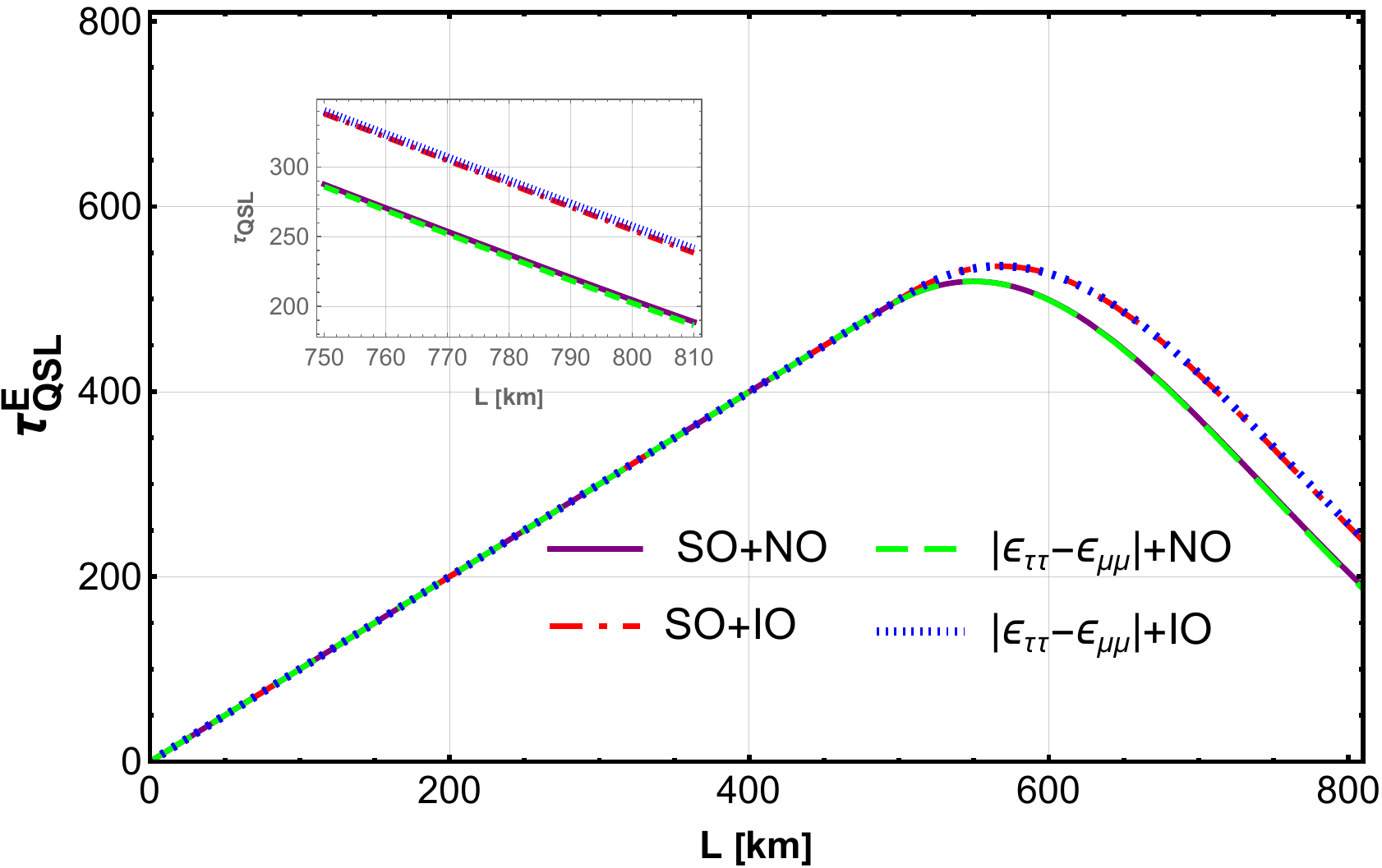}
        \caption{NO$\nu$A}
        \label{11a_fig:sub4}
    \end{subfigure}
    
    \medskip
    
    \begin{subfigure}{0.49\textwidth}
        \centering
        \includegraphics[width=\textwidth]{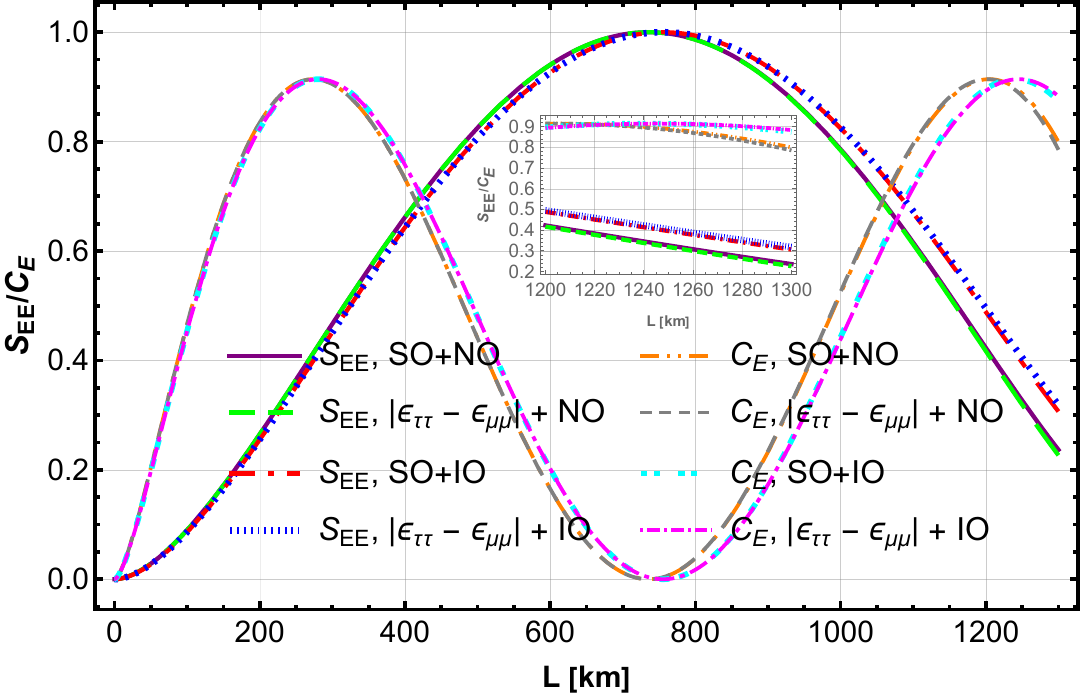}
        \caption{DUNE}
        \label{11a_fig:sub5}
    \end{subfigure}
    \hfill
    \begin{subfigure}{0.49\textwidth}
        \centering
        \includegraphics[width=\textwidth]{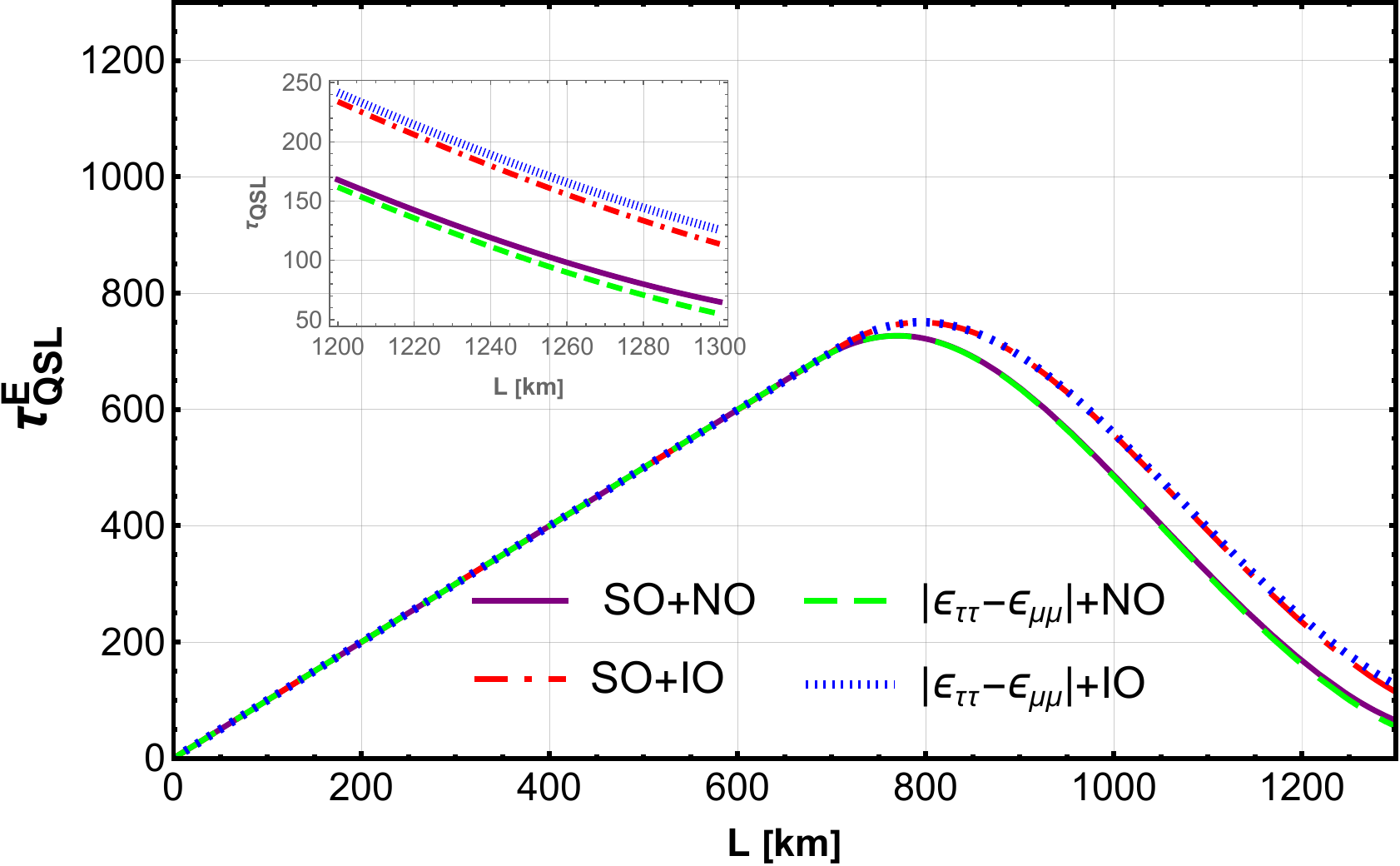}
        \caption{DUNE}
        \label{11a_fig:sub6}
    \end{subfigure}
    \caption{\justifying{Same as Fig\,\ref{fig5}, except that the diagonal NSI parameter $\left |\epsilon_{\tau \tau}-\epsilon_{\mu\mu}\right |$ is used in the analysis, with both the SO and NSI parameters fixed at their best-fit values. (a),(b) T2K, (c),(d) NO$\nu$A, and (e),(f) DUNE }}
    \label{fig11}
\end{figure*}

\end{document}